\documentclass[prd,groupedaddress, showpacs, doublecolumn,nofootinbib]{revtex4} %
\usepackage{graphicx}
\usepackage{dcolumn}
\usepackage{bm}
\usepackage{amssymb}
\usepackage{amsmath}
\usepackage{epsfig}
\usepackage{hhline}
\usepackage{color}
\usepackage{multirow}
\usepackage{cancel}
\usepackage{tikz}
\usetikzlibrary{snakes}
\usepackage{slashed}
\usepackage{epstopdf}
\usepackage[normalem]{ulem}

\def\lapp{\mathrel{\rlap{\raise.5ex\hbox{$<$}}
                    {\lower.5ex\hbox{$\sim$}}}}
\def\gapp{\mathrel{\rlap{\raise.5ex\hbox{$>$}}
                    {\lower.5ex\hbox{$\sim$}}}}

{\newcommand{\lsim}{\mbox{\raisebox{-.6ex}{~$\stackrel{<}{\sim}$~}}}
{

\newcommand{\bmt}{\begin{pmatrix}}
\newcommand{\emt}{\end{pmatrix}}
\newcommand{\ba}{\begin{array}{c}}
\newcommand{\ea}{\end{array}}
\newcommand{\be}{\begin{equation}}
\newcommand{\ee}{\end{equation}}
\newcommand{\bea}{\begin{eqnarray}}
\newcommand{\eea}{\end{eqnarray}}

\newcommand{\bi}{\begin{itemize}}
\newcommand{\ei}{\end{itemize}}

\newcommand{\baz}{\begin{array}{cc}}

\newcommand{\mathsym}[1]{{}}

\newcommand{\bt}{\begin{tabular}}
\newcommand{\et}{\end{tabular}}

\newcommand{\benu}{\begin{enumerate}}
\newcommand{\eenu}{\end{enumerate}}
\newcommand{\bav}{\begin{array}{cccc}}

\begin{document}
\title{\bf Mini Review on Vector-like Leptonic Dark Matter, Neutrino Mass and Collider Signatures}
\author{Subhaditya Bhattacharya}
\email{subhab@iitg.ac.in}
\affiliation{Department of Physics, Indian Institute of Technology Guwahati, North Guwahati, Assam- 781039, India}
\author{Purusottam Ghosh,}
\email{pghoshiitg@gmail.com}
\affiliation{Department of Physics, Indian Institute of Technology Guwahati, North Guwahati, Assam- 781039, India}
\author{Nirakar Sahoo}
\email{nirakar.pintu.sahoo@gmail.com}
\affiliation{Institute of Physics, Sachivalaya Marg, Bhubaneswar, Odisha 751005, India \\ \it and Homi Bhabha National Institute, Training School Complex, Anushakti Nagar, Mumbai 400085, India}
\author{Narendra Sahu}
\email{nsahu@iith.ac.in}
\affiliation{Department of Physics, Indian Institute of Technology,
  Hyderabad, Kandi, Sangareddy, 502285, 
Telangana, India}

\begin{abstract}
We review a class of models in which the Standard Model (SM) is augmented by vector-like leptons: one 
doublet and a singlet, which are odd under an unbroken discrete $Z_2$ symmetry. As a result, the neutral 
component of these additional vector-like leptons are stable and behave as dark matter. We study the phenomenological 
constraints on the model parameters and elucidate the parameter space for relic density, direct detection and collider 
signatures of dark matter. In such models, we further add a scalar triplet of hypercharge two and study the consequences. In 
particular, after electro weak symmetry breaking (EWSB), the triplet scalar gets an induced vacuum expectation value (vev), 
which yield Majorana masses not only to the light neutrinos but also to vector-like leptonic doublet DM. Due to the Majorana mass 
of DM, the $Z$ mediated elastic scattering with nucleon is forbidden and hence allowing the model to survive from stringent 
direct search bound. The DM without scalar triplet lives in a small singlet-doublet leptonic mixing region ($\sin\theta \le 0.1$) 
due to large contribution from singlet component and have small mass difference ($\Delta m \sim 10$ GeV) with charged companion, 
the NLSP (next to lightest stable particle), to aid co-annihilation for yielding correct relic density. Both these 
observations change to certain extent in presence of scalar triplet to aid observability of 
hadronically quiet leptonic final states at LHC, while one may also confirm/rule-out the model through displaced vertex signal 
of NLSP, a characteristic signature of the model in relic density and direct search allowed parameter space.
\end{abstract}

\maketitle

\newpage

\section{Introduction}

The existence of dark matter (DM) in a large scale ($ >$ a few kpc) has been proven irrefutably by various astrophysical 
observations. The prime among them are galaxy rotation curves~\cite{rotation_curve}, gravitational lensing~\cite{bullet_cluster} 
and large scale structure of the Universe. See for a review ~\cite{DM_review1, DM_review2}.
In the recent years the satellite borne experiments like Wilkinson Microwave Anisotropy Probe (WMAP)~\cite{wmap} and PLANCK~\cite{PLANCK} 
precisely determined the relic abundance of DM by measuring the temperature fluctuation in the cosmic microwave background 
radiation (CMBR). All the above said evidences of DM emerge via gravitational interaction in astrophysical environments, 
which make a challenge to probe the existence of DM in a terrestrial laboratory where density of DM is feeble. 

Alternatively one can explore other elementary properties of DM which can be probed at an earth based laboratory. In fact, 
one can assign a weak interaction property to DM through which it can be thermalised in the early Universe at a temperature 
above its mass scale. As the temperature falls due to adiabatic expansion of the Universe, the DM gets decoupled from the 
thermal bath below its mass scale. As a result the ratio: $n_{\rm DM}/s$, where $s$ is the entropy density, remains constant 
and is precisely measured by PLANCK in terms of $\Omega_{\rm DM}h^2=0.1186\pm 0.0020$~\cite{PLANCK}. 

The particle nature of DM, apart from relic abundance, is completely unknown. In particular, the mass, spin and interaction apart 
from gravity {\it etc}. This leads to huge uncertainty in search of DM. Despite this, many experiments are currently operational, that  
uses direct, indirect and collider search methods. Xenon-100~\cite{xenon100}, LUX~\cite{lux}, Xenon-1T~\cite{xenon_1T}, PANDA~\cite{panda} 
are some of the direct DM search experiments which are looking for signature of DM via nuclear scattering, while PAMELA~\cite{pamela}, 
AMS-2~\cite{ams_2}, Fermi gamma ray space telescope~\cite{fermi_gamma_telescope}, IceCube~\cite{ice_cube} {\it etc }
are some of the indirect DM search experiments which are looking for signature of DM in the sky. The search of DM is also 
going on at collider experiments like large Hadron Collider (LHC)~\cite{atlas_paper_DM, cms_paper_DM}. Except some excess in the 
antiparticle flux in the indirect search data, direct and collider searches for DM has produced null observation so far. This in turn put a strong 
bound on the DM mass and coupling with which it can interact to the visible sector of the universe. 

After the Higgs discovery in 2012 at CERN LHC, the standard model (SM) seems to be complete. However, it is found that none of the particles 
in SM can be a candidate of DM, which is required to be stable on cosmological time scale. While neutrinos in SM are stable, but their
relic density is far less than the required DM abundance and is also disfavored from the structure formation. 
Moreover, neutrinos are massless within the SM. A tiny but non-zero neutrino mass generation requires the SM to be 
extended. This opens up the possibility of exploring new models of DM, while explaining non-zero masses for neutrinos in the same 
framework and thus predict a measurable alternation to DM phenomenology, which can be examined in some of the above said experiments. 

In this review we explore the possibility of leptonic DM and non-zero neutrino masses~\cite{Fukuda:2001nk} in a framework beyond the SM. 
The simplest leptonic DM can arise by augmenting the SM with an additional singlet fermion~\cite{singlet_fermion_DM} $\chi$, stabilized by a 
$\mathcal{Z}_2$ symmetry. However, unless we assume the presence of an additional scalar singlet, which acquires vacuum expectation 
value (vev) and thus mixes with SM Higgs, the lepton singlet DM can not possess renormalisable interaction with visible sector. The next 
possibility is to introduce a vector-like leptonic doublet: $N=~(N^0  ~~N^-~)^T$, which is also odd under the $\mathcal{Z}_2$ symmetry. The 
annihilation cross-section of such fermions are large due to $Z$ mediation and correct relic density can only be achieved at 
a very high DM mass. However, the combination of singlet $\chi$ with the doublet vector-like lepton $N$ provides a good candidate of 
DM~\cite{singlet_doublet_DM}, which has been discussed in details here. We discuss the phenomenological constraints on model parameters 
and then elucidate the allowed parameter space of such models from relic density and direct detection constraints. We also indicate 
collider search strategies for such DM. It turns out that the displaced vertex of the charged fermion: $N^\pm$ (also called next to lightest 
stable particle (NLSP)) is a natural signature of such DM model.

In an attempt to address neutrino mass generation in the same framework, we further add a scalar triplet $\Delta$ of hypercharge 2 and 
study the consequences~\cite{bhattacharyaetal}. In particular, after electro weak symmetry breaking (EWSB), the triplet scalar acquires  
an induced vacuum expectation value (vev) which give rise sub-eV Majorana masses to light neutrinos through the Type II Seesaw 
mechanism~\cite{type_II}. The scalar triplet also generates a Majorana mass for the neutral component of the vector-like lepton 
doublet: $N^0$, which constitues a minor component of the DM. Due to Majorana mass of DM, the $Z$ mediated elastic scattering with 
nucleon is forbidden~\cite{inelastic_DM_papers}. As a result, the model survives from the stringent direct search bound. In absence 
of scalar triplet, the singlet-doublet DM is allowed to have only a tiny fraction of doublet component ($\sin\theta \le 0.1$) to 
evade direct search bound. In this limit, due to large contribution from the singlet component, the annihilation cross section 
of the DM becomes smaller than what it requires to achieve correct relic density. To make it up for correct relic density, the DM 
additionally requires to co-annihilate with its charged and heavy neutral companions and therefore requires small mass difference 
($\Delta m \sim 10$ GeV) with charged companion or NLSP. However, in presence of the scalar triplet, we show that both the 
singlet-doublet mixing ($\sin\theta$) and the mass difference with NLSP ($\Delta m$) can be relaxed and larger parameter 
space is available for correct relic density and being compatible with the latest direct detection bounds. Moreover, the scalar 
triplet aid observability of hadronically quiet leptonic final states at LHC while one may also confirm/rule-out the model through 
displaced vertex of NLSP, a characteristic signature of the model in relic density and direct search allowed parameter space.

The paper is arranged as follows. In section-II, we briefly discuss about a vector-like singlet leptonic DM. In section-III, we discuss 
the viable parameter space of a vector-like inert doublet lepton DM. It is shown that an inert lepton doublet DM alone is ruled out 
due to large Z-mediated elastic scattering with the nucleus. However, in presence of a scalar triplet of hyper charge-2, the inert 
lepton doublet DM can be reinstated in a limited parameter space, which we discuss in section-IV. Moreover in section-IV, we discuss 
how the scalar triplet can give rise non-zero masses to active neutrinos via type-II seesaw~\cite{type_II}. In section-V we discuss 
how an appropriate combination of singlet and doublet vector like leptons can give rise a nice possibility of DM in a wide range 
of parameter space. A triplet extension of singlet-doublet leptonic DM is further discussed in section-VI. We discuss collider 
signature of singlet-doublet leptonic DM in presence of a scalar triplet in section-VII and conclude in section-VIII. We provide 
some vertices of inert lepton doublet (ILD) DM in presence of scalar triplet in Appendix-A.     

\section{Vector-like leptonic singlet dark matter}
\label{sec:VLLS}
A simplest possibility to explain DM content of the present Universe is to augment the SM by adding a 
vector-like singlet lepton $\chi$. The stability of $\chi$ can be ensured by imposing an additional discrete $Z_2$ symmetry, 
under which $\chi$ is odd while all other particles are even. In fact, a singlet DM has been discussed extensively 
in the literature~\cite{singlet_fermion_DM}. Here we briefly recapitulate the main features to show the allowed parameter space 
by observed relic density and latest constraint from direct detection experiments. 

The Lagrangian describing the singlet leptonic DM $\chi$ can be given as:
\begin{eqnarray}\label{lag:lagVFSiglet}
\mathcal{L}
= \overline{\chi}~(i\gamma^\mu \partial_{\mu}-m_{\chi})~\chi - \frac{1}{\Lambda} \Big(H^\dagger H - \frac{v^2}{2}\Big)\overline{\chi}~\chi.
\end{eqnarray}

\begin{figure}[htb!]
$$
 \includegraphics[height=5.0cm]{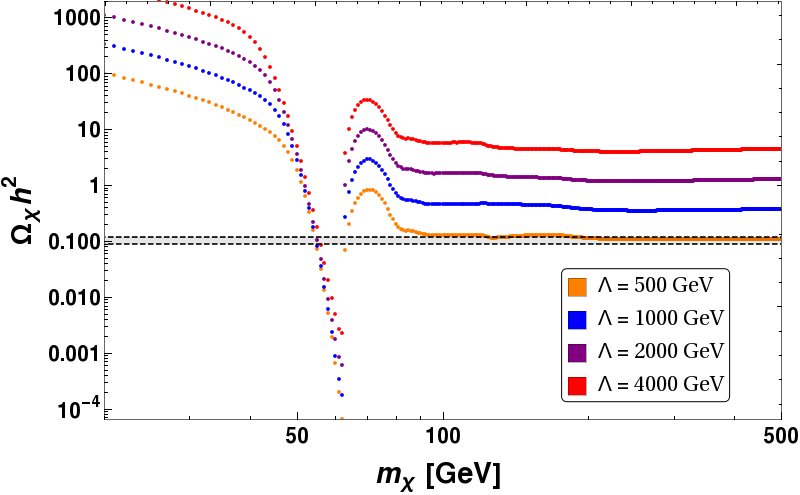} ~~
  \includegraphics[height=5.0cm]{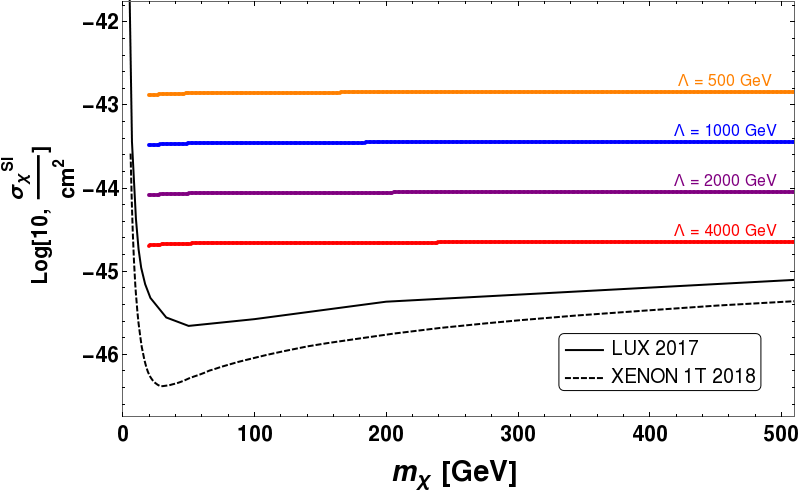} 
$$
 \caption{[Left] Relic density as a function of singlet DM mass, $m_\chi$  for different values of $\Lambda$ mentioned in the figure inset. 
[Right] SI DD cross-section vs DM mass, $m_{\chi}$ for different values of $\Lambda$. }
 \label{fig:omega_mN_fN}
\end{figure}

Notice that the Lagrangian introduces two new parameters: DM mass, $m_\chi$ and the new physics scale $\Lambda$ connecting DM to the SM through 
effective dimension five operator, on which the 
DM phenomenology depends. In the early Universe, $\chi$ freezes out via the interaction $\overline{\chi} \chi \to {\rm SM~particles}$ 
to give rise a net relic density that we observe today. We use {\bf micrOmegas}~\cite{micromega} to calculate the relic density as well as 
spin independent elastic cross-section with nucleon of $\chi$. In Fig.~\ref{fig:omega_mN_fN}, we show relic 
abundance (left-panel) and spin independent direct detection (SIDD) cross section (right panel) as a function of DM mass ($m_\chi$) for 
different values $\Lambda \sim \{500-4000\}$ GeV \footnote{The scale $\Lambda$ is a priori unknown and should be validated from experimental constraints. 
In effective theory consideration, we ensure that $m_\chi < \Lambda$.}. We observe that the constraint on SIDD cross section favors large values of $\Lambda$, while 
large $\Lambda$ values yield over abundance of DM. We also note in the right panel figure \ref{fig:omega_mN_fN}, that the SIDD cross section is very less 
sensitive to DM mass. This is because the direct search cross-section is proportional to the effective DM-nucleon reduced mass square ($\mu_r=\frac{m_N m_\chi}{m_N+m_\chi}$) (see Eq. 58 or Eq. 59), where $m_N<m_\chi$ yields a mild dependence on DM mass. This feature is observed in rest of the analysis as well. 

Therefore, a singlet leptonic DM alone is almost ruled out. However, the dark sector of the Universe may not be simple as in the case of singlet leptonic DM. In the following we discuss a 
few more models with larger number of parameters, yet predictive. We end this section by noting that one can think of a pseudoscalar propagator 
to yield an effective DM-SM interaction of the form $(\bar{\chi}\gamma_5\chi) (H^\dagger H)/\Lambda$. In this case, the relic density and direct search cross-sections become velocity 
dependent (see for example in \cite{Ghorbani:2014qpa}). Please also see section V below Eq. 35 for more details.   
%

\section{Inert lepton doublet dark matter}

Let us assume that the dark sector is composed of a vector-like lepton doublet: $N=~(N^0  ~~N^-~)^T$, which is odd under 
an extended $\mathcal{Z}_2$ symmetry (hence called inert lepton doublet (ILD)), while all the Standard Model (SM) fields are even. 
As a result the neutral component of the ILD $N$ is stable. The quantum numbers of dark sector fields and that of SM Higgs under 
the SM gauge group, augmented by a $\mathcal{Z}_2$ symmetry, are given in Table \ref{tab:tab0}. We will check if $N^0$ can be a 
viable candidate of DM with correct relic abundance while satisfying the direct detection constraints from the null 
observation at various terrestrial laboratories.  

\begin{table}[htb!]
\begin{center}
\begin{tabular}{|p{3cm}|p{6cm}|}
 \hline
  \hspace{0.2cm}  Fields  &  $\hspace{0.8cm} SU(3)_C \times SU(2)_L \times U(1)_Y\times \mathcal{Z}_2 $  \\
 \hline
  \hspace{0.1cm}$N=\left(\begin{matrix}
 N^0 \\  N^- 
\end{matrix}\right)$ & \hspace{1cm} 1 \hspace{1.3cm} 2 \hspace{1cm} -1 \hspace{0.5cm} - \\
 \hline
 \hline
 \hspace{0.1cm}$H=\left(\begin{matrix}
 H^+ \\  H^0 
\end{matrix}\right)$  & \hspace{1cm} 1 \hspace{1.3cm} 2 \hspace{1cm} 1 \hspace{0.5cm} + \\
 \hline
\end{tabular}
\end{center}
\caption{Quantum numbers of additional dark sector fermion and SM Higgs under $ \mathcal{G}\equiv SU(3)_C \times SU(2)_L \times U(1)_Y\times \mathcal{Z}_2 $ .}
  \label{tab:tab0}
\end{table}

The Lagrangian of the model is given as:
\bea\label{lag:IFD}
\mathcal{L}^{IL} &=& \overline{N}~[i\gamma^{\mu}(\partial_{\mu} - i g \frac{\sigma^a}{2}W_{\mu}^a - i g^{\prime}\frac{Y}{2}B_{\mu})-m_N]~N\,.
\eea
Thus the only new parameter introduced in the above Lagrangian is the mass of $N$, {\it i.e.} $m_N$. Expanding the covariant derivative of 
the above Lagrangian $\mathcal{L}^{IL}$, we get the interaction terms of $N^0$ and $N^\pm$ with the SM gauge bosons as: 
\bea
\mathcal{L}^{IL}_{int}&=&\overline{N}i\gamma^\mu(-i g \frac{\sigma^a}{2}W_{\mu}^a + i \frac{g^{\prime}}{2}B_{\mu})N  \nonumber \\
&=& \Big(\frac{e_0}{2 \sin\theta_W \cos\theta_W}\Big) \overline{N^0} \gamma^{\mu} Z_{\mu} N^0 + \frac{e_0}{\sqrt2 \sin\theta_W}\overline{N^0}\gamma^{\mu}W_{\mu}^+N^-
 +\frac{e_0}{\sqrt2\sin\theta_W}{N^+}\gamma^{\mu}W_{\mu}^-N^0 \nonumber \\
&&- e_0 {N^+}\gamma^{\mu}A_{\mu}N^-  - \Big(\frac{e_0}{2 \sin\theta_W \cos\theta_W}\Big) \cos2\theta_W {N^+}\gamma^{\mu}Z_{\mu}N^-  . 
\eea
where $g=e_0/\sin\theta_W$ and $g'=e_0/\cos\theta_W$ with $e_0$ being the electromagnetic coupling constant and $\theta_W$ being the Weinberg angle. 

Since $N$ is a doublet under $SU(2)_L$, it can contribute to invisible $Z$-decay width if its mass is less than 45 GeV which is strongly constrained. 
Therefore, in our analysis we will assume $m_N > 45$ GeV.   

\subsection{Relic abundance of ILD Dark Matter}\label{ILD_relic}
\begin{figure}[htb!]
\begin{center}
    \begin{tikzpicture}[line width=0.5 pt, scale=0.9]
        \draw[solid] (-3,1.0)--(-1.0,1.0);
        \draw[solid] (-3,-1.0)--(-1.0,-1.0);
        \draw[solid] (-1.0,1.0)--(-1.0,-1.0);
        \draw[solid] (-1.0,1.0)--(1.0,1.0);
        \draw[solid] (-1.0,-1.0)--(1.0,-1.0);
        \node at (-3.4,1.0) {$\overline{N^0}$};
        \node at (-3.4,-1.0) {$N^-$};
        \node [right] at (-1.05,0.0) {$N^-$};
        \node at (1.6,1.0) {$W^-$};
        \node at (1.6,-1.0) {$Z/A$};
        \draw[solid] (5.0,1.0)--(6.5,0.0);
        \draw[solid] (5.0,-1.0)--(6.5,0.0);
        \draw[snake] (6.5,0.0)--(8.5,0.0);
        \draw[solid] (8.5,0.0)--(10.0,1.0);
        \draw[solid] (8.5,0.0)--(10.0,-1.0);
        \node at (4.7,1.0) {$\overline{N^0}$};
        \node at (4.7,-1.0) {$N^-$};
        \node [above] at (7.4,0.05) {$W$};
        \node at (11.1,1.0) {$f/h/W/W$};
        \node at (11.1,-1.0) {$\overline{f^\prime}/W/A/Z$};
     \end{tikzpicture}
 \end{center}
\begin{center}
    \begin{tikzpicture}[line width=0.5 pt, scale=0.9]
        \draw[solid] (-3,1.0)--(-1.0,1.0);
        \draw[solid] (-3,-1.0)--(-1.0,-1.0);
        \draw[solid](-1.0,1.0)--(-1.0,-1.0);
        \draw[solid] (-1.0,1.0)--(1.0,1.0);
        \draw[solid] (-1.0,-1.0)--(1.0,-1.0);
        \node at (-3.4,1.0) {$\overline{N^0}$};
        \node at (-3.4,-1.0) {$N^-$};
        \node [right] at (-1.05,0.0) {$N^0$};
        \node at (1.5,1.0) {$Z$};
        \node at (1.5,-1.0) {$W^-$};
     \end{tikzpicture}
 \end{center}
\caption{ Co-annihilation processes of DM $N^0$ with its charged partner $N^\pm$ to SM particles. }
\label{fd:coann_ILD}
 \end{figure}
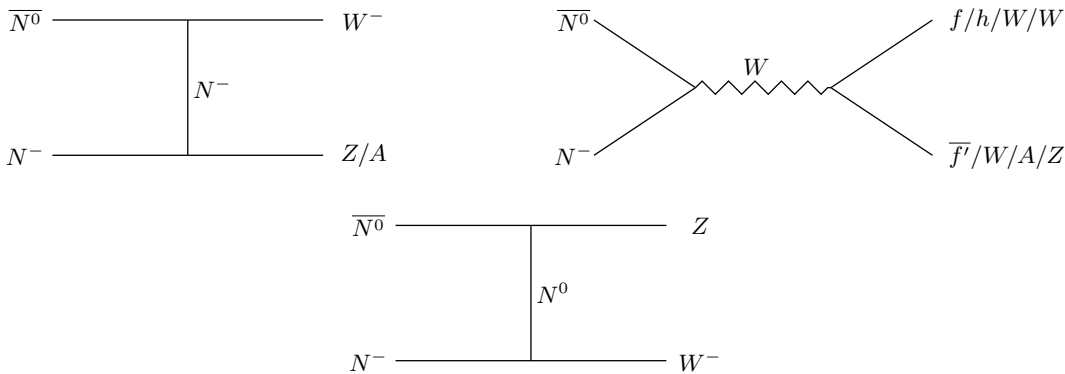

%
The number changing annihilation and co-annihilation processes which control freeze-out and hence relic density of DM $N^0$ are shown 
in Figs. \ref{fd:ann_ILD}, \ref{fd:ann_ILD_charge} and \ref{fd:coann_ILD}.

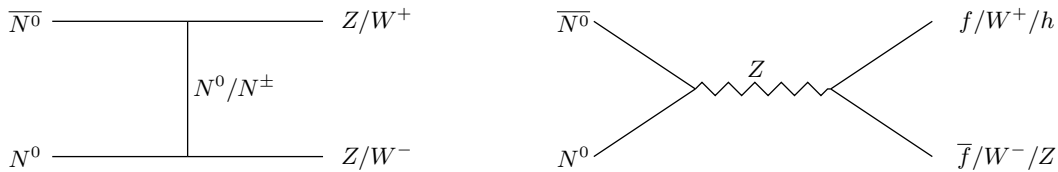
\begin{figure}[htb!]
\begin{center}
    \begin{tikzpicture}[line width=0.5 pt, scale=0.9]
        \draw[solid] (-3,1.0)--(-1.0,1.0);
        \draw[solid] (-3,-1.0)--(-1.0,-1.0);
        \draw[solid](-1.0,1.0)--(-1.0,-1.0);
        \draw[solid] (-1.0,1.0)--(1.0,1.0);
        \draw[solid] (-1.0,-1.0)--(1.0,-1.0);
        \node at (-3.4,1.0) {$\overline{N^0}$};
        \node at (-3.4,-1.0) {$N^0$};
        \node [right] at (-1.05,0.0) {$N^0/ N^\pm $};
        \node at (1.8,1.0) {$Z / W^+$};
        \node at (1.8,-1.0) {$Z / W^-$};
        \draw[solid] (5.0,1.0)--(6.5,0.0);
        \draw[solid] (5.0,-1.0)--(6.5,0.0);
        \draw[snake] (6.5,0.0)--(8.5,0.0);
        \draw[solid] (8.5,0.0)--(10.0,1.0);
        \draw[solid] (8.5,0.0)--(10.0,-1.0);
        \node at (4.7,1.0) {$\overline{N^0}$};
        \node at (4.7,-1.0) {$N^0$};
        \node [above] at (7.4,0.05) {$Z$};
        \node at (11.1,1.0) {$f/W^+/h$};
        \node at (11.1,-1.0) {$\overline{f}/W^-/Z$};
     \end{tikzpicture}
 \end{center}
\caption{Annihilation of ILD DM to SM particles . }
\label{fd:ann_ILD}
 \end{figure}
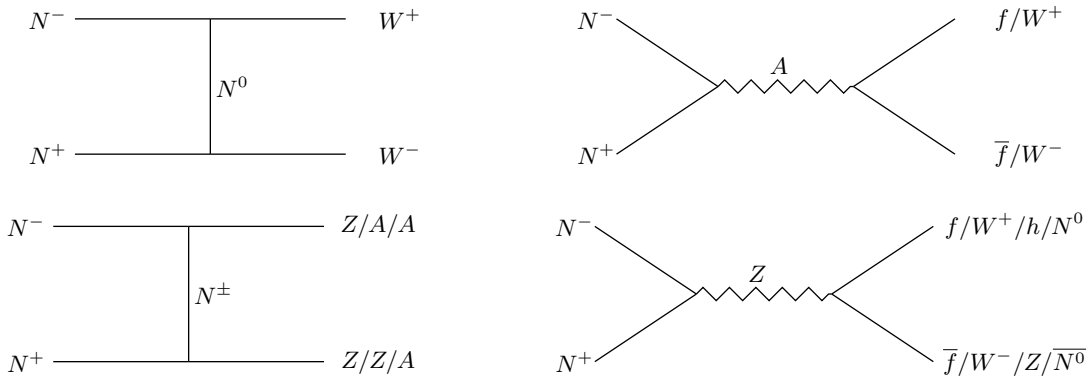
\begin{figure}[htb!]
\begin{center}
    \begin{tikzpicture}[line width=0.5 pt, scale=0.9]
        \draw[solid] (-3,1.0)--(-1.0,1.0);
        \draw[solid] (-3,-1.0)--(-1.0,-1.0);
        \draw[solid] (-1.0,1.0)--(-1.0,-1.0);
        \draw[solid] (-1.0,1.0)--(1.0,1.0);
        \draw[solid] (-1.0,-1.0)--(1.0,-1.0);
        \node at (-3.4,1.0) {$N^-$};
        \node at (-3.4,-1.0) {$N^+$};
        \node [right] at (-1.05,0.0) {$N^0$};
        \node at (1.8,1.0) {$W^+$};
        \node at (1.8,-1.0) {$W^-$};
        \draw[solid] (5.0,1.0)--(6.5,0.0);
        \draw[solid] (5.0,-1.0)--(6.5,0.0);
        \draw[snake] (6.5,0.0)--(8.5,0.0);
        \draw[solid] (8.5,0.0)--(10.0,1.0);
        \draw[solid] (8.5,0.0)--(10.0,-1.0);
        \node at (4.7,1.0) {$N^-$};
        \node at (4.7,-1.0) {$N^+$};
        \node [above] at (7.4,0.05) {$A$};
        \node at (11.1,1.0) {$f/W^+$};
        \node at (11.1,-1.0) {$\overline{f}/W^-$};
     \end{tikzpicture}
 \end{center}
\begin{center}
    \begin{tikzpicture}[line width=0.5 pt, scale=0.9]
        \draw[solid] (-3,1.0)--(-1.0,1.0);
        \draw[solid] (-3,-1.0)--(-1.0,-1.0);
        \draw[solid](-1.0,1.0)--(-1.0,-1.0);
        \draw[solid] (-1.0,1.0)--(1.0,1.0);
        \draw[solid] (-1.0,-1.0)--(1.0,-1.0);
        \node at (-3.4,1.0) {$N^-$};
        \node at (-3.4,-1.0) {$N^+$};
        \node [right] at (-1.05,0.0) {$N^\pm$};
        \node at (1.8,1.0) {$ Z / A / A $};
        \node at (1.8,-1.0) {$ Z / Z / A$};
        \draw[solid] (5.0,1.0)--(6.5,0.0);
        \draw[solid] (5.0,-1.0)--(6.5,0.0);
        \draw[snake] (6.5,0.0)--(8.5,0.0);
        \draw[solid] (8.5,0.0)--(10.0,1.0);
        \draw[solid] (8.5,0.0)--(10.0,-1.0);
        \node at (4.7,1.0) {$N^-$};
        \node at (4.7,-1.0) {$N^+$};
        \node [above] at (7.4,0.05) {$Z$};
        \node at (11.2,1.0) {$f/W^+/h  /N^0$};
        \node at (11.2,-1.0) {$\overline{f}/W^-/Z /\overline{N^0}$};
     \end{tikzpicture}
 \end{center}
\caption{Annihilation of charged partner of ILD DM to SM particles which contributes as co-annihilation with ILD DM.}
\label{fd:ann_ILD_charge}
 \end{figure}

To estimate the relic density of DM in this framework one needs to solve the relevant Boltzmann equation: 
 \bea
 \frac{dn_{N^0}}{dt} + 3 H n_{N^0}& =& -{\langle \sigma v\rangle}_{N^0 \overline{N^0} \rightarrow SM SM} \Big(n_{N^0}^2-{n_{N^0}^{eq}}^2\Big) \nonumber \\
 && - {\langle \sigma v\rangle}_{N^0 N^\pm \rightarrow SM SM} \Big(n_{N^0} n_{N^\pm}  -{n_{N^0}^{eq}} {n_{N^\pm}^{eq}}  \Big) .
 \eea

 \begin{figure}[htb!]
$$
 \includegraphics[height=6.0cm]{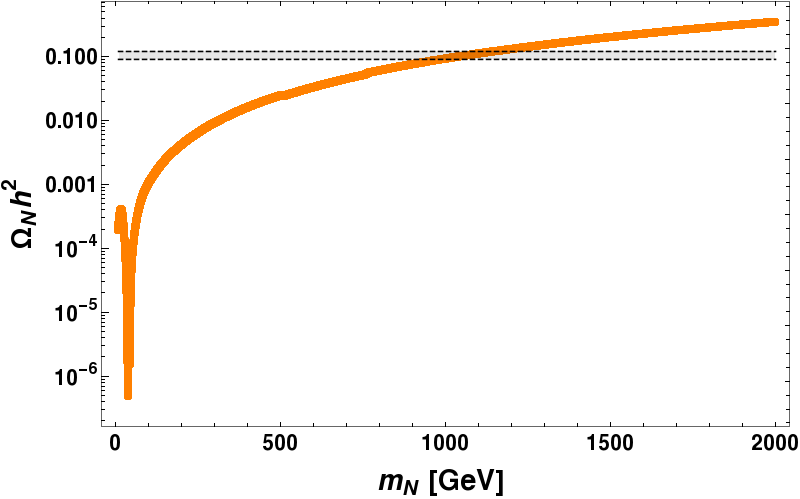} 
$$
 \caption{Variation of relic density of DM $N^0$ with mass $m_N$. The gray patch corresponds to relic density allowed limit from 
PLANCK: $0.1166 \leq \Omega h^2 \leq 0.1206$.}
\label{fig:relic_ILD}
\end{figure}
 
To find the relic density of ILD DM, here we adopted {\bf micrOmegas}~\cite{micromega} and implemented the model in it. Notice that the 
relic density of DM is mainly controlled by DM mass, $m_N$ and SM gauged couplings. Since SM gauge couplings are fixed, the only relevant 
parameter which controls the relic density is the DM mass, $m_{N}$. The behavior of relic density with DM mass shown in Fig.~\ref{fig:relic_ILD} 
along with correct relic density bound which is shown in grey patch. Note here that a sharp drop around DM mass $m_N \sim m_h/2$ due to Higgs 
resonance. From Fig.~\ref{fig:relic_ILD} we can see that the DM mass around $m_N \sim 1 ~\rm{TeV}$ only satisfy current relic density bound. For 
heavier mass of $N^0$ we get over abundance of DM (due to small cross section), while for lighter mass of $N^0$ we get under abundance of 
DM (due to large cross-section).

\subsection{Direct search constraint on ILD Dark Matter}\label{ILD_DD}
\begin{figure}[htb!]
$$
\includegraphics[height=5.0cm]{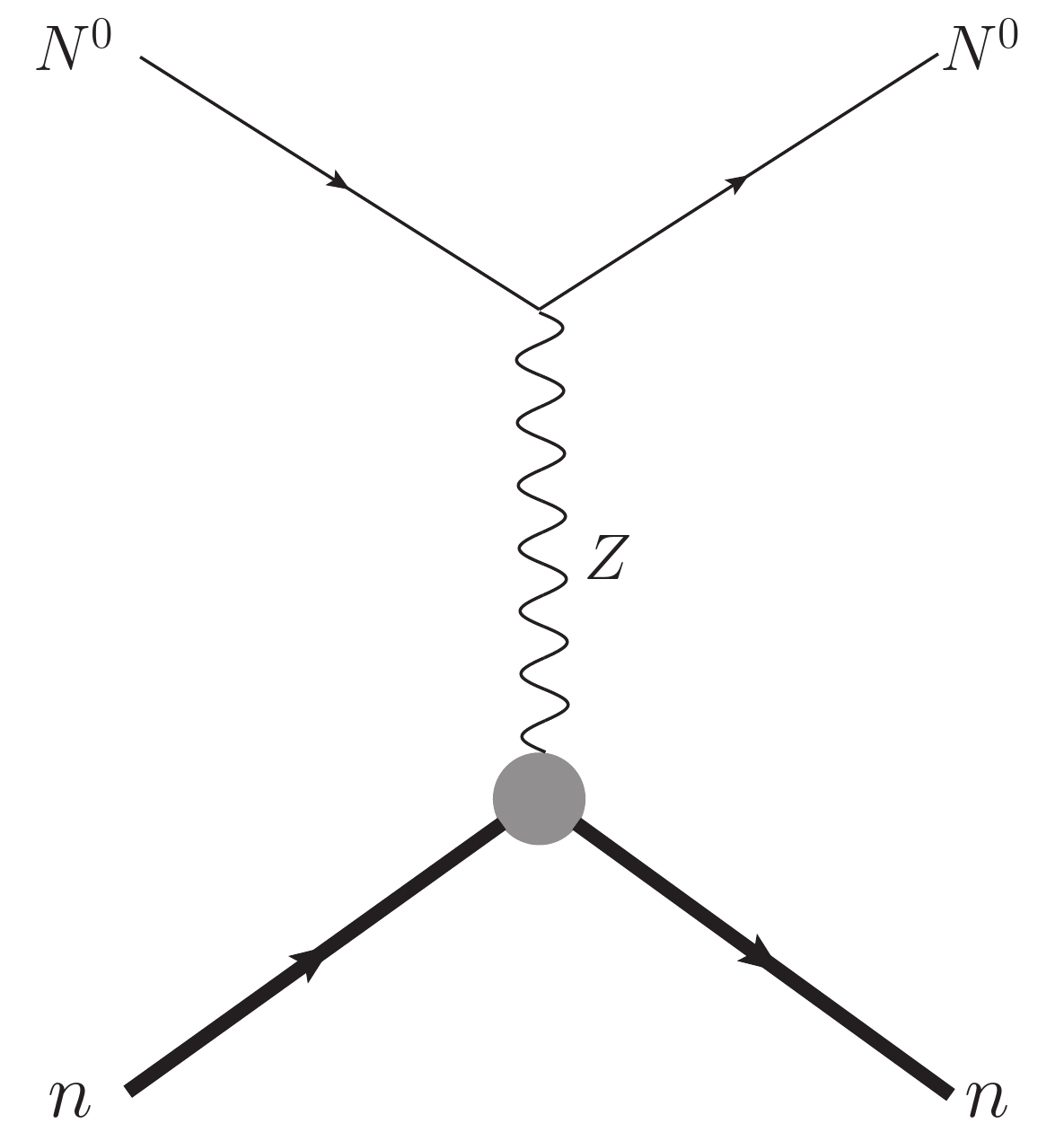} ~~~~~~
\includegraphics[height=5.0cm]{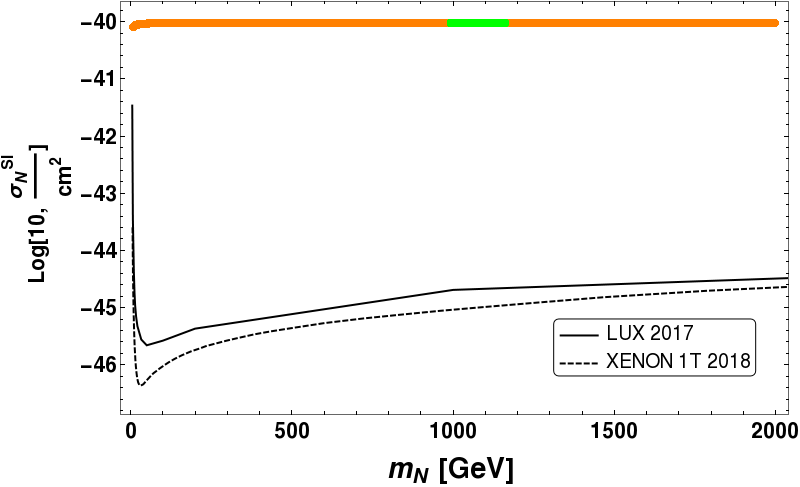}
$$
 \caption{Left: Feynman diagram for direct detection (DD) of DM, $N^0$. Right: Spin independent(SI) DM-nucleon cross-section is plotted in 
$m_N- \sigma_N^{SI}$ plane. Relic density allowed DM mass region is indicated here by green patch.}
 \label{fig:DDX_ILD}
\end{figure}

In a direct detection experiment, the DM $N^0$ scatters with the nucleon through t-channel $Z$ mediated diagram, as shown schematically in the 
left panel of Fig.~\ref{fig:DDX_ILD}. Like relic density we obtain the DM-nucleon cross-section using {\bf micrOmegas}~\cite{micromega}. Since 
$\overline{N^0} N^0 Z$ interaction is coming from SM gauge coupling and which is large, so the outcome of spin independent direct detection (SIDD)
cross section becomes large. The SIDD cross-section in this case is plotted with DM mass, $m_N$, which is shown in orange colored patch in right 
panel of Fig.~\ref{fig:DDX_ILD}. The green patch in this Fig.~\ref{fig:DDX_ILD} indicates relic density allowed mass range in the same plane. 
LUX 2017~\cite{lux} and XENON 1T~\cite{xenon_1T} direct detection limits are also plotted in the same figure (right panel of Fig.~\ref{fig:DDX_ILD}). 
Thus we see that an ILD DM is completely ruled out by direct detection bound. However, as we discuss in section \ref{triplet-ILD} the ILD DM can be 
resurrected in presence of a scalar triplet of hyper charge 2. Moreover, the scalar triplet will generate sub-eV masses of active neutrinos 
through type-II seesaw.  
%

\section{Triplet extension of the ILD dark matter}\label{triplet-ILD}

We now extend the ILD dark matter model with a scalar triplet, $\Delta$ ($Y_\Delta = 2$) which is even under 
the discrete $\mathcal{Z}_2$ symmetry. The Lagrangian of this extended sector is given as: 
\bea \label{triplet-lag}
\mathcal{L}^{II}= {\rm Tr}[(D_\mu \Delta)^\dagger (D^\mu \Delta)]-V(H,\Delta)+\mathcal{L}^{II}_{\rm Yuk},
\eea
 where the covariant derivative is defined as 
\begin{equation*}
D_\mu \Delta = \partial_\mu \Delta -ig \big[\frac{\sigma^a}{2}W_\mu^a,\Delta\big] - i g^\prime \frac{Y_\Delta}{2} B_\mu \Delta 
\end{equation*}
and in the adjoint representation the triplet $\Delta$ can be expressed in a $2\times 2$ matrix form: 
 $\Delta = \left(\begin{matrix}
\frac{H^+}{\sqrt{2}} & H^{++}  \\
\delta^0 & -\frac{H^+}{\sqrt{2}}
\end{matrix}\right)$.\\
Similarly the scalar doublet $H$ can be written in component form as: 
\begin{equation}
H= \begin{pmatrix} \phi^+ \\ \phi^0 \end{pmatrix}\,.
\end{equation}
The modified scalar potential including $\Delta$ and $H$ can be given as: 
\bea \label{eq:ScalarPotential}
 V(H,\Delta)&= &-{\mu_H^2}(H^\dagger H)+\frac{\lambda_{H}}{4}\big(H^\dagger H\big)^2+ M_\Delta^2 {\rm Tr}[\Delta^\dagger \Delta]
+\lambda_1 (H^\dagger H){\rm Tr}[\Delta^\dagger \Delta] \nonumber \\
 && + \lambda_2 \big({\rm Tr}[\Delta^\dagger \Delta]\big)^2+ \lambda_3 {\rm Tr}[(\Delta^\dagger \Delta)^2]
+\lambda_4 H^\dagger \Delta \Delta^\dagger H +\big[\mu \big( H^T i {\sigma}^2 \Delta^\dagger H\big)+ {\rm h.c.} \big] \,,
 \eea
where we assume that $M_\Delta^2$ is positive. So the scalar triplet $\Delta$ does not acquire any vev. However, it acquires an 
induced vev after EW phase transition. The vevs of the scalar fields are given by: 
\begin{equation}
\langle \Delta \rangle= \begin{pmatrix} 0 & 0\\ v_t/\sqrt{2} & 0 \end{pmatrix} ~~~~{\rm and} ~~~~\langle H \rangle= \begin{pmatrix} 
0 \\ v/\sqrt {2} \end{pmatrix}.
\end{equation}
Since the addition of a scalar triplet can modify the $\rho$ parameter, whose observed value: $\rho=1.00037 \pm 0.00023$ \cite{pdg}, 
does not differ much from SM prediction: $\rho=1$, so we have a constraint on the vev $v_t$ as:
\begin{equation}\label{const}
 v_t \leq 3.64 \rm GeV\,.
\end{equation}
On the other hand electroweak symmetry breaking gives $\sqrt{v^2 + 2 v_t^2} = 246 {\rm GeV}$. This implies that $v_t << v$. 

Now minimizing the scalar potential, $V(H,\Delta)$ (in Eq.~\ref{eq:ScalarPotential}) we get: 
\begin{eqnarray}\label{mass-parameter}
M_\Delta^2 &=& \frac{2 \mu v^2 -\sqrt{2} (\lambda_1 + \lambda_4)v^2 v_t - 2\sqrt{2} (\lambda_2 + \lambda_3)v_t^3}{2 \sqrt{2} v_t}  ,\nonumber\\
\mu_H^2 &=& \frac{\lambda_H v^2}{4} + \frac{(\lambda_1 + \lambda_4)v_t^2}{2} - \sqrt{2} \mu v_t\,.
\end{eqnarray} 
In the limit $v_t << v$, from Eq.~\ref{mass-parameter} we get the vevs,
\begin{equation}
v_t = \frac{\mu v^2}{M_\Delta^2 + (\lambda_1 + \lambda_4)v^2/2} ~~~ {\rm and} ~~~ v \approx \frac{2\mu_H}{\sqrt{\lambda_H}}\,.
 \label{vev}
\end{equation} 

In presence of the scalar triplet $\Delta$, the Yukawa interactions in Eq.~\ref{triplet-lag} are given by:
\bea\label{yukawa_coupling}
\mathcal{L}^{II}_{Yuk}=\frac{1}{\sqrt2}\big[(f_L)_{\alpha \beta}\overline{L^c_{\alpha}}i\sigma^2 \Delta L_\beta 
+ f_N \overline{N^c}i\sigma^2 \Delta N + h.c.]\,,
\eea
where $L$ is the SM lepton doublet and $\alpha,\beta$ denote family indices. The Yukawa interactions importantly inherit the source 
of neutrino masses ( first term in square bracket) and DM-SM interactions (second term in square bracket).

\subsection{Scalar doublet-triplet mixing}
The quantum fluctuations around the minimum of scalar potential, $V(H,\Delta)$ can be given as: \\ 
\begin{equation} 
\Delta = \left(\begin{matrix}
\frac{H^+}{\sqrt2} & H^{++}  \\
\frac{v_t+ h_t + i A^0}{\sqrt2} & -\frac{H^+}{\sqrt2}
\end{matrix}\right) ~~~{\rm and}~~~ 
H = \left(\begin{matrix} 0 \\ \frac{v + h}{\sqrt{2}} \end{matrix} \right)  .
\end{equation}
Thus the scalar sector constitute two CP-even Higgses: $h$ and $h_t$, one CP-odd Higgs: $A^0$, one singly charged scalar: $H^\pm$ 
and one doubly charged scalar: $H^{\pm\pm}$. In the limit $v_t << v$, the mass matrix of the CP-even Higgses: $h$ and $h_t$, is 
given by:
\begin{equation}
\mathcal{M}^2 = \begin{pmatrix} m_h^2 & -\sqrt{2} \mu v \cr \\
-\sqrt{2}\mu v & m_T^ 2 \,
\end{pmatrix} ,
\end{equation} 
where $m_h^2 \approx   \lambda_H v^2/2$ and $m_T^2 = \mu v^2/\sqrt{2} v_t$. Diagonalizing the above mass matrix we get 
two neutral physical Higgses: $H_1$ and $H_2$: 
\begin{equation}
H_1 =\cos\alpha ~h + \sin\alpha~ h_t ,~~~~ H_2= -\sin\alpha~ h + \cos\alpha ~h_t\,,
\end{equation}
 where $H_1$ is the standard model like Higgs and $H_2$ is the triplet like scalar. 
The corresponding mass eigenvalues are $m_{H_1}$ (SM like Higgs) and $m_{H_2}$ (triplet like scalar) are given by:
 \begin{eqnarray}
m_{H_1}^2 \approx m_h^2-\frac{(\mu v/\sqrt{2})^2}{m_T^2-m_h^2} , \nonumber\\
m_{H_2}^2 \approx m_T^2+\frac{(\mu v/\sqrt{2})^2}{m_T^2-m_h^2}\,.
 \end{eqnarray}  
The mixing angle is given by
 \begin{equation}\label{mix}
 \tan 2\alpha = \frac{ -\sqrt{2} \mu v}{(m_T^2 - m_h^2)}\,.
 \end{equation}
From Eqs.(\ref{mix}), (\ref{vev}) and (\ref{const}) we see that there exist an upper bound on the mixing angle 
\begin{equation}
\sin \alpha < 0.02 \left( \frac{174 {~\rm GeV} }{v} \right) \left(
  \frac{1}{1 - 0.39 \frac{(m_h/125~ \rm GeV)^2}{(m_T/200 ~\rm GeV)^2}}  \right)\,.
\end{equation}
We also get a constraint on $\sin \alpha $ from SM Higgs phenomenology, since the mixing can change the strength of the Higgs
coupling to different SM particles. See for example~\cite {Cheung:2015dta, Hartling:2014aga}, in which the global 
fit yields a constraint on mixing angle $\sin \alpha \lesssim 0.5$, which is much larger than the above constraint obtained 
using $\rho$ parameter.

From Eq. \ref{eq:ScalarPotential}, all the couplings $\lambda_H, \lambda_1, \lambda_2, \lambda_3,\lambda_4$ and $\mu$ can be expressed 
in terms of physical scalar masses: $m_{H_1}$, $m_{H_2}$, $m_{H^\pm}$, $m_{H^{\pm\pm}}$, $m_{A^0}$ and the vevs $v$ and $v_t$ as~\cite{Arhrib:2011uy} : 
\bea
\lambda_H &=& \frac{2}{v^2}\Big( m_{H_1}^2 \cos^2\alpha + m_{H_2}^2 \sin^2\alpha\Big), \nonumber \\
\lambda_1 &=& \frac{4 m_{H^\pm}^2}{v^2+4 v_t^2} - \frac{2 m_{A^0}^2}{v^2+4 v_t^2} + \frac{\sin2\alpha}{2 v_t v} \big(m_{H_1}^2-m_{H_2}^2\big), \nonumber \\
\lambda_2 &=& \frac{1}{v_t^2} \Big[ \frac{1}{2}\Big( m_{H_1}^2 \sin^2\alpha + m_{H_2}^2 \cos^2\alpha\Big)+ \frac{v^2}{2(v^2+4 v_t^2)} m_{A^0}^2
-\frac{2 v^2}{v^2+4 v_t^2} m_{H^\pm}^2+m_{H^{\pm\pm}}^2  \Big] ,  \nonumber \\
\lambda_3&=& \frac{1}{v_t^2} \Big[\frac{2 v^2}{v^2+2v_t^2}m_{H^\pm}^2 - \frac{v^2}{v^2+4v_t^2} m_{A^0}^2-m_{H^{\pm\pm}}^2\Big] ,\nonumber \\
\lambda_4 &=& \frac{4 m_{A^0}^2}{v^2+4v_t^2} -\frac{4 m_{H^\pm}^2}{v^2+2v_t^2} ,\nonumber \\
\mu &=& \frac{\sqrt2 v_t}{v^2+4v_t^2} m_{A^0}^2. 
\label{eq:triplet-coupling}
\eea
where $m_{A^0}$ is the mass of pseudo scalar. It is important to note that the quartic couplings $\lambda_2$ and $\lambda_3$ are 
inversely proportional to the triplet vev $v_t$ which has important consequences for dark matter relic abundance that we discuss in 
section \ref{scalar_effect_DM}.

\subsection{Non-zero neutrino masses}\label{neutrino-mass-section}

The coupling of scalar triplet $\Delta$ to SM lepton and Higgs doublet combinely break the lepton number by two units as given in 
Eq.~\ref{yukawa_coupling}. As a result the $\Delta L_\alpha  L_\beta$ coupling yields Majorana masses to three flavors 
of active neutrinos as~\cite{type_II}:
\begin{equation}\label{neutrino-mass}
(M_\nu)_{\alpha \beta}= \sqrt{2} (f_L)_{\alpha\beta}\langle \Delta \rangle \approx (f_L)_{\alpha\beta} \frac{-\mu v^2}{\sqrt{2} M_{\Delta}^2 }\,.
\end{equation}
Assuming $\mu \simeq M_\Delta \simeq \mathcal{O}(10^{13})$ GeV, we can explain neutrino masses of order $1 \rm eV$ with a coupling strength 
$f_L \simeq 1$. However, the scale of $M_\Delta$ can be brought down to $\sim$ TeV by taking the coupling to be much smaller $f_L \simeq 10^{-11}$, 
and indeed represents a bit of fine tuning in the neutrino sector. 
 
To get the neutrino mass eigen values, the above mass matrix can be diagonalized  by the usual $U_{PMNS}$ matrix as :
\begin{equation}\label{eq:neu_mass}
M_\nu= U_{\text {PMNS}} \, M_\nu^{diag} \, U^T_{\text{ PMNS}} \,,
\end{equation}
where $U_{PMNS}$ is given by 
\begin{equation}\label{eq:pmns}
U_{\text PMNS} =
\begin{pmatrix}
c_{12}c_{13}  &  s_{12}c_{13}  & s_{13}e^{-i\delta_{13}}\\
-s_{12}c_{23}-c_{12}s_{23}s_{13}e^{i\delta_{13}} &
c_{12}c_{23}-s_{12}s_{23}s_{13}e^{i\delta_{13}}  & s_{23}c_{13} \\
s_{12}s_{23}-c_{12}c_{23}s_{13}e^{i\delta_{13}}  &
-c_{12}s_{23}-s_{12}c_{23}s_{13}e^{i\delta_{13}}  & c_{23}c_{13}
\end{pmatrix}
. U_{ph}\,,
\end{equation}
with $c_{ij}$, $s_{ij}$ stand for $\cos\theta_{ij}$ and
$\sin\theta_{ij}$ respectively and $U_{ph}$ is given by 
\begin{equation}
U_{ph}= \text {Diag} \left ( e^{-i\gamma_1} , e^{-i\gamma_2} , 1
\right ) \,.
\end{equation}
Where $\gamma_{1}$, $\gamma_2$ are two Majorana phases. The diagonal matrix $M_\nu^{diag}$ = Diag $(m_1,m_2,m_3)$ 
with diagonal entries are the mass eigen values for the neutrinos.  The current neutrino oscillation data at $3\sigma$
confidence level give the constraint on mixing angles ~\cite{pdg} :
\begin{equation}\label{eq:osc_angle}
0.259 < \sin^2\theta_{12} < 0.359,\,\,  0.374 <  \sin^2\theta_{23} <
0.628, \,\, 0.0176 <  \sin^2\theta_{13} < 0.0295\,.
\end{equation}
However little information is available about the CP violating Dirac phase $\delta$ as well as the Majorana phases.  Although 
the absolute mass of neutrinos is not measured yet, the mass square differences have already been measured to a good degree of 
accuracy :
\begin{eqnarray}\label{eq:mass_sqr}
\Delta m^2_0\equiv m_2^2-m_1^2= (6.99 - 8.18) \times 10^{-5} \rm eV^2 \nonumber \\
|\Delta m^2_{\rm atm} |\equiv |m_3^2-m_1^2|= (2.23 - 2.61) \times 10^{-3} \rm eV^2\,.
\end{eqnarray}
One of the main issues of neutrino physics lies in the sign of the atmospheric mass square difference $|\Delta m^2_{\rm atm} |\equiv |m_3^2-m_1^2|$, 
which is still unknown. This yields two possibilities: normal hierarchy (NH) ($m_1 < m_2 < m_3$) or inverted hierarchy (IH) ($m_3 < m_1 < m_2$). 
Another possibility, yet allowed, is to have a degenerate (DG) neutrino mass spectrum ($m_1 \sim m_2 \sim m_3$). 
Assuming that the neutrinos are Majorana, the mass matrix can be written as :
\begin{equation}\label{eq:mass_mat}
M_\nu = \begin{pmatrix}
 a &  b  & c \\
b &  d  & e \\
c &  e  & f
\end{pmatrix} .
\end{equation}
Using equations (\ref{eq:neu_mass}), (\ref{eq:pmns}), (\ref{eq:osc_angle}) and (\ref{eq:mass_sqr}),
we can estimate the unknown parameters in neutrino mass matrix of Eq.~\ref{eq:mass_mat}. To estimate the parameters in NH, we use 
the best fit values of the oscillation parameters. For a typical value of the lightest neutrino mass of $m_1 =0.0001$ eV, we get the mass parameters (in eV) as :
\begin{eqnarray}
a=0.003833, \, b=0.00759, \, c=0.002691 ,\nonumber \\
d=0.023865,  \,e=0.02083, \, f=0.03038   .
\end{eqnarray}
Similarly for IH case, choosing the lightest neutrino mass $m_3 =0.001$ eV, we get the mass parameters (in eV) as :
\begin{eqnarray}
a=0.0484, \, b=-0.00459, \, c=-0.00573 ,\nonumber \\
d=0.02893,  \,e=-0.02366, \, f=0.02303  .
\end{eqnarray}
In both the cases, we put the Dirac and Majorana phases to be zero for simplicity.

The analysis of neutrino mass is more indicative here than being exhaustive. This is essentially to build the connection between 
the dark sector and neutrino sector advocated in the model set up. One can easily perform a scan over the mass matrix parameters to 
obtain correct ranges of the neutrino observables, and that of course lies in the vicinity of the aforementioned values. But, we do 
not aim to elaborate that in this draft. We have not also adhered to a specific lepton mixing matrix pattern (say tri-bi-maximal mixing) 
coming from a defined underlying flavour symmetry (say $A_4$), which will be able to correlate different parameters of the mass matrix.   

The mass of the scalar triplet can also be brought down to TeV scale by choosing appropriate Yukawa coupling as explained before. If the 
triplet mass is order of a few hundreds of GeV,  then it can give interesting dilepton signals in the collider. See for example, \cite{Dilepton} 
for a detailed discussion regarding the dilepton signatures of the scalar triplet at collider. 

\subsection{Pseudo-Dirac nature of ILD  Dark Matter}\label{subsec:ILD_triplet}

From  Eq.~(\ref{yukawa_coupling}) we see that the vev of $\Delta$ induces a Majorana mass to 
$N^0$ which is given by:
\begin{equation}\label{majorana-mass-1}
m=\sqrt{2} f_N  \langle \Delta \rangle \approx f_N  \frac{-\mu v^2}{\sqrt{2} M_{\Delta}^2 }\,.
\end{equation}  
Thus the $N^0$ has a large Dirac mass $M_{N}$ (as in Eq. 2) and a small Majorana mass $m$ as shown in the above Eq.~\ref{majorana-mass-1}. 
Therefore, we get a mass matrix in the basis $\{N^0_L, (N^0_R)^c\} $ as:
\begin{equation}
{\mathcal M} =
\begin{pmatrix}
  m  &  M_{N} \\
       M_{N}   & m
\end{pmatrix} .
\end{equation}
Thus the Majorana mass $m$ splits the Dirac spinor $N^0$ into two pseudo-Dirac states $N^0_{1,2}$ with mass eigenvalues 
$M_{N} \pm m$. The mass splitting between the two pseudo-Dirac states $N^0_{1,2}$ is given by
\begin{equation}
\delta m = 2 m = 2\sqrt{2} f_N \langle \Delta \rangle \,.
\end{equation} 
Note that $\delta m << M_{N}$ from the estimate of induced vev of the triplet and hence does not play any role in the 
relic abundance calculation. However, the sub-GeV order mass splitting plays a crucial role in direct detection by forbidding 
the Z-boson mediated DM-nucleon elastic scattering. Now from Eq.~(\ref{neutrino-mass}) and (\ref{majorana-mass-1}) we see that the 
ratio: 
\begin{equation}\label{improved-coupling-ratio}
R=\frac{(M_\nu)} {m} = \frac{f_L} {f_N} \lesssim 10^{-5}\,,
\end{equation}
where we assume $M_\nu \sim  1$ eV and $m \sim 100$ keV. Here the mass splitting between the two states $N^0_1$ and $N^0_2$ is chosen to 
be ${\cal O}(100)$ keV in order to forbid the $Z$-mediated inelastic scattering with the nucleons in direct detection. Thus we see that 
the ratio $R \lesssim  10^{-5}$ is heavily fine tuned. In other words, the scalar triplet strongly decay to ILD dark matter, while 
its decay to SM leptons is suppressed.

\subsection{Effect of scalar triplet on relic abundance of ILD dark matter} \label{scalar_effect_DM}

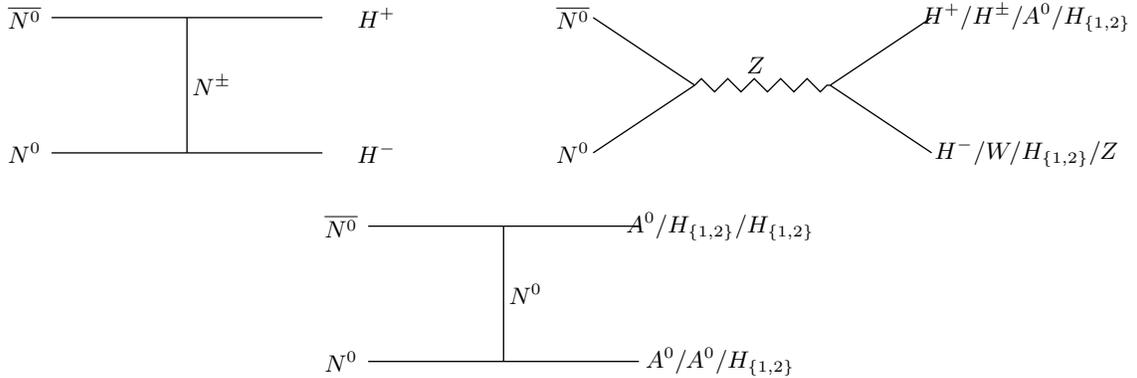
\begin{figure}[htb!]
\begin{center}
    \begin{tikzpicture}[line width=0.5 pt, scale=0.9]
        \draw[solid] (-3,1.0)--(-1.0,1.0);
        \draw[solid] (-3,-1.0)--(-1.0,-1.0);
        \draw[solid](-1.0,1.0)--(-1.0,-1.0);
        \draw[solid] (-1.0,1.0)--(1.0,1.0);
        \draw[solid] (-1.0,-1.0)--(1.0,-1.0);
        \node at (-3.4,1.0) {$\overline{N^0}$};
        \node at (-3.4,-1.0) {$N^0$};
        \node [right] at (-1.05,0.0) {$ N^\pm $};
        \node at (1.8,1.0) {$H^+$};
        \node at (1.8,-1.0) {$H^-$};
        \draw[solid] (5.0,1.0)--(6.5,0.0);
        \draw[solid] (5.0,-1.0)--(6.5,0.0);
        \draw[snake] (6.5,0.0)--(8.5,0.0);
        \draw[solid] (8.5,0.0)--(10.0,1.0);
        \draw[solid] (8.5,0.0)--(10.0,-1.0);
        \node at (4.7,1.0) {$\overline{N^0}$};
        \node at (4.7,-1.0) {$N^0$};
        \node [above] at (7.4,0.05) {$Z$};
        \node at (11.4,1.0) {$H^+/H^\pm/ A^0/H_{\{1,2\}}$};
        \node at (11.4,-1.0) {$H^-/W/H_{\{1,2\}}/Z$};
     \end{tikzpicture}
 \end{center}
\begin{center}
    \begin{tikzpicture}[line width=0.5 pt, scale=0.9]
        \draw[solid] (-3,1.0)--(-1.0,1.0);
        \draw[solid] (-3,-1.0)--(-1.0,-1.0);
        \draw[solid](-1.0,1.0)--(-1.0,-1.0);
        \draw[solid] (-1.0,1.0)--(1.0,1.0);
        \draw[solid] (-1.0,-1.0)--(1.0,-1.0);
        \node at (-3.4,1.0) {$\overline{N^0}$};
        \node at (-3.4,-1.0) {$N^0$};
        \node [right] at (-1.05,0.0) {$N^0 $};
        \node at (2.2,1.0) {$A^0/ H_{\{1,2\}} / H_{\{1,2\}} $};
        \node at (2.2,-1.0) {$A^0 / A^0 / H_{\{1,2\}}  $};
     \end{tikzpicture}
 \end{center}

\caption{Additional annihilation $N^0 \overline{N^0} $, in presence of scalar triplet. }
\label{fd:ann_ILD_triplet}
 \end{figure}

\begin{figure}[htb!]
\begin{center}
    \begin{tikzpicture}[line width=0.5 pt, scale=0.9]
        \draw[solid] (-3,1.0)--(-1.0,1.0);
        \draw[solid] (-3,-1.0)--(-1.0,-1.0);
        \draw[solid](-1.0,1.0)--(-1.0,-1.0);
        \draw[solid] (-1.0,1.0)--(1.0,1.0);
        \draw[solid] (-1.0,-1.0)--(1.0,-1.0);
        \node at (-3.4,1.0) {$N^-$};
        \node at (-3.4,-1.0) {$N^+$};
        \node [right] at (-1.05,0.0) {$ N^-/N^0 $};
        \node at (1.8,1.0) {$H^{--}/ H^-$};
        \node at (1.8,-1.0) {$H^{++}/H^+$};
        \draw[solid] (5.0,1.0)--(6.5,0.0);
        \draw[solid] (5.0,-1.0)--(6.5,0.0);
        \draw[snake] (6.5,0.0)--(8.5,0.0);
        \draw[solid] (8.5,0.0)--(10.0,1.0);
        \draw[solid] (8.5,0.0)--(10.0,-1.0);
        \node at (4.7,1.0) {$N^-$};
        \node at (4.7,-1.0) {$N^+$};
        \node [above] at (7.4,0.05) {$A/Z$};
        \node at (11.2,1.2) {$H^{++},H^\pm,A^0,H_{\{1,2\}}$};
        \node at (11.1,-1.2) {$H^{--},H^-,W,H_{\{1,2\}},Z$};
     \end{tikzpicture}
 \end{center}
\caption{Co-annihilation processes involving only charged partner of ILD DM, $N^\pm$ in presence of scalar triplet. }
\label{fd:coann_ILD_triplet}
 \end{figure}
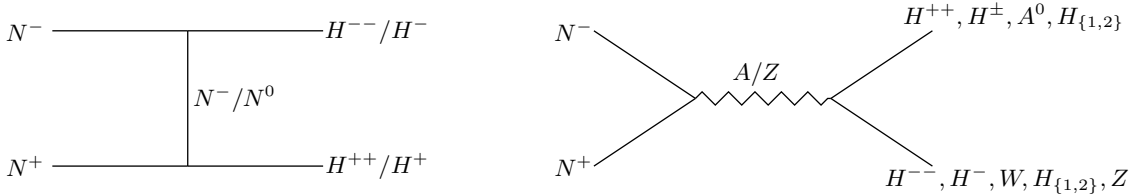

In presence of a scalar triplet, when the DM mass is larger than the triplet mass, a few additional annihilation and co-annihilation channels 
open up as shown in Figs.~\ref{fd:ann_ILD_triplet}, \ref{fd:annco_ILD_triplet}, \ref{fd:Chargedann_ILD_triplet} and \ref{fd:coann_ILD_triplet} in 
addition to the previously mentioned Feynman diagrams given in Figs. \ref{fd:ann_ILD},\ref{fd:ann_ILD_charge} and \ref{fd:coann_ILD}. These additional 
channels also play a key role in number changing processes of DM, $N^0$ to yield a modified freeze-out abundance. We numerically calculate relic density 
of $N^0$ DM once again by implementing the model in the code {\bf micrOmegas}~\cite{micromega}. The parameter space, in comparison to the ILD dark matter alone, 
is enhanced due to the additional coupling of $N^0$ with $\Delta$. In particular, the new parameters are: triplet scalar masses $m_{H_2},
m_{A^0},m_{H^\pm}, m_{H^{\pm\pm}}$, vev of scalar triplet $v_t$, coupling of scalar triplet with ILD dark matter $N^0$, {\it i.e.} $f_N$, scalar 
doublet-triplet mixing $\sin \alpha$. 

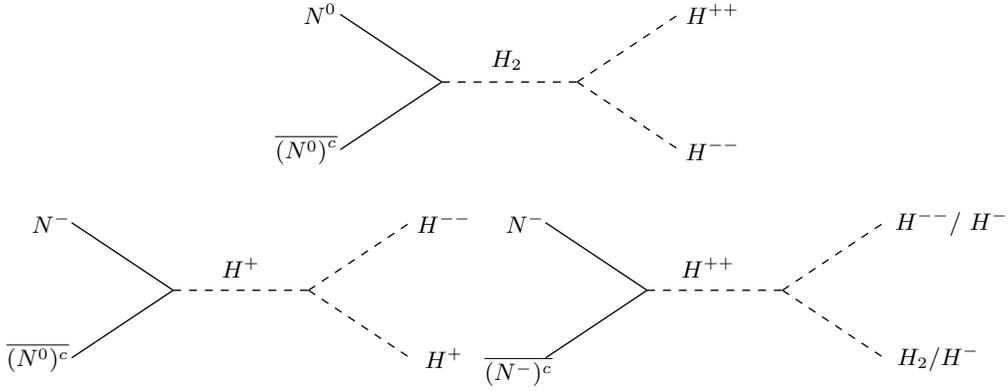
\begin{figure}[htb!]
\begin{center}
    \begin{tikzpicture}[line width=0.5 pt, scale=0.9]
        \draw[solid] (5.0,1.0)--(6.5,0.0);
        \draw[solid] (5.0,-1.0)--(6.5,0.0);
        \draw[dashed] (6.5,0.0)--(8.5,0.0);
        \draw[dashed] (8.5,0.0)--(10.0,1.0);
        \draw[dashed] (8.5,0.0)--(10.0,-1.0);
        \node at (4.7,1.0) {$N^0$};
        \node at (4.5,-1.0) {$\overline{{(N^0)}^c}$};
        \node [above] at (7.4,0.05) {$~H_2$};
        \node at (10.5,1.0) {$H^{++}$};
        \node at (10.5,-1.0) {$H^{--}$};
     \end{tikzpicture}
 \end{center}
\begin{center}
    \begin{tikzpicture}[line width=0.5 pt, scale=0.9]
        \draw[solid] (5.0,1.0)--(6.5,0.0);
        \draw[solid] (5.0,-1.0)--(6.5,0.0);
        \draw[dashed] (6.5,0.0)--(8.5,0.0);
        \draw[dashed] (8.5,0.0)--(10.0,1.0);
        \draw[dashed] (8.5,0.0)--(10.0,-1.0);
        \node at (4.7,1.0) {$N^-$};
        \node at (4.5,-1.0) {$\overline{(N^0)^c}$};
        \node [above] at (7.5,0.05) {$H^{+}$};
        \node at (10.5,1.0) {$H^{--}$};
        \node at (10.5,-1.0) {$H^{+}$};
        \draw[solid] (12.0,1.0)--(13.5,0.0);
        \draw[solid] (12.0,-1.0)--(13.5,0.0);
        \draw[dashed] (13.5,0.0)--(15.5,0.0);
        \draw[dashed] (15.5,0.0)--(17.0,1.0);
        \draw[dashed] (15.5,0.0)--(17.0,-1.0);
        \node at (11.7,1.0) {$N^-$};
        \node at (11.6,-1.2) {$\overline{(N^-)^c}$};
        \node [above] at (14.4,0.05) {$H^{++}$};
        \node at (18,1.0) {$H^{--}/~H^{-}$};
        \node at (17.8,-1.0) {$H_2/H^{-}$};
        \label{fig:dominF}
     \end{tikzpicture}
 \end{center}
\caption{Dominant annihilation ($N^0 \overline{(N^0)^c}$)and co-annihilation ($N^- \overline{(N^0)^c},~N^- \overline{(N^-)^c}$) processes of ILD DM ($N^0$) to scalar triplet in final states.}
\label{fd:annco_ILD_triplet}
 \end{figure}
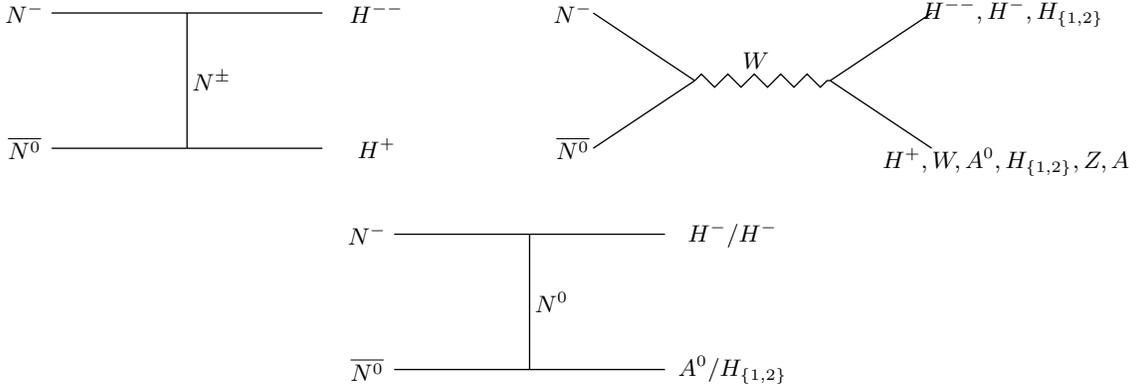
\begin{figure}[htb!]
\begin{center}
    \begin{tikzpicture}[line width=0.5 pt, scale=0.9]
        \draw[solid] (-3,1.0)--(-1.0,1.0);
        \draw[solid] (-3,-1.0)--(-1.0,-1.0);
        \draw[solid](-1.0,1.0)--(-1.0,-1.0);
        \draw[solid] (-1.0,1.0)--(1.0,1.0);
        \draw[solid] (-1.0,-1.0)--(1.0,-1.0);
        \node at (-3.4,1.0) {$N^-$};
        \node at (-3.4,-1.0) {$\overline{N^0}$};
        \node [right] at (-1.05,0.0) {$ N^\pm $};
        \node at (1.8,1.0) {$H^{--}$};
        \node at (1.8,-1.0) {$H^+$};
        \draw[solid] (5.0,1.0)--(6.5,0.0);
        \draw[solid] (5.0,-1.0)--(6.5,0.0);
        \draw[snake] (6.5,0.0)--(8.5,0.0);
        \draw[solid] (8.5,0.0)--(10.0,1.0);
        \draw[solid] (8.5,0.0)--(10.0,-1.0);
        \node at (4.7,1.0) {$N^-$};
        \node at (4.7,-1.0) {$\overline{N^0}$};
        \node [above] at (7.4,0.05) {$W$};
        \node at (11.2,1.0) {$H^{--},H^-,H_{\{1,2\}}$};
        \node at (11.1,-1.2) {$H^+,W,A^0,H_{\{1,2\}},Z,A$};
     \end{tikzpicture}
 \end{center}
\begin{center}
    \begin{tikzpicture}[line width=0.5 pt, scale=0.9]
        \draw[solid] (-3,1.0)--(-1.0,1.0);
        \draw[solid] (-3,-1.0)--(-1.0,-1.0);
        \draw[solid](-1.0,1.0)--(-1.0,-1.0);
        \draw[solid] (-1.0,1.0)--(1.0,1.0);
        \draw[solid] (-1.0,-1.0)--(1.0,-1.0);
        \node at (-3.4,1.0) {$N^-$};
        \node at (-3.4,-1.0) {$\overline{N^0}$};
        \node [right] at (-1.05,0.0) {$N^0 $};
        \node at (2.0,1.0) {$H^-/ H^- $};
        \node at (2.0,-1.0) {$A^0/ H_{\{1,2\}} $};
     \end{tikzpicture}
 \end{center}
\caption{Co-annihilation channels of ILD DM ($N^0$), with charged fermions $N^-$ in presence of scalar triplet. }
\label{fd:Chargedann_ILD_triplet}
 \end{figure}

\begin{figure}[htb!]
$$
 \includegraphics[height=5.0cm]{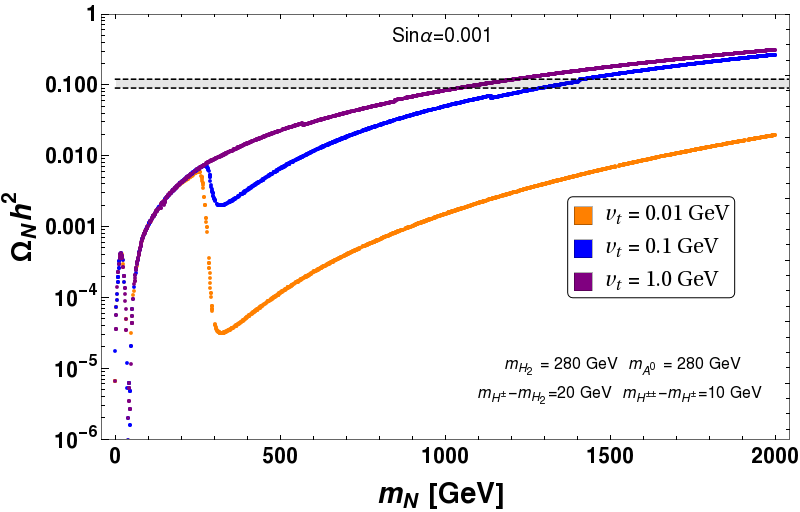} 
 \includegraphics[height=5.0cm]{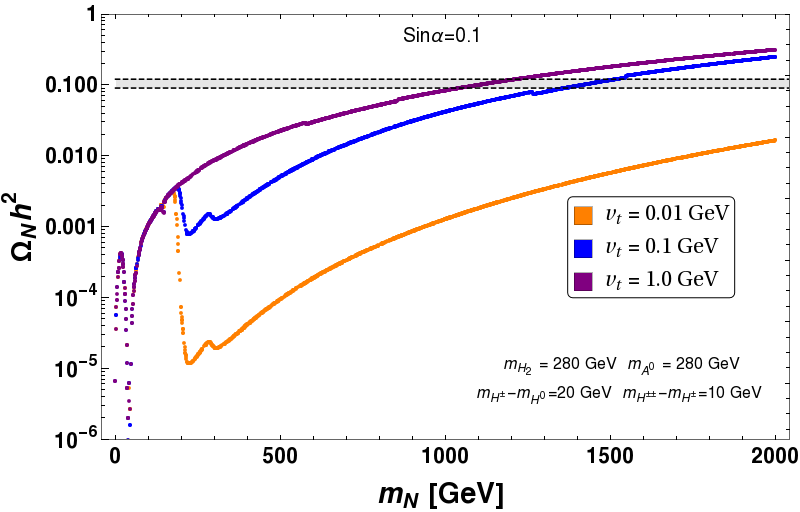}
$$
 \caption{Relic density vs $m_N$ plot for different choices triplet vev, $v_t$, keeping $\sin\alpha=0.001$ (left) and $\sin\alpha=0.1$ (right). Others parameters are mentioned in the figure. The gray patch indicates  relic density bounds. In both cases we set $f_N=0.1$.}
 \label{fig:omega_mN_vt}
\end{figure}

To understand the effect of the triplet scalar on relic density of ILD DM, we show in Fig.~\ref{fig:omega_mN_vt} 
the variation of relic density as a function of DM mass ($m_N$) with different choices of triplet vev ($v_t$) while keeping a fixed $f_N$ and $\sin \alpha$. 
In the left panel of Fig.~\ref{fig:omega_mN_vt} we choose the scalar doublet-triplet mixing to be $\sin\alpha = 0.001$ while in the right-panel 
of Fig.~\ref{fig:omega_mN_vt} we choose $\sin\alpha = 0.1$. In both cases we set the physical CP even, CP odd and charged triplet scalar 
masses respectively at $m_{H_2}= 280, ~m_{A^0}= 280, m_{H^\pm}=300$ and $m_{H^{\pm\pm}}=310$ GeV and $f_N=0.1$. We note that there is a resonance drop 
near $m_N \sim m_Z/2$ as usual for $Z$ mediated s-channel diagrams. Additionally we find that near $m_N \sim 280$ GeV (which is the mass of $H_2$), relic density 
drops suddenly because of new annihilation processes $N N \rightarrow \Delta \Delta$ start to contribute (see diagrams in Figs. \ref{fd:ann_ILD_triplet}, 
\ref{fd:annco_ILD_triplet}, \ref{fd:Chargedann_ILD_triplet} and \ref{fd:coann_ILD_triplet}). With larger triplet vev $v_t \sim$ GeV, the effect of annihilation 
to scalar triplet becomes subdued. This can be understood as follows. First of all, we see that the quartic couplings involving the triplet, as given 
in  Eq.~\ref{eq:triplet-coupling}, are inversely proportional to the triplet vev ($v_t$). In a typical annihilation process: $N^-\overline{{N^-}^c} \to H^-H^-$, 
mediated by $H^{--}$, the vertex $H^{--}H^-H^-$ is proportional to $\sqrt 2 v_t \lambda_3\sim 1/v_t$, which diminishes with larger $v_t$. On the other hand, let 
us consider the process: $N^0\overline{{N^0}^c} \to H^{++}H^{--}$, which has a significant contribution to the total relic density. This process is mediated 
by $H_1$ and $H_2$. In small $\sin\alpha$ limit, the $H_1$ mediated diagram is vanishingly small as the $N^0\bar{N^0}H_1 \sim \sin \alpha$. So, $H_2$ 
mediation dominates here. However, the vertex involving $H_2H^{++}H^{--}$ is proportional to $(2\cos\alpha \lambda_2 v_t-\sin\alpha\lambda_1v_d)$. One 
can see that for small $v_t$, the first term is negligible, while for a larger $v_t$, the first term becomes comparable to that of the second one and has a cancellation. 
This cancellation therefore decreases the annihilation cross-section to the chosen final state. Such a phenomena is also present for 
co-annihilation processes like $N^-\overline{{N^-}^c} \to H^{--}H_2$ etc., where the vertices involve a combination of $\lambda_1, ~\rm{and}~\lambda_2$. 
On the other hand, for smaller values of triplet vev ($v_t \sim 0.01$ GeV), there is a larger drop in relic density due to the additional annihilation channels (to 
the triplet scalars as mentioned). Therefore, the DM $N^0$ achieves correct relic density for larger DM mass $m_N$ (as compared to that of the case 
in absence of triplet). Moreover, we set $f_N=0.1$ in both cases. It is straightforward to see that annihilation to the triplet final states are 
proportional to $f_N$. Therefore with larger $f_N$, the drop in relic density in the vicinity of triplet scalar mass decreases even further. In summary, 
the presence of scalar triplet shifts the relic density of ILD DM to a higher DM mass region which crucially depends on the choice of the triplet vev 
as well as $\Delta NN$ coupling $f_N$. 

An important conclusion about ILD dark matter is that the mass of DM ($m_{N}$) is around 1 TeV which satisfies the observed relic abundance. This implies 
the mass of $N^-$, the charged partner of $N^0$, is about 1 TeV as well. However, the electroweak correction induces a small mass splitting between $N^0$ and 
$N^-$ to be around 162 MeV. Therefore, $N^-$ can give rise a displaced vertex signature through the 3-body decay 
$N^- \to N^0 \ell^- \bar{\nu_{\ell}}$ ~\cite{Thomas:1998wy}. But the main drawback is that the production cross-section of $N^\pm$ of mass $\sim {\rm TeV}$ 
is highly suppressed at LHC as this can only be possible through Drell-Yan. 
Therefore, in section-\ref{sec:model} we discuss a more predictive model by enlarging the dark sector with an additional singlet fermion $\chi$. 
 
\subsection{Effect of scalar triplet on direct detection of ILD dark matter}

As discussed in section \ref{ILD_DD}, the ILD dark matter alone is ruled out due to large Z-mediated elastic scattering with nucleus. 
However, it can be reinstated in presence of the scalar triplet, which not only forbids the Z-mediated elastic 
scattering~\cite{inelastic_DM_papers} but also provides a new portal for the detection of ILD dark matter via the 
doublet-triplet mixing as we discuss below. 

The interaction of DM with the $Z$ boson with the kinetic term is given as 
\begin{equation}
\mathcal{L}_{Z-DM} \supset i \bar{N^0}\left( \gamma^\mu \partial_\mu -ig_Z  \gamma^\mu Z_\mu \right) N^0 \, ,
\end{equation}
where $ig_Z = \frac{g}{2 \cos \theta_W}$. After the symmetry breaking the scalar triplet $\Delta$ gets an induced vev and hence gives Majorana 
mass to the ILD dark matter $N_0$ as shown in Eq.~\ref{majorana-mass-1}.  The presence of such Majorana mass term splits the Dirac DM state into two real 
Majorana states $N^0_1$ and $N^0_2$ with a mass splitting of $\delta m$ as discussed in sec~\ref{subsec:ILD_triplet}. Now we rewrite the Lagrangian 
involving DM-$Z$ interaction in terms of the new Majorana states as:

\begin{equation}\label{eq:in_dm}
{\mathcal L}_{\rm Z-DM} \supset \overline{N^0_1} i\gamma^\mu \partial_\mu N^0_1 + \overline{N^0_2} i\gamma^\mu \partial_\mu N^0_2 
- i g_z \overline{N^0_1} \gamma^\mu  N^0_2 Z_\mu \,.
\end{equation}
\begin{figure}[h]
\includegraphics[scale=0.5]{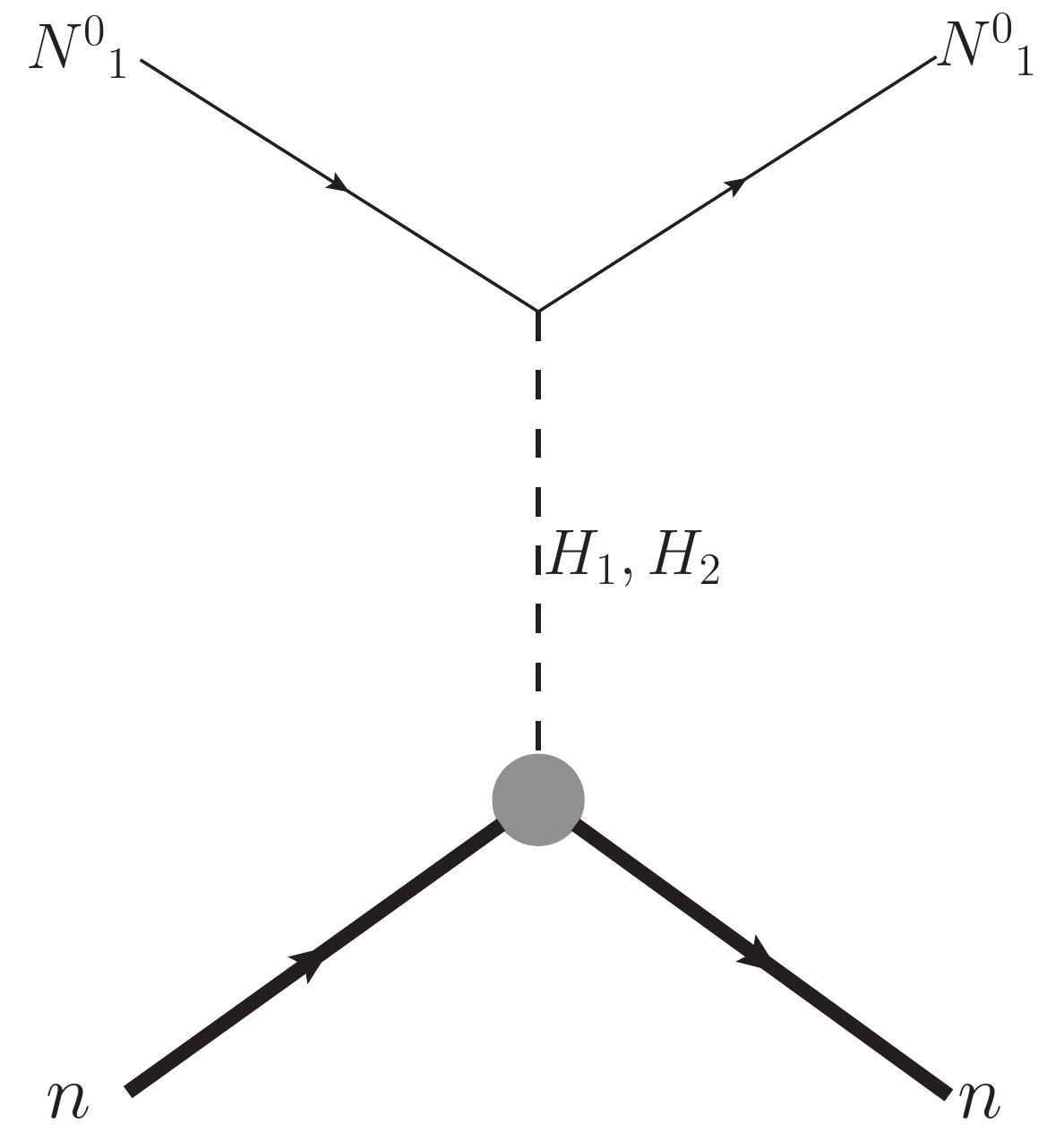}
\caption{Elastic scattering of the ILD DM($N^0$) with the nucleon through scalar mediation due to doublet triplet mixing.}
\label{fey:dd_ild}
\end{figure}
We can see that the dominant gauge interaction becomes off-diagonal.  The absence of diagonal interaction term for the DM-$Z$ vertex leads 
to the vanishing contribution to elastic scattering of the DM with the nucleus.  However there could be an inelastic scattering through $Z$ 
mediation, which is suppressed if the mass splitting between two states is of the order ${\cal O}(100)$ keV or less.  But the Yukawa term involving 
DM and $\Delta$ is still diagonal in the new basis and hence can lead to elastic scattering through a mixing between the doublet-triplet 
Higgs.  Assuming $N^0_1$ to be the lightest among the two Majorana states, hence being the DM, the relevant diagram for the elastic scattering 
is shown in Fig~\ref{fey:dd_ild}.
\begin{figure}
\includegraphics[scale=0.3]{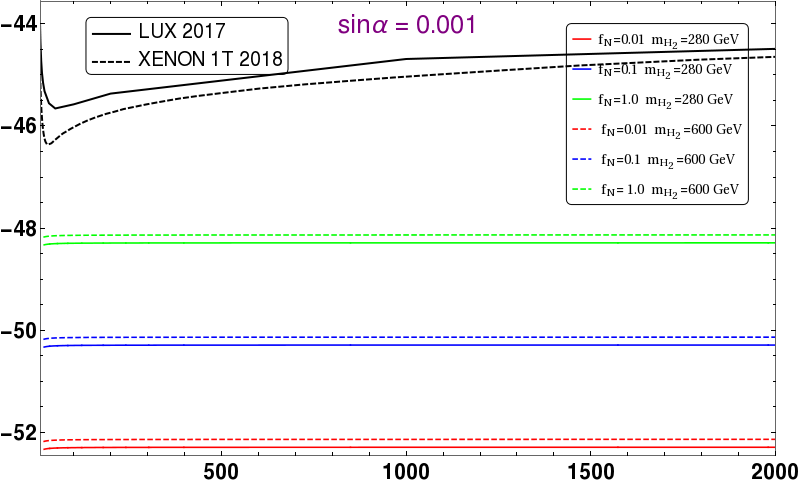}
\includegraphics[scale=0.3]{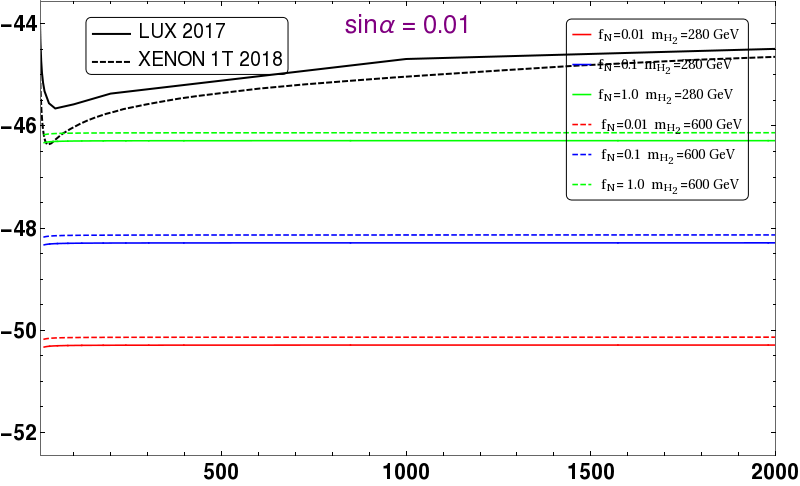}
\caption{ SI direct detection cross-section $\sigma ^{SI}$ as a function of DM mass $m_{N^0_1}$ for $\sin \alpha =0.001$ (Left) and $\sin \alpha =0.01$ (Right). 
Different choices of the coupling $\{f_N= 0.01,0.1,1\}$ are shown in Red, Blue and Green color respectively. Dashed and solid lines corresponds to different values 
of the heavy Higgs $m_{H_2}=600,280$ GeVs respectively. The bound from LUX 2017 and XENON1T 2018 are shown.}
\label{fig:ILD_DD}
\end{figure}

The direct detection cross-section mainly depends on $m_{H_2}$, $f_N$ and $\sin \alpha$.  We have plotted the spin independent direct detection 
cross-section as a function of the DM mass $m_{N^0_1}$ in Fig~\ref{fig:ILD_DD}.  Keeping $m_{H_2}=280$ GeV (Solid) and $m_{H_2}=600$ GeV (Dashed) 
fixed, we have shown the cross-section for three different values $\{f_N= 0.01,0.1,1\}$ in Red, Blue and Green color respectively.  Since there is 
a relative negative sign between the two amplitudes,  the destructive interference is more for $m_{H_2}$ comparable to SM Higgs.  Hence cross 
section for $m_{H_2}=280$ GeV turns out to be smaller than $m_{H_2}=600$ GeV.  But if we increase the mass of $m_{H_2}$ to TeV scale then $H_2$ 
mediated process will be suppressed due to the large mass in the propagator and only $H_1$ mediated process will contribute.  The direct search 
cross-section increases with larger Yukawa coupling $f_N$ from $0.01$ to $1$ and can be is easily seen from the Fig~\ref{fig:ILD_DD}.  Since the 
DM couples dominantly to the triplet scalar, the more the mixing angle ($\sin \alpha$), the more is the cross-section which can be clearly seen 
from the left and right panel of Fig~\ref{fig:ILD_DD}.  But all these cross-sections are well below the present experimental bound of LUX and 
Xenon-1T. Note that the relic density allowed parameter space of ILD DM in presence of a scalar triplet live in a very high DM mass region $\sim TeV$ with 
moderate $f_N$ and Higgs data unambiguously indicates that the mixing angle ($\sin\alpha$) should be kept small, as we have shown in 
Fig~\ref{fig:ILD_DD}. Therefore, the ILD becomes a viable DM candidate in the presence of a triplet scalar allowed by 
both relic density and direct search constraints.

\section{Singlet-doublet leptonic dark matter}\label{sec:model}

Now let us assume that the dark sector is composed of two vector like leptons : a doublet, $N=~(N^0  ~~N^-~)^T$ and a singlet $\chi$, which are 
odd under an extended $\mathcal{Z}_2$ symmetry while all the Standard Model (SM) fields are even. As a result the lightest odd particle 
in the dark sector is stable and behave as a candidate of DM. The quantum numbers of dark sector fields and that of SM Higgs 
under the SM gauge group, augmented by a $\mathcal{Z}_2$ symmetry, are given in Table \ref{tab:tab0}.


The Lagrangian of the model can be given as follows:
\bea\label{lag:lagVF}
\mathcal{L}^{VF} &=& \overline{N}~[i\gamma^{\mu}(\partial_{\mu} - i g \frac{\sigma^a}{2}W_{\mu}^a - i g^{\prime}\frac{Y}{2}B_{\mu})-m_N]~N
\nonumber \\
&&+ \overline{\chi}~(i\gamma^\mu \partial_{\mu}-m_{\chi})~\chi - (Y_1\overline{N}\widetilde{H}\chi+h.c) .
\eea
Note that here we have assumed to have a CP conserving interaction between the additional vector like fermion to the SM Higgs. One may also 
think of a coupling $- (Y_1\overline{N} \gamma_5 \widetilde{H}\chi+h.c)$ that will violate CP. Now, it is a bit intriguing to think of 
such interactions before the debate on the SM Higgs to be a scalar or a pseudoscalar is settled. The outcome of such an interaction will alter the subsequent phenomenology 
significantly. For example, it is known that advocating a pseudoscalar ($S$) interaction to a vector like DM ($\psi$), for example, with a term like 
$-yS\bar\psi\gamma_5\psi$ indicates that DM-nucleon scattering becomes velocity dependent and therefore reduces the direct search constraint significantly 
to allow the model live in a larger allowed parameter space. However, the Yukawa interaction term itself ($-yS\bar\psi\gamma_5\psi$) do not violate parity as $S$ is 
assumed to be pseudoscalar by itself (see for example in \cite{Ghorbani:2014qpa}).    

After Electroweak symmetry breaking (EWSB) the SM Higgs acquires a vacuum expectation value $v$. The quantum field around the vacuum
can be given as: 
$
H=
\left(\begin{matrix}
 0 && \frac{1}{\sqrt{2}} (v+h)
\end{matrix}\right)^T
$ 
where $v= 246$ GeV. The presence of the Yukawa term: $Y_1\overline{N}\widetilde{H}\chi$ term in the Lagrangian (Eq.~\ref{lag:lagVF}), arises an 
admixture between $N^0$ and $\chi$. The bare mass terms of the vector like fermions in $\mathcal{L}^{VF}$ then take the following form:

\bea
-\mathcal{L}^{VF}_{mass}&=&m_N\overline{N^0}N^0+m_N{N^+}N^-+m_{\chi} \overline{\chi}\chi +\frac{Y_1 v}{\sqrt2} \overline{N^0}\chi 
+\frac{Y_1 v}{\sqrt2} \overline{\chi}N^0 \nonumber \\
&=&
\overline{\left(\begin{matrix}
\chi & N^0 
\end{matrix}\right)}
{\left(\begin{matrix}
m_{\chi} & \frac{Y_1 v}{\sqrt2}\\
\frac{Y_1 v}{\sqrt2} & m_N
\end{matrix}\right)}
{\left(\begin{matrix}
\chi \\ N^0
\end{matrix}\right)}
+m_N{N^+}N^- . 
\eea

 The unphysical basis,$\left(\begin{matrix}
 \chi &&  N^0 
\end{matrix}\right)^T$ is related to physical basis, $\left(\begin{matrix}
 N_1 &&  N_2 
\end{matrix}\right)^T$ through the following unitary transformation:

\bea
 \left(\begin{matrix}
 \chi \\ N^0
\end{matrix}\right) 
=\mathcal U \left(\begin{matrix}
N_1 \\  N_2 
\end{matrix}\right)=\left(\begin{matrix}
 \cos\theta & -\sin\theta \\
 \sin\theta & \cos\theta
 \end{matrix}\right) 
\left(\begin{matrix}
N_1 \\  N_2 
\end{matrix}\right),
\eea

where the mixing angle
\bea\label{ref:mixang}
\tan{2\theta}= - \frac{\sqrt2 Y_1 v}{m_N-m_{\chi}} .
\eea
The mass eigenvalues of the physical states $N_1$ and $N_2$ are respectively given by:
\bea\label{ref:phymass}
m_{N_1} &=& m_{\chi} \cos^2\theta + m_N \sin^2\theta + \frac{Y_1 v}{\sqrt2}\sin 2\theta \nonumber \\
m_{N_2} &=& m_{\chi} \sin^2\theta + m_N \cos^2\theta - \frac{Y_1 v}{\sqrt2}\sin 2\theta
\eea

For small $\sin\theta$ ($\sin\theta \rightarrow 0$) limit, $m_{N_1}$  and $m_{N_2}$  can be further expressed as:
\bea
m_{N_1} &\simeq& m_{\chi}+\frac{Y_1 v}{\sqrt 2}\sin{2\theta} \equiv  m_{\chi}-\frac{(Y_1 v)^2}{(m_N-m_{\chi})} , \nonumber \\ 
m_{N_2} &\simeq& m_{N}-\frac{Y_1 v}{\sqrt 2}\sin{2\theta} \equiv  m_{N}+\frac{(Y_1 v)^2}{(m_N-m_{\chi})} .
\eea

Here we have considered $Y_1 v/\sqrt2 \ll m _{\chi} < m_N$. Hence $m_{N_1} < m_{N_2}$. Therefore $N_1$ becomes the stable DM 
candidate. From Eqs.\ref{ref:mixang} and \ref{ref:phymass}, one can write :
\bea\label{ref:reltn}
 Y_1 &=& - \frac{\Delta{m} \sin{2\theta}}{\sqrt2 v}, \nonumber \\
 m_N &=& m_{N_1}\sin^2\theta + m_{N_2} \cos^2\theta .
\eea
where $\Delta m = m_{N_2} - m_{N_1}$ \footnote{We would like to remind the readers that $N_1$ and $N_2$ are not same as $N_1^0$ and $N_2^0$. $N_{1,2}$ are the physical eigenstates 
arising out of the singlet ($\chi$) doublet ($N$) admixture. Here, $N_1$ is the DM while $N_2$ is the NLSP. Where as, $N_1^0$ and $N_2^0$ are the 
two pseudo Dirac states that emerge from the neutral component ($N^0$) of the vector-like lepton doublet ($N$) due to the majorana mass term 
acquired by $N^0$ in presence of the scalar triplet $\Delta$.} 
is the mass difference between the two mass eigenstates and $m_N$ is the mass of electrically charged 
component of vector like fermion doublet $N^-$.

Therefore, one can express interaction terms of $\mathcal{L}^{VF}$ in mass basis of $N_1$ and $N_2$ as
\bea
\mathcal{L}^{VF}_{int} &=& \Big(\frac{e_0}{2 \sin\theta_W \cos\theta_W}\Big) \Big[\sin^2\theta \overline{N_1}\gamma^{\mu}Z_{\mu}N_1+\cos^2\theta \overline{N_2}\gamma^{\mu}Z_{\mu}N_2 \nonumber \\
&&~~~~~~~~~~~~~~~~~~~~~~~~~~~~~~~~~~~~~+
\sin\theta \cos\theta(\overline{N_1}\gamma^{\mu}Z_{\mu}N_2+\overline{N_2}\gamma^{\mu}Z_{\mu}N_1)\Big]   \nonumber \\
&&+\frac{e_0}{\sqrt2\sin\theta_W}\sin\theta \overline{N_1}\gamma^\mu W_\mu^+ N^- +\frac{e_0}{\sqrt2 \sin\theta_W} \cos\theta \overline{N_2}\gamma^\mu W_\mu^+ N^-  \nonumber \\
&&  +\frac{e_0}{\sqrt2 \sin\theta_W} \sin\theta{N^+}\gamma^\mu W_\mu^- N_1 + \frac{e_0}{\sqrt2\sin\theta_W}\cos\theta {N^+}\gamma^\mu W_\mu^- N_2  \nonumber \\
&&- \Big(\frac{e_0}{2 \sin\theta_W\cos\theta_W}\Big) \cos2\theta_W {N^+}\gamma^{\mu}Z_{\mu}N^- - e_0 {N^+}\gamma^{\mu}A_{\mu}N^- \nonumber \\
&& -\frac{Y_1}{\sqrt2}h\Big[\sin2\theta(\overline{N_1}N_1-\overline{N_2}N_2)+\cos2\theta(\overline{N_1}N_2+\overline{N_2}N_1)\Big]
\eea


The relevant DM phenomenology of the model then mainly depend on following three independent parameters :
\bea\label{parms}
\{~m_{N_1},~\Delta m , ~\sin\theta \}
\eea

\subsection{Constraints on the model parameters }\label{constraints}
The model parameters are not totally free from theoretical and experimental bounds. Here we would like to discuss briefly the constraints 
coming from Perturbativity, invisible decay widths of $Z$ and $H$, and corrections to electroweak parameters.
\begin{itemize} 
 \item {\bf Perturbativity}: The upper limit of perturbativity bound on quartic and Yukawa couplings of the model are given 
by, 
 \begin{align}
  ~~ |Y_1|  < \sqrt{4 \pi}\,.
 \end{align}

\item {\bf Invisible decay width of Higgs } :
If the mass of DM is below $m_h/2$, then Higgs can decay to two invisible particles in final state and will yield 
invisible decay width. Recent Large Hadron Collider (LHC) data put strong constraint on the invisible branching fraction of Higgs to be 
$Br(h\rightarrow inv) \leq 0.24$ ~\cite{higgs_inv}, which can be expressed as: 

\bea
\frac{\Gamma(h \rightarrow inv. )}{\Gamma(h \rightarrow SM) + \Gamma(h \rightarrow inv.  )} \leq 0.24\,,  
\eea
where $\Gamma(h\to SM)=4.2$ MeV for Higgs mass $m_h=125.09~ {\rm GeV}$, obtained from recent measured LHC data~\cite{pdg}. Therefore, the 
invisible Higgs decay width is given by
\bea
\Gamma(h \rightarrow inv.  ) \leq 1.32 ~\rm~ MeV, 
\eea
where
\bea 
\Gamma(h \rightarrow inv.  )&=& \Gamma(h \rightarrow \overline{N_1} N_1  ). \nonumber \\
&=& \frac{1}{16 \pi} \Big(Y_1 \sin2\theta\Big)^2  m_h~~ \Big(1-\frac{4 m_{N_1}^2}{m_h^2}\Big)^\frac{3}{2} ~~\Theta(m_h-  2m_{N_1})\,,
\eea
where the step function $\Theta(m_h-  2m_{N_1})=1$ if $m_{N_1} \le m_h/2$ and is 0 if $m_{N_1} > m_h/2$. The decay width of Higgs 
to DM is proportional to $Y_1=- \frac{\Delta{m} \sin{2\theta}}{\sqrt2 v}$. Therefore, it depends on the mass splitting with NLSP ($\Delta m$) 
as well as on the doublet singlet mixing ($\sin\theta$). The invisible Higgs decay constraint on the model for $m_{N_1} < m_h/2$ is shown in 
Fig.~\ref{fig:Hinv}, where we have shown the exclusion in $M_{N_1}-\sin\theta$ plane for different choices of $\Delta m$ ranging from 10 GeV 
to 500 GeV. The inner side of the contour is excluded. This essentially shows that with large $\Delta m \sim 500$ GeV, essentially all of 
$m_{N_1} \le m_h/2$  is excluded, while for a small $\Delta m \sim 10$ GeV, the constraint is milder due to less Higgs decay width and 
excludes only regions for DM masses $M_{N_1} \le$ 30 GeV within $\sin\theta \sim\{0.6-0.8\}$. Therefore, even in $m_{N_1} \le m_h/2$ region, 
if $\Delta m$ and $\sin\theta$ are small, which turns out to be the case for satisfying relic density and direct search as we demonstrate 
later, then are allowed by the Higgs invisible decay constraint.

\begin{figure}[htb!]
$$
 \includegraphics[height=8.0cm]{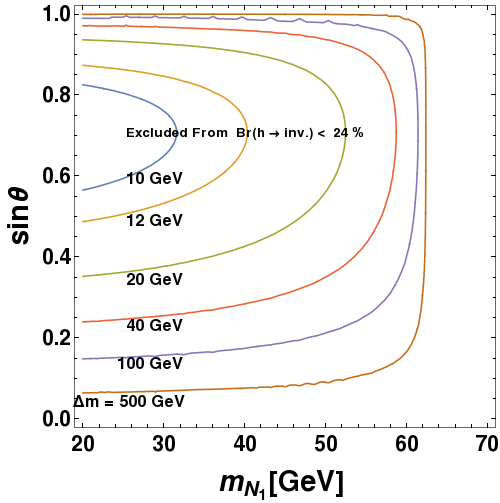} 
$$
 \caption{Constraints from Higgs invisible decay width is shown in $m_{N_1}-\sin\theta$ plane. Here each contour line correspond to 
different value of $\Delta m$ depicted in the figure. Inner region of each contour line excluded from Higgs invisible decay width, 
$\Gamma(h \rightarrow inv.  ) \leq 1.32 ~\rm~ MeV$ for a particular value of $\Delta m $.  }
\label{fig:Hinv}
\end{figure}

\item {\bf Invisible decay width of Z }: If the masses of dark sector particles are below $m_Z/2$, then $Z$ can decay to dark sector particles leading to 
an increase of $Z$-decay width. However, from current observation, invisible decay width of $Z$ boson is strongly bounded. The upper limit on invisible 
$Z$-decay width is given by~\cite{pdg}:
\bea
\Gamma(Z \rightarrow inv.) = 499.0\pm 1.5 ~~ \rm ~MeV \,,
\eea
where the decay of $Z$ to $N_1$ DM is given as: 
\bea
\Gamma(Z\rightarrow \overline{N_1} N_1) = \frac{1}{48 \pi} \Big(\frac{g~ \sin^2\theta}{\cos\theta_W}\Big)^2 ~m_Z~~ \Big(1+\frac{2 m_{N_1}^2}{m_Z^2}\Big)\sqrt{1-\frac{4 m_{N_1}^2}{m_Z^2}}~~\Theta(m_Z-2 m_{N_1})~.
\eea

The $Z-$ invisible decay does not play a crucial role in small $\sin\theta$ regions, which are required for the DM to achieve correct relic density, 
thus allowing almost all of the $M_{N_1} \le m_Z/2$ parameter space of the model.  
\end{itemize}

\subsection{Corrections to the electroweak precision  parameters}

Addition of a vector like fermion doublet to the SM gives correction to the electroweak precision test parameters $S,T$ and $U$\cite{peskin, EW_constraint1}.  The values of these parameters are tightly constrained by experiments. The new observed parameters are infect four in number  $\hat{S}$, $\hat{T}$, $W$ and $Y$~\cite{EW_constraint2}, where the $\hat{S}$, $\hat{T}$ are related to Peskin-Takeuchi parameters $S$, $T$ as $\hat{S}=\alpha S/4 \sin^2 \theta_w$, $\hat{T}=\alpha T$, 
while $W$ and $Y$ are two new set of parameters. The measured values of these parameters at LEP-I and LEP-II put a lower bound on the mass scale of vector like fermions.  The result of a global fit of the parameters is presented in the table \ref{tab:EWPP} for a light Higgs~\cite{EW_constraint2}~\footnote{The value 
$\hat{S}$, $\hat{T}$, $W$ and $Y$ are obtained using a Higgs mass $m_h=115$ GeV. However, we now know that the SM Higgs mass is 125 GeV. Therefore, 
the value of $\hat{S}$, $\hat{T}$, $W$ and $Y$ are expected to change. But this effect is nullified by the small values of $\sin \theta$.}.
\begin{table}[h]
\begin{center}
\begin{tabular}{|c|c|c|c|c|}
\hline 
  & $10^3 \hat{S}$ & $10^3 \hat{T}$ & $10^3 W$ & $10^3 Y$ \\\hline
Light Higgs & $0.0\pm 1.3$ & $0.1 \pm 0.9$ & $0.1 \pm 1.2$ & $-0.4\pm 0.8$ \\ \hline
\end{tabular}
\end{center}
\caption{Global fit for the electroweak precision parameters taken from ref.~\cite{EW_constraint2}.}
\label{tab:EWPP}
\end{table} 

In the present scenario, we have a vector like doublet and a singlet fermion field are added to the SM.  But the physical states are a charged fermion $N^-$, and two singlet doublet mixed neutral fermions $N_1$ (dominant singlet component) and $N_2$ (dominant doublet component).  Therefore, the contribution to the precision parameters also depends on the mixing angle $\sin\theta$. The expression for $\hat{S}$ in terms $m_{N_1}$, $m_{N_2}$, $m_N$ and $\sin \theta$ of is given as~\cite{Cynolter:2008ea}:
\begin{eqnarray}
\hat{S} &=& \frac{g^2}{16\pi^2} \left[ \frac{1}{3} \left\{  \ln \left(\frac{\mu_{ew}^2}{m_N^2} \right) -\cos^4 \theta \ln \left(\frac{\mu_{ew}^2}{m_{N_2}^2} \right)
- \sin^4 \theta \ln \left(\frac{\mu_{ew}^2}{m_{N_1}^2} \right) \right\} - 2\sin^2 \theta \cos^2 \theta \left\{   
\ln \left(\frac{\mu_{ew}^2}{m_{N_1} m_{N_2}} \right) \right.\right.\nonumber\\ 
 && \left. \left. + \frac{m_{N_1}^4-8 m_{N_1}^2 m_{N_2}^2 +m_{N_2}^4}{9 (m_{N_1}^2-m_{N_2}^2)^2} + \frac{(m_{N_1}^2+m_{N_2}^2)(m_{N_1}^4-4m_{N_1}^2m_{N_2}^2+m_{N_2}^4)}{6(m_{N_1}^2-m_{N_2}^2)^3} 
\ln \left(\frac{m_{N_2}^2}{m_{N_1}^2} \right) \right.\right.\nonumber\\
&& \left. \left. +   \frac{m_{N_1} m_{N_2}(m_{N_1}^2 + m_{N_2}^2)}{2(m_{N_1}^2-m_{N_2}^2)^2} + \frac{m_{N_1}^3m_{N_2}^3}{(m_{N_1}^2-m_{N_2}^2)^3} \ln \left(\frac{m_{N_2}^2}{m_{N_1}^2} \right)      \right\}          \right]
\end{eqnarray}
where $\mu_{ew}$ is at the EW scale.

\begin{figure}[htbp]
\includegraphics[scale=.3]{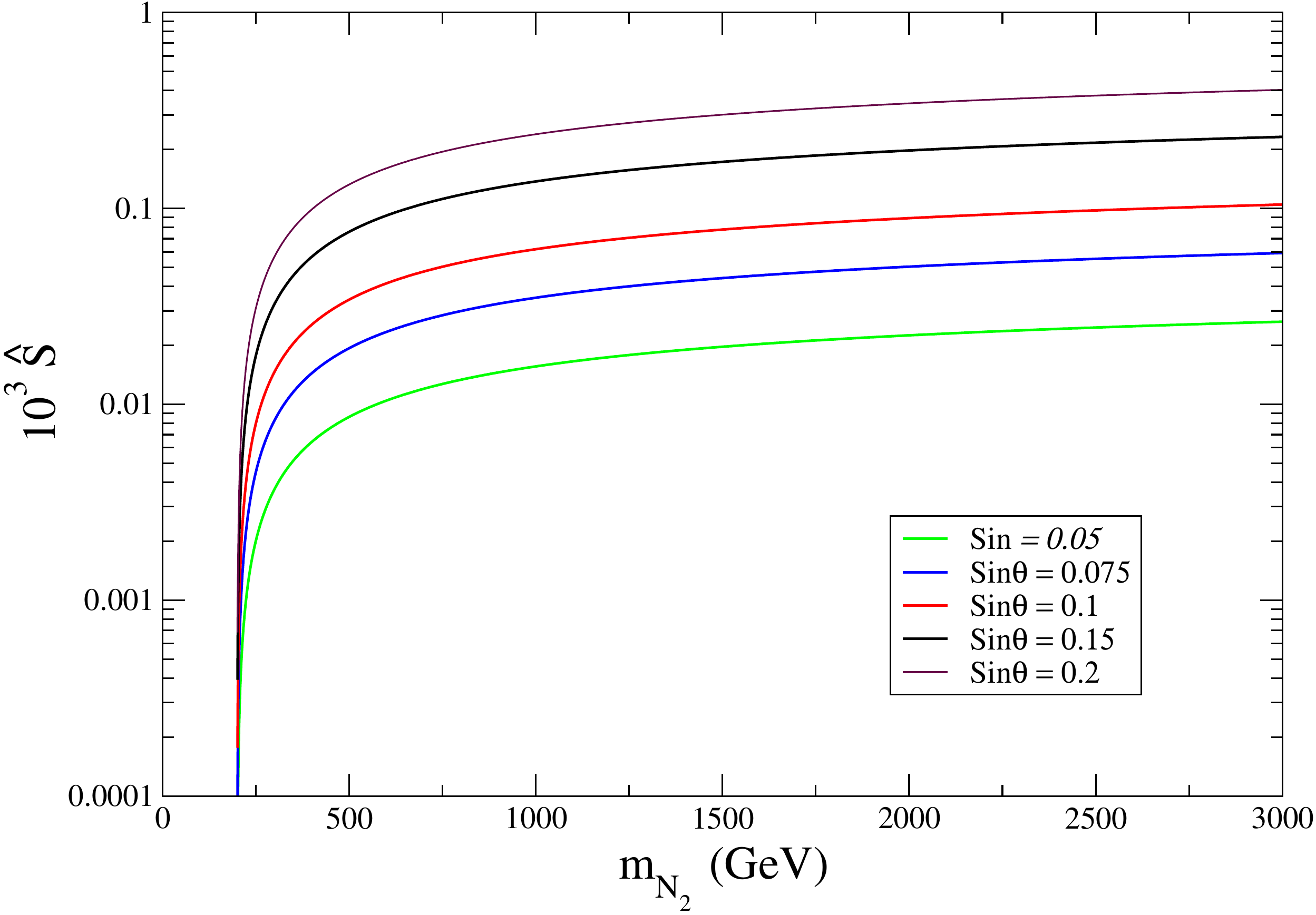}
\includegraphics[scale=0.3]{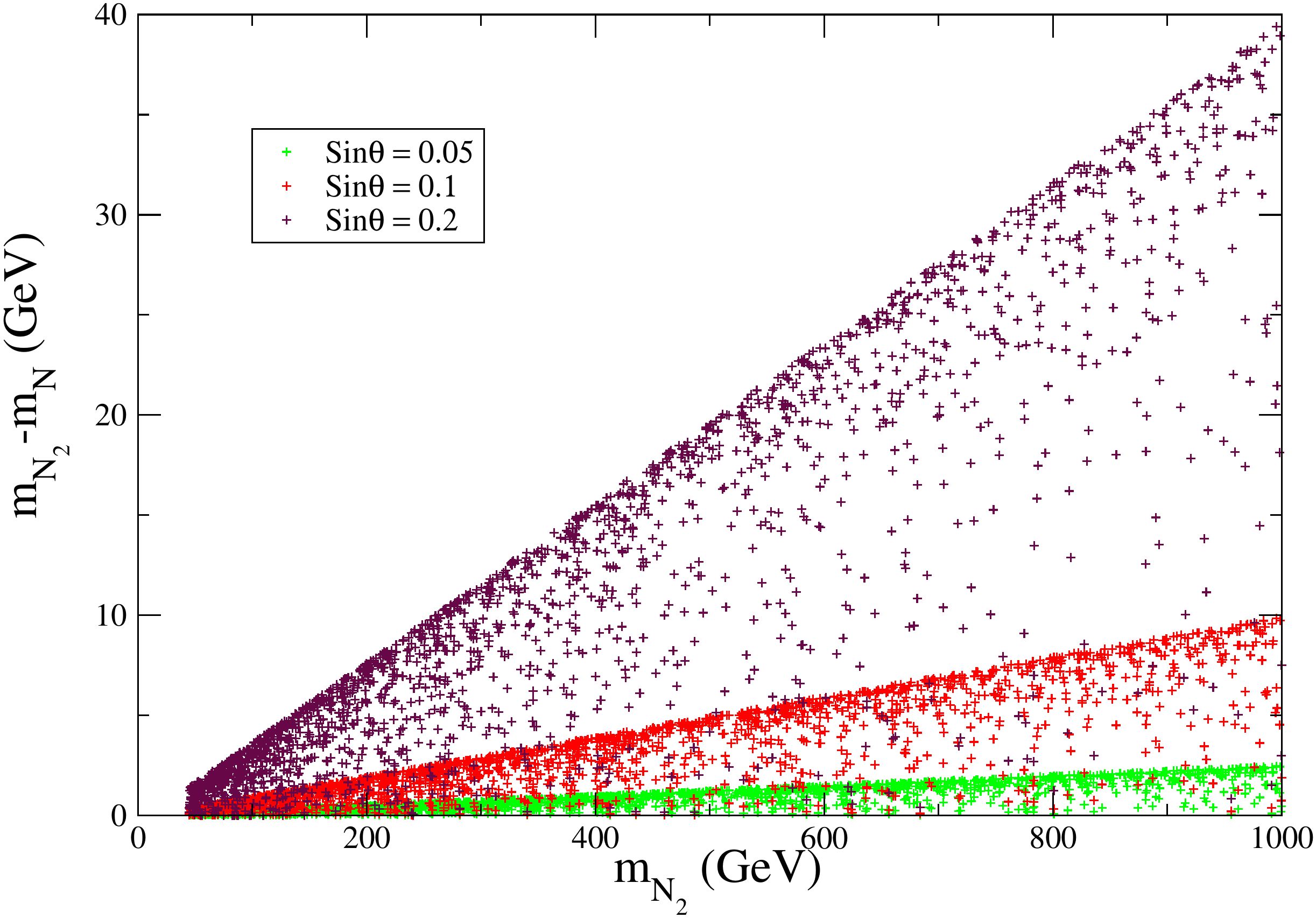}
\caption{In the left panel, $\hat{S}$ is shown as a function of $m_{N_2}$ for $m_{N_1}=200$ GeV and $\rm sin\theta = 0.05$ (Green colour), 
$\rm sin\theta = 0.075$ (Blue color), $\rm sin\theta = 0.1$ (Red color),  $\rm sin\theta = 0.15$ (Black color) and  $\rm sin\theta = 0.2$ (Maroon color).  In the right panel, allowed values of $\hat{S}$ is plotted in
 $m_{N_2}-m_N$ versus $m_{N_2}$ plane for $\sin \theta=0.05$ (Green Color),  $\rm sin\theta = 0.1$ (Red color) and  $\rm sin\theta = 0.2$ (Maroon color).}\label{shat_values}
\end{figure}
In the left panel of Fig.~\ref{shat_values}, we have plotted $\hat{S}$ as a function of $m_{N_2}$ keeping $m_{N_1}=200$ GeV for different values of the mixing angle.  In the right panel, we have shown the allowed values of $\hat{S}$ in the plane of $m_{N_2}-m_N$ versus $m_{N_2}$ for $\sin \theta=0.05$ (Green Color), $\rm sin\theta = 0.1$ (Red color) and  $\rm sin\theta = 0.2$ (Maroon color). We observed 
that $\hat{S}$ does not put strong constraints on $m_{N_1}$ and $m_{N_2}$. Moreover, small values of $\sin \theta$ allows a small mass splitting between $N_2$ and $N^-$ 
which relaxes the constraint on $\hat{T}$ parameter as we discuss below. The expression for $\hat{T}$ is given as~\cite{Cynolter:2008ea}:  
\begin{equation}
\hat{T}=\frac{g^2}{16 \pi^2 M_W^2}\left[2 \sin^2 \theta \cos^2 \theta ~\Pi(m_{N_1},m_{N_2},0)-2\cos^2 \theta ~\Pi(m_N,m_{N_2},0)-2\sin^2\theta ~\Pi(m_N,m_{N_1},0)\right]\,,
\end{equation}
 where $\Pi(a,b,0)$ is given by: 
\begin{eqnarray}
\Pi (a,b,0) &=& -\frac{1}{2}(M_a^2+M_b^2)\left({\rm Div}+\ln \left(\frac{\mu_{ew}^2}{M_a M_b} \right)  \right)-\frac{1}{4} (M_a^2+ M_b^2)-\frac{(M_a^4+M_b^4)}
{4(M_a^2-M_b^2)} \ln \frac{M_b^2}{M_a^2} \nonumber\\
&& + M_a M_b \left\{{\rm Div}+ \ln  \left(\frac{\mu_{ew}^2}{M_a M_b} \right)+1+ \frac{(M_a^2 +M_b^2)}{2(M_a^2-M_b^2)} \ln  \frac{M_b^2}{M_a^2}  \right\} \,,
\end{eqnarray}
with Div$=\frac{1}{\epsilon} + \rm ln 4 \pi - \gamma_\epsilon $ contains the divergent term in dimensional regularisation method.
\begin{figure}[!htb]
\includegraphics[scale=.3]{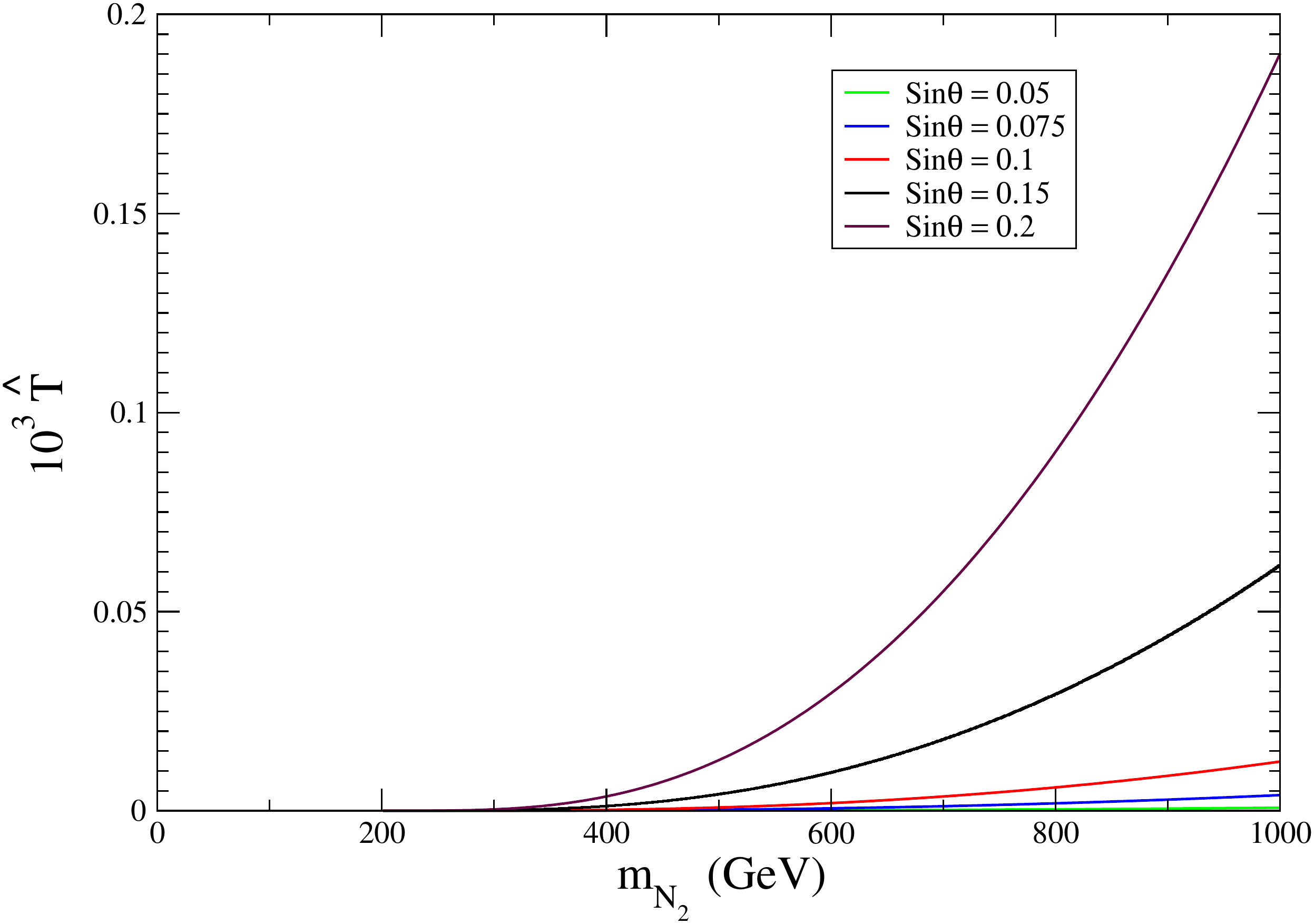}
\includegraphics[scale=0.3]{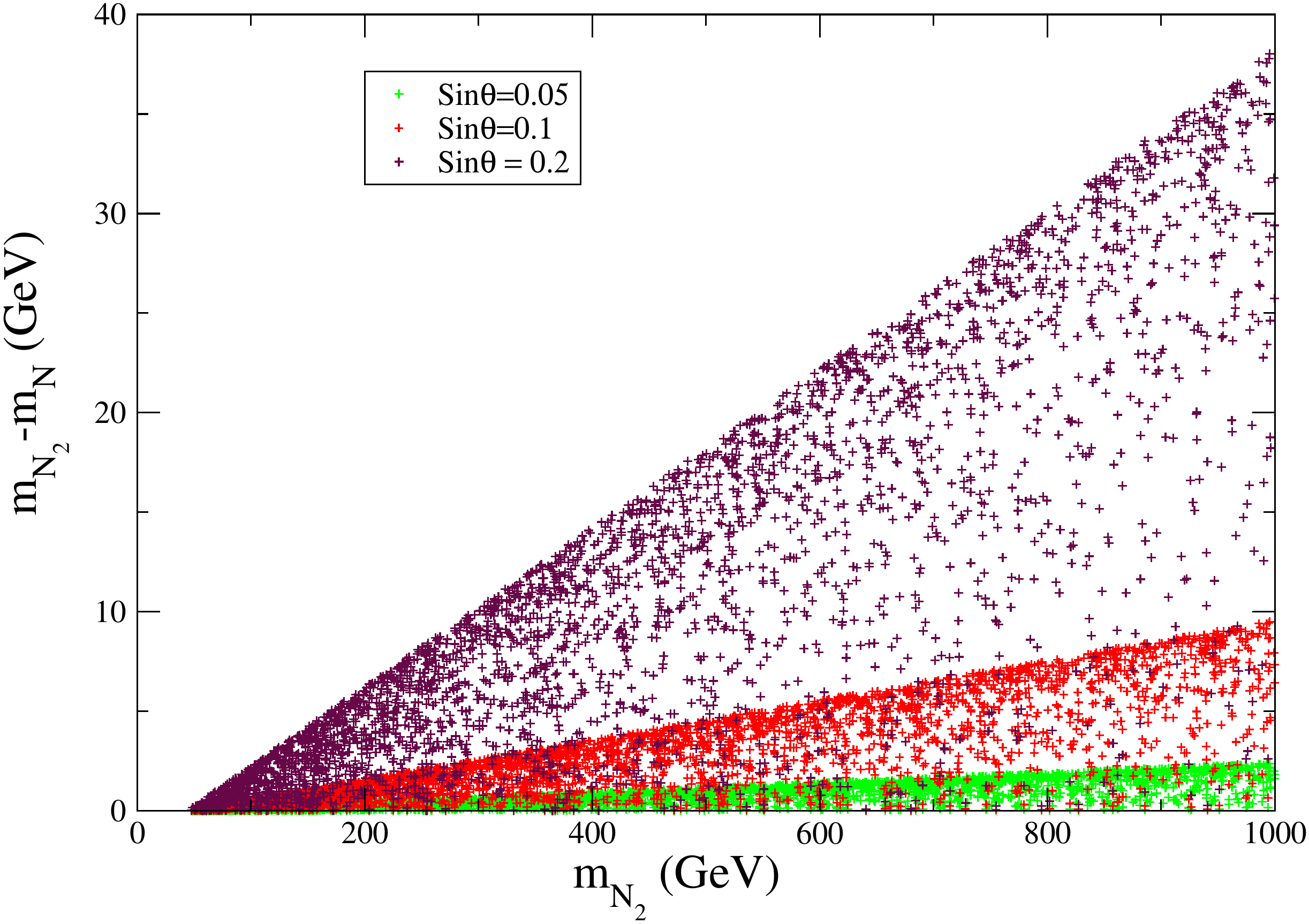}
\caption{In the left panel, $\hat{T}$ is shown as a function of $m_{N_2}$ for $m_{N_1}=200$ GeV and $\rm sin\theta = 0.05$ (Green colour), 
$\rm sin\theta = 0.075$ (Blue color), $\rm sin\theta = 0.1$ (Red color),  $\rm sin\theta = 0.15$ (Black color) and  $\rm sin\theta = 0.2$ (Maroon color). In the right panel, allowed values of $\hat{T}$ is plotted in 
 $m_{N_2}-m_N$ versus $m_{N_2}$ plane for $\sin \theta=0.05$ (Green color), $\sin \theta=0.1$ (Red color) $\rm sin\theta = 0.2$ (Maroon color).}\label{that_values}
\end{figure}
From the left panel of Fig.~(\ref{that_values}) we see that for $\sin \theta < 0.05$ we don't get strong constraints on $m_{N_2}$ and $m_{N_1}$. Moreover, 
small values of $\sin \theta$ restricts the value of $m_{N_2}-m_N$ to be less than a GeV. As a result large $m_{N_2}$ values are also allowed. Near $m_{N_2} 
\approx m_N$, $\hat{T}$ vanishes as expected. The value of $Y$ and $W$ are usually suppressed by the masses of new fermions. Since the allowed masses 
of $N_1$, $N_2$ and $N^\pm$ are above 100 GeV by the relic density constraint (to be discussed later), so $Y$ and $W$ are naturally suppressed.

\subsection{Relic density of singlet-doublet leptonic dark matter}\label{relic_density}
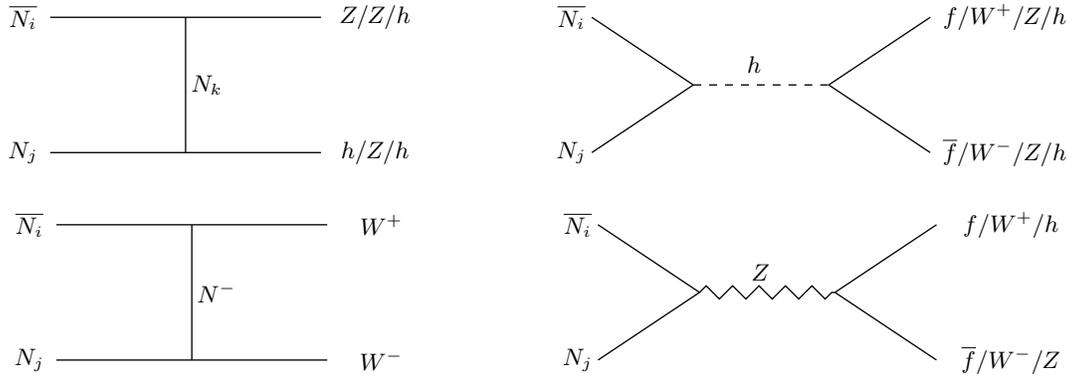
\begin{figure}[htb!]
\begin{center}
    \begin{tikzpicture}[line width=0.5 pt, scale=0.9]
        \draw[solid] (-3,1.0)--(-1.0,1.0);
        \draw[solid] (-3,-1.0)--(-1.0,-1.0);
        \draw[solid] (-1.0,1.0)--(-1.0,-1.0);
        \draw[solid] (-1.0,1.0)--(1.0,1.0);
        \draw[solid] (-1.0,-1.0)--(1.0,-1.0);
        \node at (-3.4,1.0) {$\overline{N_i}$};
        \node at (-3.4,-1.0) {$N_j$};
        \node [right] at (-1.05,0.0) {$N_k$};
        \node at (1.8,1.0) {$Z/Z/h$};
        \node at (1.8,-1.0) {$h/Z/h$};
        \draw[solid] (5.0,1.0)--(6.5,0.0);
        \draw[solid] (5.0,-1.0)--(6.5,0.0);
        \draw[dashed] (6.5,0.0)--(8.5,0.0);
        \draw[solid] (8.5,0.0)--(10.0,1.0);
        \draw[solid] (8.5,0.0)--(10.0,-1.0);
        \node at (4.7,1.0) {$\overline{N_i}$};
        \node at (4.7,-1.0) {$N_j$};
        \node [above] at (7.4,0.05) {$h$};
        \node at (11.1,1.0) {$f/W^+/Z/h$};
        \node at (11.1,-1.0) {$\overline{f}/W^-/Z/h$};
     \end{tikzpicture}
 \end{center}
\begin{center}
    \begin{tikzpicture}[line width=0.5 pt, scale=0.9]
        \draw[solid] (-3,1.0)--(-1.0,1.0);
        \draw[solid] (-3,-1.0)--(-1.0,-1.0);
        \draw[solid](-1.0,1.0)--(-1.0,-1.0);
        \draw[solid] (-1.0,1.0)--(1.0,1.0);
        \draw[solid] (-1.0,-1.0)--(1.0,-1.0);
        \node at (-3.4,1.0) {$\overline{N_i}$};
        \node at (-3.4,-1.0) {$N_j$};
        \node [right] at (-1.05,0.0) {$N^-$};
        \node at (1.8,1.0) {$W^+$};
        \node at (1.8,-1.0) {$W^-$};
        \draw[solid] (5.0,1.0)--(6.5,0.0);
        \draw[solid] (5.0,-1.0)--(6.5,0.0);
        \draw[snake] (6.5,0.0)--(8.5,0.0);
        \draw[solid] (8.5,0.0)--(10.0,1.0);
        \draw[solid] (8.5,0.0)--(10.0,-1.0);
        \node at (4.7,1.0) {$\overline{N_i}$};
        \node at (4.7,-1.0) {$N_j$};
        \node [above] at (7.4,0.05) {$Z$};
        \node at (11.1,1.0) {$f/W^+/h$};
        \node at (11.1,-1.0) {$\overline{f}/W^-/Z$};
     \end{tikzpicture}
 \end{center}
\caption{Annihilation ($i=j$) and co-annihilation ($i\neq j$) of vector-like fermion DM. Here $(i,j=1,2)$. }
\label{fd:an-coan}
 \end{figure}

As stated earlier, the lightest stable physical state $N_1$ is the DM, which is an admixture of a singlet vector-like fermion $(\chi)$ and 
the neutral component of a vector-like fermionic doublet $(N)$. Due to presence of mass hierarchy between dark sector particles 
$N_1,~N_2 ~~{\rm and} ~~N^-$, the lightest component $N_1$ not only annihilate with itself but also co-annihilate with $N_2$ and $N^-$ 
to yield a net a relic density. The relevant diagrams are shown in Figs.~\ref{fd:an-coan}, \ref{co-ann-2}, \ref{co-ann-3}. 
\begin{figure}[htb!]
\begin{center}
    \begin{tikzpicture}[line width=0.5 pt, scale=0.9]
        \draw[solid] (-3,1.0)--(-1.0,1.0);
        \draw[solid] (-3,-1.0)--(-1.0,-1.0);
        \draw[solid] (-1.0,1.0)--(-1.0,-1.0);
        \draw[solid] (-1.0,1.0)--(1.0,1.0);
        \draw[solid] (-1.0,-1.0)--(1.0,-1.0);
        \node at (-3.4,1.0) {$\overline{N_i}$};
        \node at (-3.4,-1.0) {$N^-$};
        \node [right] at (-1.05,0.0) {$N_j$};
        \node at (1.8,1.0) {$Z/h$};
        \node at (1.8,-1.0) {$W^-/W^-$};
        \draw[solid] (5.0,1.0)--(6.5,0.0);
        \draw[solid] (5.0,-1.0)--(6.5,0.0);
        \draw[snake] (6.5,0.0)--(8.5,0.0);
        \draw[solid] (8.5,0.0)--(10.0,1.0);
        \draw[solid] (8.5,0.0)--(10.0,-1.0);
        \node at (4.7,1.0) {$\overline{N_i}$};
        \node at (4.7,-1.0) {$N^-$};
        \node [above] at (7.4,0.05) {$W$};
        \node at (11.1,1.0) {$f/h/W/W$};
        \node at (11.1,-1.0) {$\overline{f^\prime}/W/A/Z$};
     \end{tikzpicture}
 \end{center}
\begin{center}
    \begin{tikzpicture}[line width=0.5 pt, scale=0.9]
        \draw[solid] (-3,1.0)--(-1.0,1.0);
        \draw[solid] (-3,-1.0)--(-1.0,-1.0);
        \draw[solid](-1.0,1.0)--(-1.0,-1.0);
        \draw[solid] (-1.0,1.0)--(1.0,1.0);
        \draw[solid] (-1.0,-1.0)--(1.0,-1.0);
        \node at (-3.4,1.0) {$\overline{N_i}$};
        \node at (-3.4,-1.0) {$N^-$};
        \node [right] at (-1.05,0.0) {$N^-$};
        \node at (1.8,1.0) {$W^-$};
        \node at (1.8,-1.0) {$A/Z$};
     \end{tikzpicture}
 \end{center}
\caption{ Co-annihilation process of $N_i ~(i=1,2)$ with the charge component $N^-$ to SM particles. }
\label{co-ann-2}
 \end{figure}
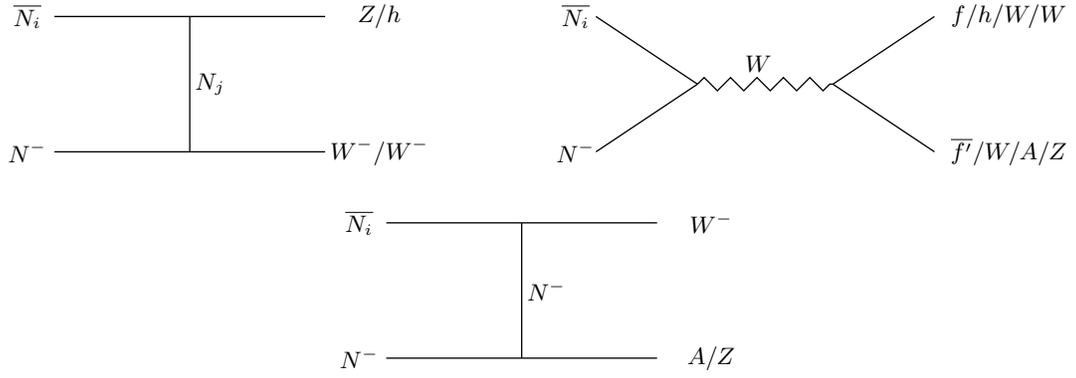
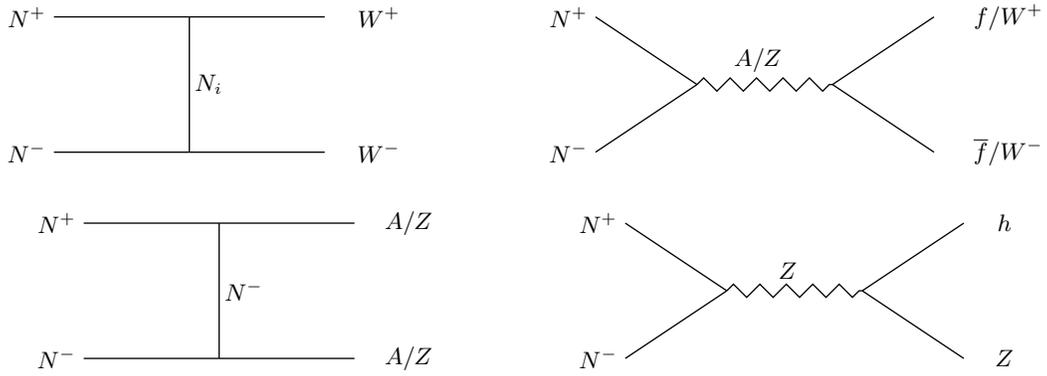
\begin{figure}[htb!]
\begin{center}
    \begin{tikzpicture}[line width=0.5 pt, scale=0.9]
        \draw[solid] (-3,1.0)--(-1.0,1.0);
        \draw[solid] (-3,-1.0)--(-1.0,-1.0);
        \draw[solid] (-1.0,1.0)--(-1.0,-1.0);
        \draw[solid] (-1.0,1.0)--(1.0,1.0);
        \draw[solid] (-1.0,-1.0)--(1.0,-1.0);
        \node at (-3.4,1.0) {${N^+}$};
        \node at (-3.4,-1.0) {$N^-$};
        \node [right] at (-1.05,0.0) {$N_i$};
        \node at (1.8,1.0) {$W^+$};
        \node at (1.8,-1.0) {$W^-$};
        \draw[solid] (5.0,1.0)--(6.5,0.0);
        \draw[solid] (5.0,-1.0)--(6.5,0.0);
        \draw[snake] (6.5,0.0)--(8.5,0.0);
        \draw[solid] (8.5,0.0)--(10.0,1.0);
        \draw[solid] (8.5,0.0)--(10.0,-1.0);
        \node at (4.6,1.0) {${N^+}$};
        \node at (4.6,-1.0) {$N^-$};
        \node [above] at (7.4,0.05) {$A/Z$};
        \node at (11.1,1.0) {$f/W^+$};
        \node at (11.1,-1.0) {$\overline{f}/W^-$};
     \end{tikzpicture}
 \end{center}
\begin{center}
    \begin{tikzpicture}[line width=0.5 pt, scale=0.9]
        \draw[solid] (-3,1.0)--(-1.0,1.0);
        \draw[solid] (-3,-1.0)--(-1.0,-1.0);
        \draw[solid](-1.0,1.0)--(-1.0,-1.0);
        \draw[solid] (-1.0,1.0)--(1.0,1.0);
        \draw[solid] (-1.0,-1.0)--(1.0,-1.0);
        \node at (-3.4,1.0) {${N^+}$};
        \node at (-3.4,-1.0) {$N^-$};
        \node [right] at (-1.05,0.0) {$N^-$};
        \node at (1.8,1.0) {$A/Z$};
        \node at (1.8,-1.0) {$A/Z$};
        \draw[solid] (5.0,1.0)--(6.5,0.0);
        \draw[solid] (5.0,-1.0)--(6.5,0.0);
        \draw[snake] (6.5,0.0)--(8.5,0.0);
        \draw[solid] (8.5,0.0)--(10.0,1.0);
        \draw[solid] (8.5,0.0)--(10.0,-1.0);
        \node at (4.6,1.0) {${N^+}$};
        \node at (4.6,-1.0) {$N^-$};
        \node [above] at (7.4,0.05) {$Z$};
        \node at (10.6,1.0) {$h$};
        \node at (10.6,-1.0) {$Z$};
     \end{tikzpicture}
 \end{center}
\caption{Co-annihilation process of charged fermions $N^\pm$ to SM particles in final states . }
\label{co-ann-3}
 \end{figure}
 
We assume all the heavier particles: $N_2 \rm ~~and~~ N^-$ in the dark sector ultimately decay to lightest stable particle $N_1$. So in this scenario 
one can write the Boltzmann equation in terms of total number density $n=n_{N_1}+n_{N_2}+n_{N^\pm}$ as 
 \bea
 \frac{dn}{dt} + 3 H n = -{\langle \sigma v\rangle}_{eff} \Big(n^2-n_{eq}^2\Big),
 \eea
 where
\bea
{\langle \sigma v\rangle}_{eff}&&= \frac{g_1^2}{g_{eff}^2} {\langle \sigma v \rangle}_{\overline{N_1}N_1}+\frac{2 g_1 g_2}{g_{eff}^2} {\langle \sigma v \rangle}_{\overline{N_1}N_2}\Big(1+\frac{\Delta m}{m_{N_1}}\Big)^\frac{3}{2} e^{-x \frac{\Delta m}{m_{N_1}}} \nonumber \\
&&+\frac{2 g_1 g_3}{g_{eff}^2} {\langle \sigma v \rangle}_{\overline{N_1}N^-}\Big(1+\frac{\Delta m}{m_{N_1}}\Big)^\frac{3}{2} e^{-x \frac{\Delta m}{m_{N_1}}} \nonumber \\
&& +\frac{2 g_2 g_3}{g_{eff}^2} {\langle \sigma v \rangle}_{\overline{N_2}N^-}\Big(1+\frac{\Delta m}{m_{N_1}}\Big)^3 e^{- 2 x \frac{\Delta m}{m_{N_1}}} \nonumber \\
&& +\frac{g_2^2}{g_{eff}^2} {\langle \sigma v \rangle}_{\overline{N_2}N_2}\Big(1+\frac{\Delta m}{m_{N_1}}\Big)^3 e^{- 2 x \frac{\Delta m}{m_{N_1}}} \nonumber \\
&& +\frac{g_3^2}{g_{eff}^2} {\langle \sigma v \rangle}_{{N^+}N^-}\Big(1+\frac{\Delta m}{m_{N_1}}\Big)^3 e^{- 2 x \frac{\Delta m}{m_{N_1}}}. 
\label{eq:vf-ann}
\eea

In above equation, $g_{eff}$, defined as effective degrees of freedom, which is given by
\bea
g_{eff}=g_1 + g_2 \Big(1+\frac{\Delta m}{m_{N_1}}\Big)^\frac{3}{2} e^{-x \frac{\Delta m}{m_{N_1}}} + g_3\Big(1+\frac{\Delta m}{m_{N_1}}\Big)^\frac{3}{2} e^{-x \frac{\Delta m}{m_{N_1}}} ,
\eea
where $g_1 ,~ g_2 \rm ~and~ g_3$ are the degrees of freedom of $N_1, ~N_2 \rm ~and~ N^-$ respectively and $x=x_f=\frac{m_{N_1}}{T_f}$, where $T_f$ is the freeze out temperature. Then the relic density of the $N_1$ DM can be given by~\cite{griest,co_ann_papers}
\begin{equation}
\Omega_{N_1} h^2 = \frac{1.09 \times 10^9 \rm GeV^{-1}}{g_\star ^{1/2} M_{PL}} \frac{1}{J(x_f)}\,,
\end{equation}
where $J(x_f)$ is given by
\begin{equation}
J(x_f)= \int_{x_f}^ \infty \frac{\langle \sigma |v| \rangle _{eff}}{x^2} \hspace{.2cm} dx \,.
\end{equation}
We note here that the freeze-out abundance of $N_1$ DM is controlled by the annihilation and co-annihilation channels as shown in Fig.~\ref{fd:an-coan}, 
\ref{co-ann-2} and \ref{co-ann-3}. Therefore, the important parameters which decide the relic abundance of $N_1$ are mass of DM ($m_{N_1}$), the 
mass splitting ($\Delta m$) between the DM and the next-to-lightest stable particle (NLSP) and the singlet-doublet mixing angle $\sin \theta$. 
Here we use MicrOmega~\cite{micromega} to calculate the relic density of $N_1$ DM. 

\begin{figure}[htb!]
$$
 \includegraphics[height=5.0cm]{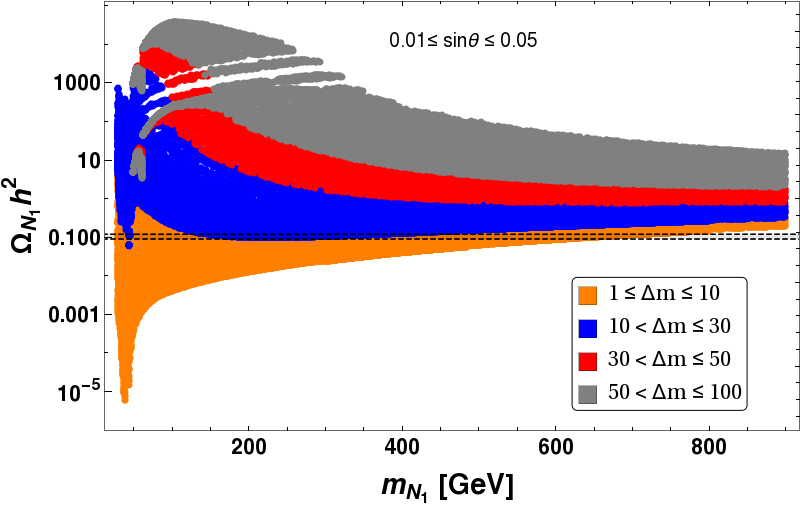} 
 \includegraphics[height=5.0cm]{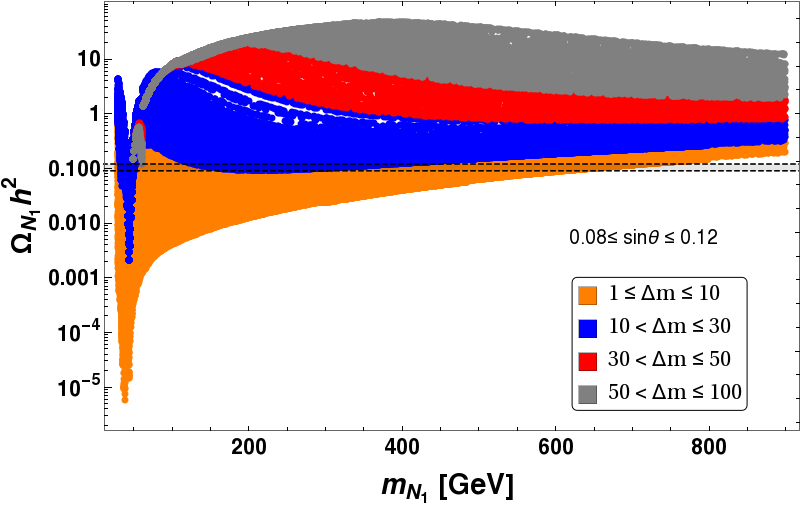}
$$
 \caption{Variation of relic density with DM mass $m_{N_1}$ keeping fixed range of $\sin\theta$: $0.01\leq \sin\theta \leq 0.05 $ (left pannel) 
and $0.08\leq \sin\theta \leq 0.12 $ (right panel). The different color patches corresponds to different $\Delta m $ region : $ 1 \leq \Delta m 
({\rm in GeV})\leq 10 $ (Orange), $ 10 < \Delta m ({\rm in GeV})\leq 30 $ (Blue), $ 30 <  \Delta m ({\rm in GeV}) \leq 50 $ (Red) and 
$ 50 < \Delta m ({\rm in GeV}) \leq 100 $ (Gray). Correct relic density, $0.1166 \leq \Omega h^2 \leq 0.1206$ is shown by black dashed line. All 
the masses are in GeVs.}
\label{fig:relicVSmN1_A}
\end{figure}

\begin{figure}[htb!]
$$
 \includegraphics[height=5.0cm]{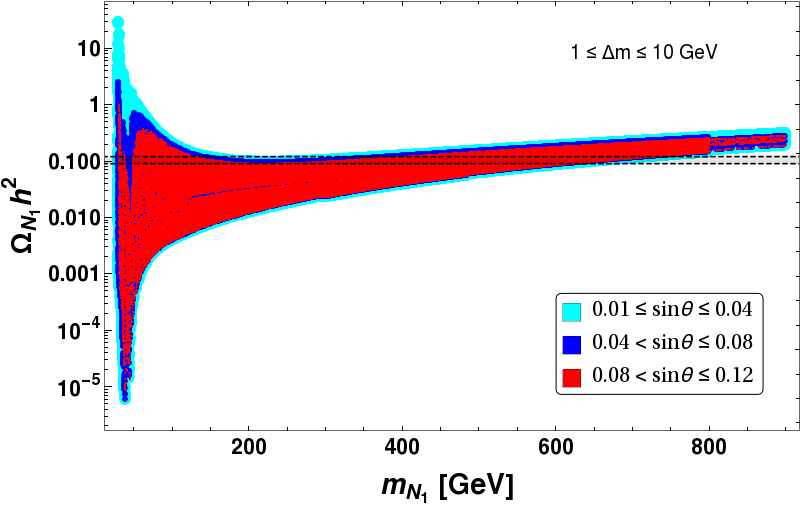} 
 \includegraphics[height=5.0cm]{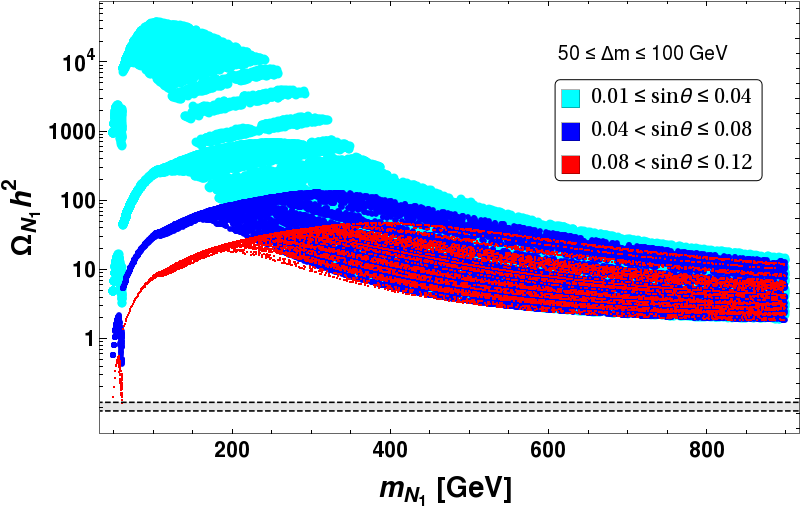}
$$
 \caption{Variation of relic density with DM mass $m_{N_1}$ keeping fixed region of $\Delta m$: $1\leq \Delta m \leq 10 $ GeV (left panel) and $50\leq \Delta m \leq 100 $ GeV (right panel). The different color patches are corresponding to different $\sin\theta $ region : $ 0.01 \leq \sin\theta \leq 0.04 $ (Cyan), $ 0.04 < \sin\theta \leq 0.08 $ (Blue), $ 0.08 < \sin\theta \leq 0.12 $ (Red). Correct relic density, $0.1166 \leq \Omega h^2 \leq 0.1206$ is shown by black dashed line. All the masses are in GeVs. }
 \label{fig:relicVSmN1_B}
\end{figure}

Variation of relic density of $N_1$ DM is shown in Fig.~\ref{fig:relicVSmN1_A} as a function of its mass, for a fixed 
$\Delta m=10 - 100$ GeV (in left and right panels of Fig.~\ref{fig:relicVSmN1_A} respectively) and different choices of mixing angle $\sin\theta$. We 
note that the annihilation cross-section increases with $\sin\theta$, due to larger $SU(2)$ component, resulting in smaller relic density. The resonance 
drop at $m_Z/2$ and at $m_h/2$ is observed due to $s$-channel $Z$ and $H$ mediated contributions to relic abundance. Another important feature of 
Fig.~\ref{fig:relicVSmN1_A} is that when $\Delta m$ is small, relic density is smaller due to large co-annihilation contribution (less Boltzmann 
suppression followed from Eq.~\ref{eq:vf-ann}). This feature can also be corroborated from Fig.~\ref{fig:relicVSmN1_A}, where we have shown relic 
density as a function of DM mass by keeping a fixed range of $\sin \theta$ and chosen different possible $\Delta m$. Alternatively in Fig.~\ref{fig:relicVSmN1_B}, 
we have shown relic density as a function of DM mass by keeping a fixed range of $\Delta m$, while varying $\sin \theta$. We see from the left panel 
of Fig.~\ref{fig:relicVSmN1_B} that for small $\Delta m$ co-annihilation dominates and hence the effect of $\sin \theta$ on relic abundance is quite 
negligible. On the other hand, from the right panel of Fig.~\ref{fig:relicVSmN1_B}, we see that for large $\Delta m$, where co-annihilation is suppressed, 
the effect of $\sin \theta$ on relic abundance is clearly visible. For small $\sin \theta$, the effective annihilation cross-section is small which leads 
to large relic abundance, while for large $\sin \theta$ the relic abundance is small provided that the $\Delta m$ is big enough to avoid co-annihilation contributions.     


\begin{figure}[htb!]
$$
 \includegraphics[height=5.2cm]{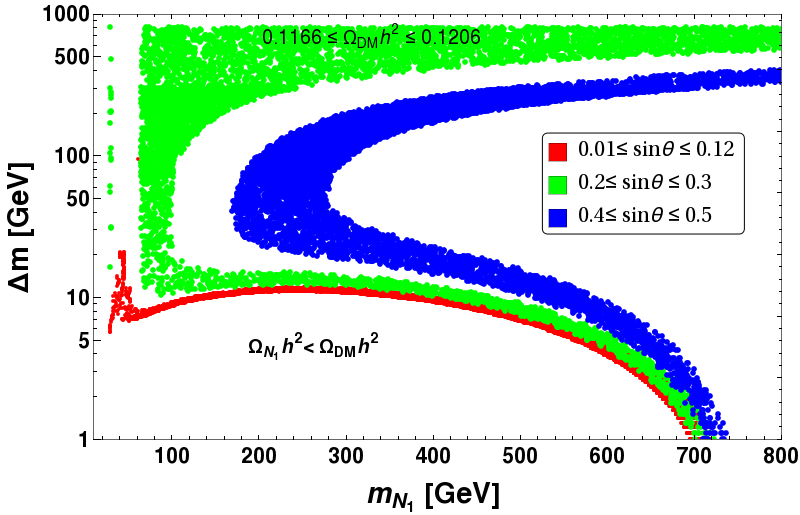}
$$
 \caption{Relic density allowed region are shown in $m_{N_1}-\Delta m$ plane for different range of range of $\sin\theta$: $\{0.01-0.12\}$ (Red),
$\{0.2-0.03\}$ (Green) and $\{0.4-0.5\}$ (Blue). }
 \label{fig:vf-relic}
\end{figure}

From Fig.~\ref{fig:vf-relic}, we see that for a wide range of singlet-doublet mixing ($\sin \theta $), we can get 
correct relic abundance in the plane of $m_{N_1}$ versus $\Delta m$. Different ranges of $\sin\theta$ are indicted by different color codes. 
To understand our result, we divide the plane of $m_{N_1}$ versus $\Delta m$ into two regions: (i) the bottom portion with small $\Delta m$, where 
$\Delta m$ decreases with larger mass of $N_1$, (ii) the top portion with larger mass splitting $\Delta m$, where $\Delta m$ increases slowly 
with larger DM mass $m_{N_1}$. In the former case, for a given range of $\sin \theta$, the annihilation cross-section decreases for large mass of 
$N_1$. Therefore, we need more co-annihilation contribution to compensate, which requires $\Delta m$ to decrease. This also imply that the region 
below to each colored zone is under abundant (small $\Delta m$ implying large co-annihilation for a given mass of $N_1$), while the region above 
is over abundant (large $\Delta m$ implying small co-annihilation for a given mass of $N_1$). To understand the allowed parameter space in region 
(ii), we first note that co-annihilation contribution is much smaller here due to large $\Delta m$, so the annihilation processes effectively 
contribute to relic density. Now, let us recall that the Yukawa coupling $Y \propto \Delta m \sin\theta$. Therefore, for a given $\sin \theta$, larger 
$\Delta m$ can lead to larger $Y$ and therefore larger annihilation cross-section to yield under abundance, which can only be tamed down to correct 
relic density by having a larger DM mass. Hence in case-(ii), the region above to each colored zone (allowed region of 
correct relic density) is under abundant, while the region below to each colored zone is over abundant. Thus the over and under 
abundant regions of both cases (i) and (ii) are consistent with each other. 

%
\subsection{Constraints on parameters from direct search of singlet-doublet leptonic dark matter}\label{direct_search}

\begin{figure}[htb!]
$$
 \includegraphics[height=3.5cm]{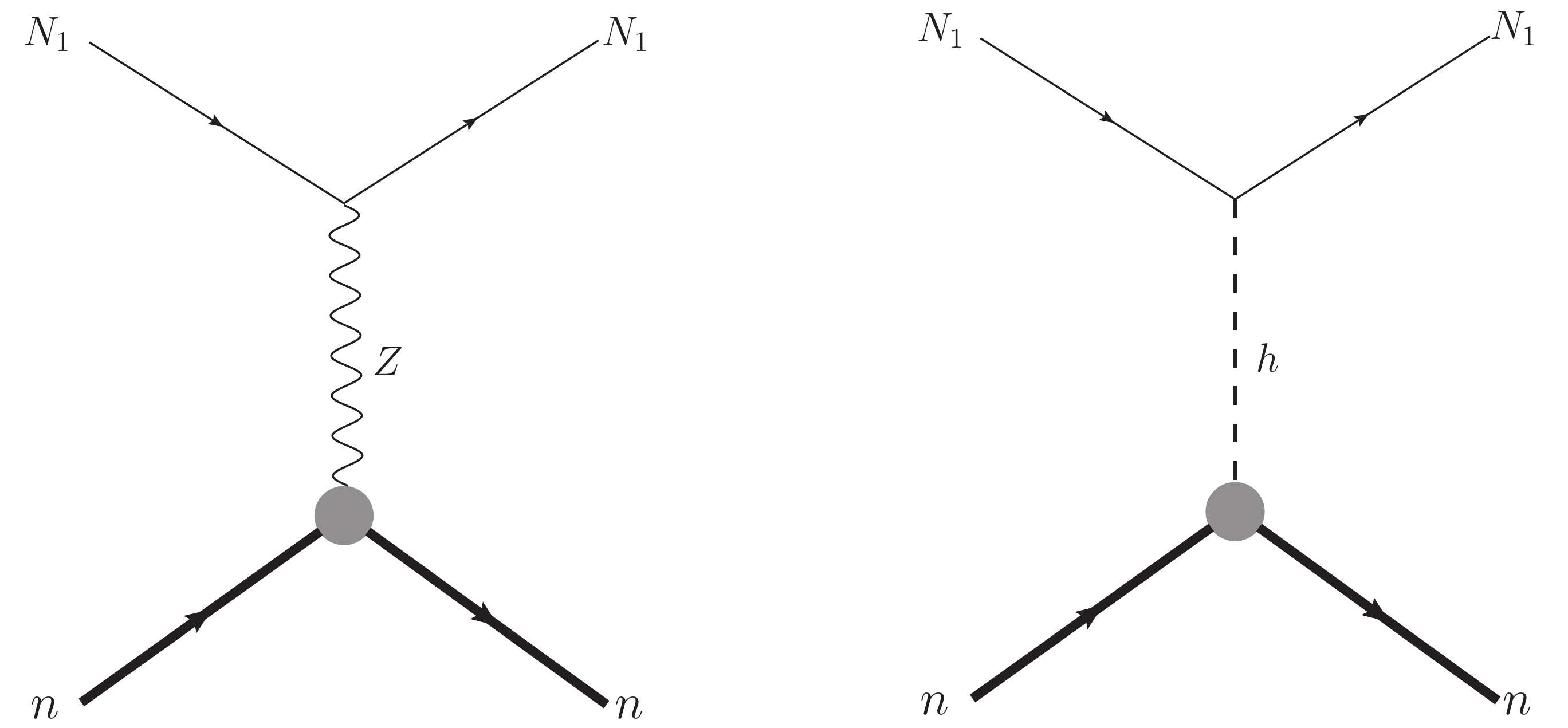}
 $$
 \caption{Feynman diagrams of SI direct detection of singlet-doublet $N_1$ DM.}
 \label{fig:fd-direct}
\end{figure}

Let us now turn to constraints on parameters from direct search of $N_1$ DM in terrestrial laboratories. Due to singlet-doublet 
mixing, the $N_1$ DM in direct search experiments can scatter off the target nucleus via Z and Higgs mediated processes as shown 
by the Feynman graphs in Fig.~\ref{fig:fd-direct}. The cross-section per nucleon for $Z$-boson mediation is given 
by~\cite{ Goodman:1984dc,Essig:2007az}
\begin{equation}\label{DM-nucleon-Z}
\sigma_{\rm SI}^Z = \frac{1}{\pi A^2 }\mu_r^2 |\mathcal{M}|^2   
\end{equation}
where $A$ is the mass number of the target nucleus, $\mu_r=M_1 m_n/(M_1 + m_n) \approx m_n$ is the reduced mass, $m_n$ is the mass of nucleon 
(proton or neutron) and $\mathcal{M}$ is the amplitude for $Z$-boson mediated DM-nucleon cross-section given by
\begin{equation}\label{Z-mediated-process}
\mathcal{M}= \sqrt{2} G_F [\tilde{Z} (f_p/f_n) + (A-\tilde{Z})] f_n \sin^2 \theta\,,
\end{equation}
where $f_p$ and $f_n$ are the interaction strengths (including hadronic uncertainties) of DM with proton and neutron respectively 
and $\tilde{Z}$ is the atomic number of the target nucleus. On the other hand, the spin-independent DM-nucleon cross-section per nucleon 
mediated via the exchange of SM Higgs is given by: 
\begin{equation}\label{scalar_mediated_crossssection}
\sigma_{\rm SI}^h=\frac{1}{\pi A^2}\mu_r^2  \left[ Z f_p + (A-Z)f_n \right]^2
\end{equation} 
where the effective interaction strengths of DM with proton and neutron are given by:
\begin{equation}\label{f-values}
f_{p,n} = \sum_{q=u,d,s}f_{Tq}^{(p.n)} \alpha_q \frac{m_{(p,n)}}{m_q} + \frac{2}{27} f_{TG}^{(p,n)}\sum_{q=c,t,b} \alpha_q \frac{m_{p.n}}{m_q}
\end{equation}
with 
\begin{equation}\label{alpha-value}
\alpha_q = \frac{ Y\sin 2\theta}{M_h^2} \left( \frac{m_q}{v}\right) \,.
\end{equation}
In Eq.~\ref{f-values}, the different coupling strengths between DM and light quarks are given by~\cite{DM_review1} $f^{(p)}_{Tu}=0.020\pm 0.004$, 
$f^{(p)}_{Td}=0.026\pm 0.005$,$f^{(p)}_{Ts}=0.118\pm 0.062$, $f^{(n)}_{Tu}=0.014\pm 0.004$,$f^{(n)}_{Td}=0.036\pm 0.008$,$f^{(n)}_{Ts}=0.118\pm 0.062$. 
The coupling of DM with the gluons in target nuclei is parameterized by 
\begin{equation}\label{Gluon-interaction}
f^{(p,n)}_{TG}=1-\sum_{q=u,,d,s}f^{(p,n)}_{Tq}\,. 
\end{equation}
Thus from Eqs. (\ref{scalar_mediated_crossssection},\ref{f-values},\ref{alpha-value},\ref{Gluon-interaction}) the spin-independent 
DM-nucleon cross-section is given to be:
\begin{equation}
\sigma_{\rm SI}^h=\frac{4}{\pi A^2}\mu_r^2 \frac{Y^2 \sin^2 2\theta}{M_h^4} 
\left[\frac{m_p}{v}\left( f_{Tu}^p + f_{Td}^p + f_{Ts}^p+ \frac{2}{9}f_{TG}^p\right) + \frac{m_n}{v} \left(f_{Tu}^n + f_{Td}^n + 
f_{Ts}^n+ \frac{2}{9}f_{TG}^n \right)\right]^2\,.
\end{equation}
In the above equation the only unknown quantity is $Y$ or $\sin 2\theta$ which can be constrained by requiring that $\sigma_{\rm SI}^h$ is 
less than the current DM-nucleon cross-sections.
\begin{figure}[htb!]
$$
 \includegraphics[height=7cm]{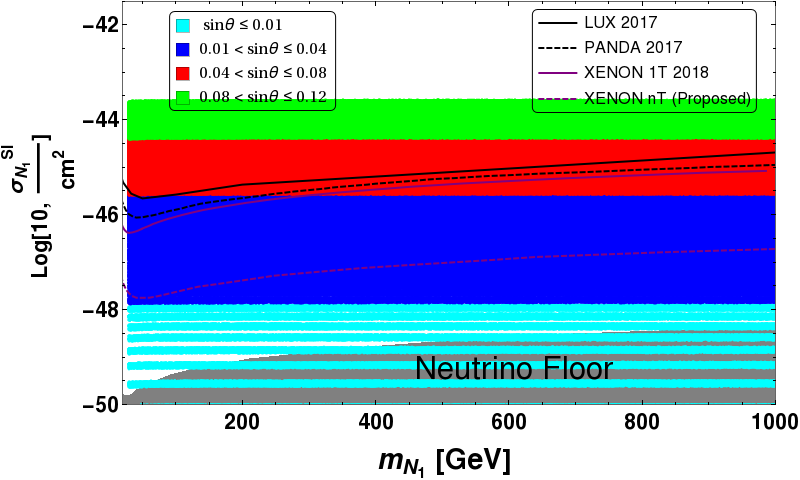}
 $$
 \caption{Spin independent direct detection cross-section $\sigma^{SI}$ as a function of DM mass $m_{N_1}$ for different 
range of $\sin\theta$: $\{0.001-0.01\}$ (Cyan),$\{0.01-0.04\}$ (Blue), $\{0.04-0.08\}$ (red) and $\{0.08-0.12\}$ (Green). The direct search limits from LUX, PANDA, XENON1T
are shown while that of the future sensitivity of XENONnT is also indicated. }
 \label{fig:directXsec}
\end{figure}
Now we make a combined analysis by taking both $Z$ and $H$ mediated diagrams taken into account together. In Fig.~\ref{fig:directXsec}, we show the 
spin-independent cross-section for $N_1$ DM within its mass range $m_{N_1}: 1-1000$ GeV. The plot is obtained by varying 
$\sin \theta$ within $\{0.001-0.12\}$ with $\sin \theta=\{0.001-0.01\}$ (Cyan), $\sin \theta=\{0.01-0.04\}$ (Blue), $\sin \theta=\{0.04-0.08\}$ (Red), 
$\sin \theta=\{0.08-0.12\}$ (Green). It clearly shows that the larger is $\sin \theta$, the stronger is the interaction strength 
(through larger contribution from $Z$ mediation) and hence the larger is the DM-nucleon cross-section.  Hence, 
it turns out that direct search experiments constraints $\sin \theta$ to a large extent. For example, we see that $\sin \theta \le 0.04$, for the DM 
mass $m_{N_1}> 300$ GeV. The effect of $\Delta m$ on DM-nucleon cross-section is less. However, we note that $\Delta m$ plays a dominant role 
in the relic abundance of DM. Approximately, $\sin \theta \le 0.05$ (Blue points) are allowed for most of the parameter space except for smaller DM 
masses. Cyan points indicate $\sin\theta <0.01$. The discrete bands correspond to specific values of $\sin\theta$ chosen for the scan and essentially one can consider 
the whole region together to fall into this category which evidently have sensitivity close to neutrino floor. Note here, the scanned points 
in Fig.~\ref{fig:directXsec}, do not satisfy relic abundance.  

\begin{figure}[htb!]
$$ 
 \includegraphics[height=5.2cm]{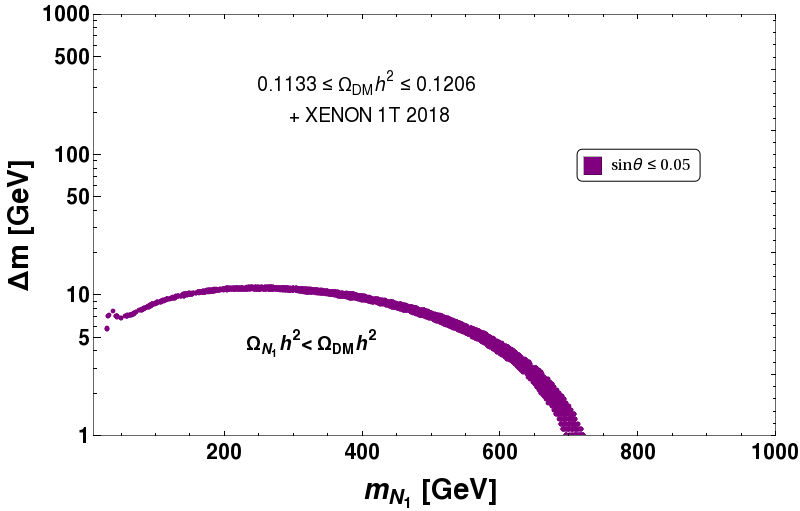} 
$$
 \caption{Relic density and direct detection ( XENON 1T 2018 ) allowed parameter space plotted in $m_{N_1}-\Delta m$ plane. }
 \label{fig:vf-relic-dd}
\end{figure}

Now let us turn to the parameter space, simultaneously allowed by observed relic density and latest constraints 
from direct DM search experiments such as Xenon-1T.  In Fig.~\ref{fig:vf-relic-dd}, we have shown the allowed parameter 
space again in $m_{N_1}-\Delta m$ plane. We see that null observation from direct search crucially tames down the relic density allowed parameter space to 
$\sin\theta < 0.05$ (Purple). Fig.~\ref{fig:vf-relic-dd} also shows that large singlet-doublet mixing, {\it i.e.} $\sin \theta \gtrsim 0.05$, allowed by correct relic density, 
is no more allowed by direct search limit in accordance with Fig.~\ref{fig:directXsec}.      

We finally summarize the DM analysis of singlet doublet case here. The model offers an interesting phenomenology to 
exploit singlet-doublet mixing ($\sin\theta$) in accordance with DM mass ($m_{N_1}$) and the splitting with charged fermion 
content ($\Delta m$) to yield a large available parameter space for correct relic density. However, due to $Z$-mediated process 
contributing to direct detection of singlet-doublet leptonic dark matter, a stringent constraint on $\sin\theta \le 0.05$ arises. 
This leads the DM to be allowed only in small $\Delta m$ region (as in Fig.~\ref{fig:vf-relic-dd}) to achieve 
correct relic density through co-annihilation processes. However, this constraint can be relaxed in presence of a scalar triplet as 
we discuss below. Moreover, the triplet can also give rise Majorana masses to light neutrinos (see section \ref{neutrino-mass-section}) 
through type-II seesaw to address DM and neutrinos in the same framework.

\section{Triplet extension of singlet-doublet leptonic dark matter }

\subsection{Pseudo-Dirac nature of singlet-doublet leptonic dark matter}
As discussed in section -\ref{sec:model}, the DM is assumed to be $N_1= \cos \theta \chi + \sin \theta N^0$ 
with a Dirac mass $m_{N_1}$. However, from  Eq.~\ref{yukawa_coupling} we see that the vev of $\Delta$ induces a Majorana mass to 
$N_1$ due to singlet-doublet mixing and is given by:
\begin{equation}\label{majorana-mass}
m_1=\sqrt{2} f_N \sin^2 \theta \langle \Delta \rangle \approx f_N \sin^2 \theta \frac{-\mu v^2}{\sqrt{2} M_{\Delta}^2 }\,.
\end{equation}  
Thus the $N^0$ has a large Dirac mass $m_{N_1}$ and a small Majorana mass $m$ as shown in the above Eq.~\ref{majorana-mass}. 
Therefore, we get a mass matrix in the basis $\{N_{1L}, (N_{1R})^c\} $ as:
\begin{equation}
{\mathcal M} =
\begin{pmatrix}
  m_1  &  m_{N_1} \\
       m_{N_1}   & m_1
\end{pmatrix} .
\end{equation}
Thus the Majorana mass $m$ splits the Dirac spinor $N_1$ into two pseudo-Dirac states $\psi_{1,2}$ with mass eigenvalues 
$m_{N_1} \pm m$. The mass splitting between the two pseudo-Dirac states $\psi_{1,2}$ is given by
\begin{equation}
\delta m_1 = 2 m_1 = 2\sqrt{2} f_N \sin^2 \theta \langle \Delta \rangle \,.
\end{equation} 
Note that $\delta m_1 << m_{N_1}$ from the estimate of induced vev of the triplet and hence does not play any role in the 
relic abundance calculation. However, the sub-GeV order mass splitting plays a crucial role in direct detection by forbidding 
the Z-boson mediated DM-nucleon elastic scattering. Now from Eq.~\ref{neutrino-mass} and (\ref{majorana-mass}) we see that the ratio: 
\begin{equation}\label{improved-coupling-ratio}
R=\frac{M_\nu} {m_1} = \frac{f_L} {f_N \sin^2 \theta}\,.
\end{equation}
Thus we see that for $R\sim 10^{-5}$ the ratio, $f_L/f_N \sim 10^{-3}$ if we assume $\sin \theta =0.1$, which is much larger than the singlet-doublet mixing 
being used in section \ref{direct_search}.

\subsection{Effect of scalar triplet on relic abundance and direct search of singlet-doublet dark matter}
We already have noted the diagrams that are present due to the addition of a scalar triplet for the ILD DM to 
freeze-out (see Section \ref{scalar_effect_DM}). The main features of having a additional scalar triplet in the singlet-doublet DM 
model is very similar to what we have discussed before in case of ILD DM. The additional freedom that we have in case of singlet-doublet 
leptonic DM is to play with the mixing parameter $\sin\theta$ and $\Delta m$.  

 \begin{figure}[htb!]
$$
 \includegraphics[height=5.0cm]{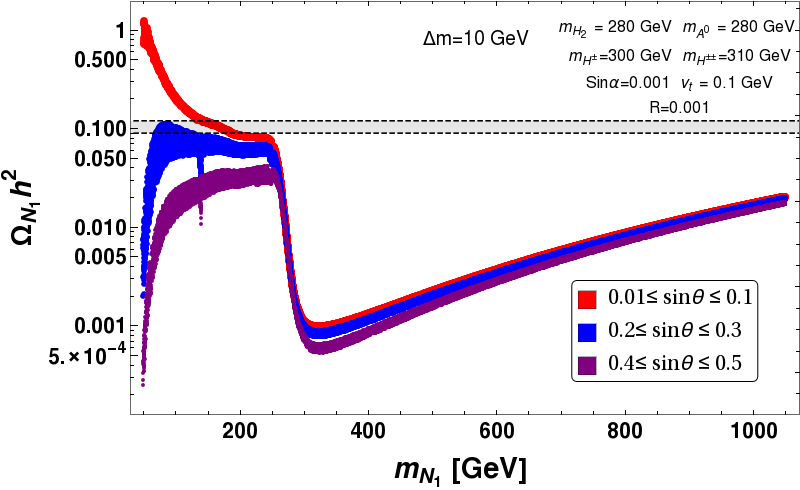} 
 \includegraphics[height=5.0cm]{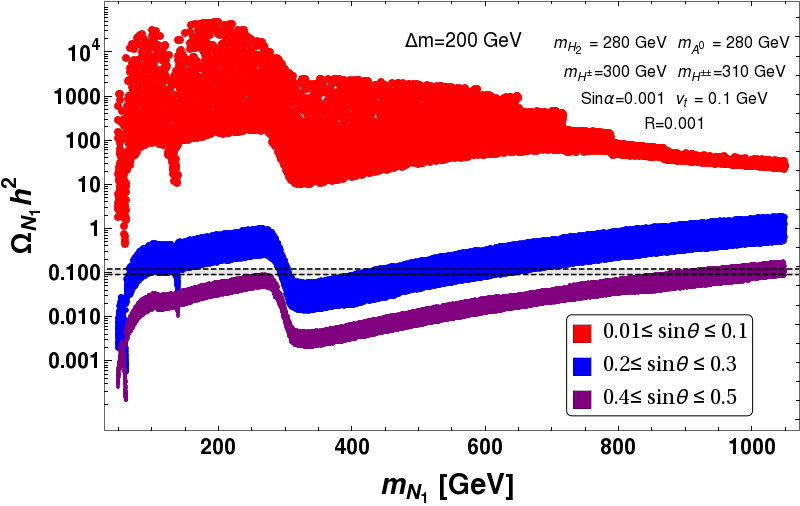}
$$
 \caption{Variation of relic density with DM mass $m_{N_1}$ keeping fixed region of $\Delta m$: $\Delta m = 10 $ GeV (left panel) and $\Delta m = 200 $ GeV (right panel) in presence of scalar triplet. Different color patches correspond to different $\sin\theta $ region : $ 0.01 \leq \sin\theta \leq 0.1 $ (Red), 
$ 0.2 < \sin\theta \leq 0.3 $ (Blue), $ 0.4 < \sin\theta \leq 0.5 $ (Purple). Correct relic density, $0.1166 \leq \Omega h^2 \leq 0.1206$ is shown by black dashed line. All the masses are in GeVs.}
 \label{fig:relic-mN-t1}
\end{figure}

 \begin{figure}[htb!]
$$
 \includegraphics[height=5.0cm]{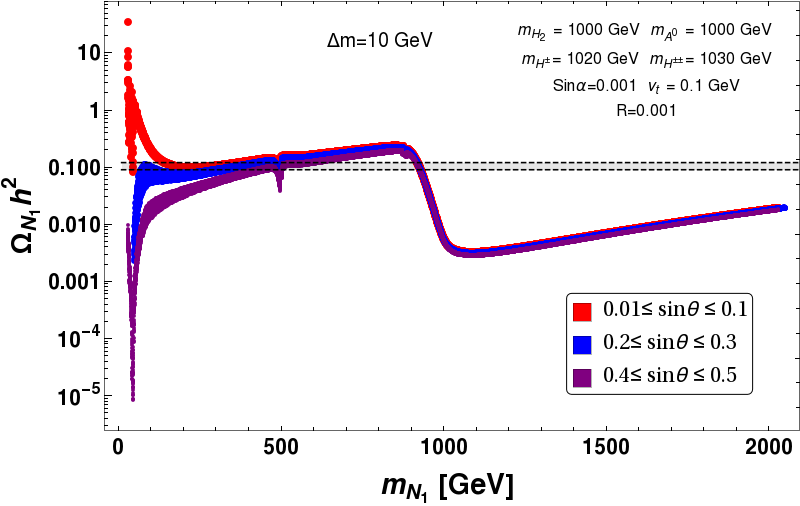} 
 \includegraphics[height=5.0cm]{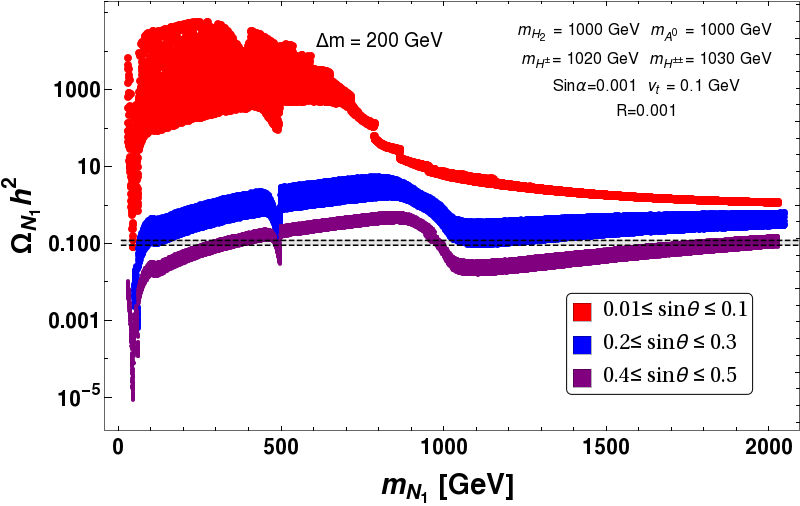}
$$
 \caption{Variation of relic density with DM mass $m_{N_1}$ keeping fixed region of $\Delta m$: $\Delta m = 10 $ GeV (left panel) and $\Delta m = 200 $ GeV (right panel) for heavy scalar triplet mass, $m_{H_2} = 1000$ GeV. Different color patches correspond to different $\sin\theta $ region : $ 0.01 \leq \sin\theta \leq 0.1 $ (Red), $ 0.2 < \sin\theta \leq 0.3 $ (Blue), $ 0.4 < \sin\theta \leq 0.5 $ (Purple). Correct relic density, $0.1166 \leq \Omega h^2 \leq 0.1206$ is shown by black dashed line. All the masses are in GeVs. }
 \label{fig:relic-mN-t2}
\end{figure}

Let us first study relic density as a function of DM mass in presence of scalar triplet. This is shown in Fig.~\ref{fig:relic-mN-t1}, where we choose two 
fixed values of $\Delta m=10, 200$ GeV in left and right panel respectively for a scalar triplet mass around 280 GeV. Different possible ranges of 
$\sin\theta$ are shown by different color codes. The main feature is again to see a drop in relic density near the value of the triplet mass, where the 
additional annihilation channel to the scalar triplet reduces relic density significantly. For small $\Delta m$, co-annihilation channels play an important 
part and therefore different mixing angles do not affect relic density significantly (compare left and right panel figures). Also due to large 
co-annihilation for small $\Delta m$ as in the left panel, the relic density turns out to be much smaller than the right panel figure where 
$\Delta m$ is large and do not offer the co-annihilation channels to be operative. An additional resonance drop at half of the triplet scalar mass is 
observed here ($\sim$ 140 GeV) due to s-channel triplet mediated processes.

A similar plot is shown in Fig.~\ref{fig:relic-mN-t2} with larger value of the scalar triplet mass $\sim$ 1000 GeV. Obviously the features from 
Fig.~\ref{fig:relic-mN-t1}, is mostly retained where the drop in relic density is observed around $\sim$ 1000 GeV and the drop is also smaller than what 
we had for smaller triplet mass.

\begin{figure}[htb!]
$$
 \includegraphics[height=5.0cm]{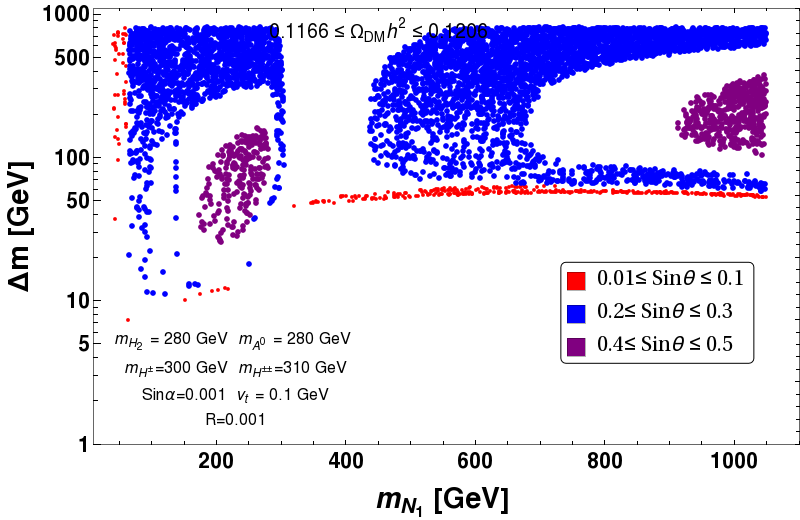} 
 \includegraphics[height=5.0cm]{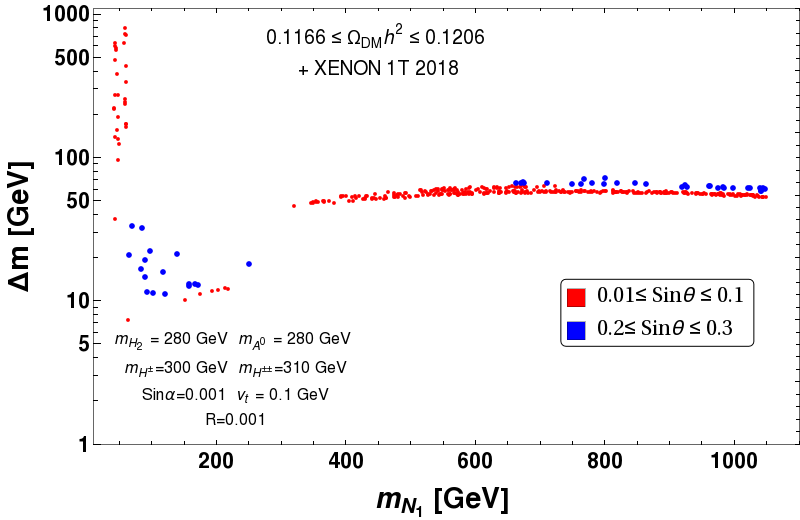}
$$
 \caption{Relic density  (left panel) and both relic density and direct search (XENON 1T 2018) (right panel) allowed parameter space plotted in $m_{N_1}-\Delta m$ plane with  
 different range of $\sin\theta $  : $ 0.01 \leq \sin\theta \leq 0.1 $ (Red), $ 0.2 < \sin\theta \leq 0.3 $ (Blue), $ 0.4 < \sin\theta \leq 0.5 $ (Purple). }
 \label{fig:RDD_VFDM_triplet}
\end{figure}

Relic density allowed parameter space of the model in $m_{N_1}-\Delta m$ plane is shown in left panel of Fig.~\ref{fig:RDD_VFDM_triplet}. 
The bottom part of the allowed parameter space is again due to co-annihilation. For small $\sin\theta \sim 0.1$ (red points), this is the only 
allowed parameter space except the resonance at the extreme left hand side. For larger $\sin\theta$, the resonance drops of doublet and triplet 
scalars also yield correct relic density. There is an under-abundant region when the triplet channel opens up, which is then reduced with larger 
DM mass. Therefore, it ends up with two different patches (both for blue and purple points) to be allowed below and above the scalar triplet mass. 
The direct search constraint in presence of scalar triplet thankfully omits the $Z$ mediated diagram due to the pseudo-Dirac splitting and allows 
a larger $\sin\theta \sim 0.3$. However, in addition to the Yukawa coupling (Y) initiated SM Higgs mediation, there is an added contribution from 
the heavy Higgs due to the doublet triplet mixing. We have already seen before that the effect of the additional contribution to direct search 
cross-section is small in the small $\sin\alpha$ limit with a moderate choice of $f_N$. Therefore, we have omitted such contributions in 
generating the direct search allowed parameter space of the model as shown in the right panel of Fig.~\ref{fig:RDD_VFDM_triplet}. This again 
depicts that the model in presence of scalar triplet earns more freedom in relaxing $\Delta m$ and $\sin\theta$ to some extent.

\section{Collider signatures}
Finally, we discuss the collider signature of the model, which can be subdivided into two categories: (i) Displaced vertex signature and 
(ii) Excess in leptonic final states. 
\subsection{Displaced Vertex signature}
In the small $\sin\theta$ limit, the charged inert fermion can show a displaced vertex signature.  If the mass difference between 
the $N^-$ and $N_1$ is greater than $W^-$ mass then $N^-$ can decay via a two body process.  But if the mass difference is smaller 
than $M_W$, then $N^-$ can decay via three body process say $N^- \rightarrow N_1 l^- \bar{\nu_l} $.  The three body decay width is 
given as \cite{bhattacharyaetal}:

\begin{equation}\label{N-decay}
\Gamma = \frac{ G_F^2 sin^2\theta}{24 \pi^3} m_N^5  I
\end{equation}
where $G_F$ is the Fermi coupling constant and $I$ is given as:
\begin{equation}\label{decay-rate}
I=\frac{1}{4}\lambda^{1/2}(1,a^2,b^2) F_1(a,b) + 6 F_2 (a,b)\ln  \left(\frac{2a}{1+a^2-b^2-\lambda^{1/2}(1,a^2,b^2)} \right) \,. 
\end{equation}
In the above Equation $F_1 (a,b)$ and $F_2 (a,b)$ are two polynomials of $a=m_{N_1}/m_N$ and $b=m_\ell/m_N$, where $m_\ell$ is the 
charged lepton mass. Up to ${\cal O}(b^2)$, these two polynomials are given by
\begin{eqnarray}
F_1 (a,b) &=& \left( a^6-2a^5-7a^4(1+b^2)+10a^3(b^2-2)+a^2(12b^2-7)+(3b^2-1)\right)\nonumber\\
F_2 (a,b) &=&  \left(a^5+a^4+a^3(1-2b^2)\right)\,.
\end{eqnarray} 
In Eq.~\ref{decay-rate}, $\lambda^{1/2}=\sqrt{1+a^4+b^4-2a^2-2b^2-2a^2b^2}$ defines the phase space. In the limit $b=m_\ell/m_N 
\to 1-a=\delta M/m_N$, $\lambda^{1/2}$ goes to zero and hence $I\to 0$. The life time of $N^-$ is then given by 
$\tau \equiv \Gamma^{-1}$. We take the freeze out temperature of DM to be $T_f= m_{N_1}/ 20$. Since the DM freezes out during radiation 
dominated era, the corresponding time of DM freeze-out is given by :
\begin{equation}
t_f= 0.301 g_\star ^{-1/2} \frac{m_{\rm pl}} {T_f^2} \, ,
\end{equation}
where $g_\star$ is the effective massless degrees of freedom at a temperature $T_f$ and $m_{\rm pl}$ is the Planck mass. Demanding 
that $N^-$ should decay before the DM freezes out (i.e. $\tau \lesssim t_f$) we get 
\begin{equation}\label{theta_constraint}
\sin \theta \gtrsim 1.1789 \times 10^{-5} \, \, \left(\frac{1.375\times
  10^{-5}} {I} \right)^{1/2} \left(  \frac{200 \rm GeV }{m_N}
\right)^{5/2} \left( \frac{g_\star}{106.75} \right)^{1/4} \left (
  \frac{m_{N_1}} {180 \rm GeV}\right)\,.
\end{equation}
The lower bound on the mixing angle depends on the mass of $N^-$ and $N_1$. For a typical value of
$m_N=200$ GeV, $m_{N_1}=180$ GeV, we get $\sin \theta \gtrsim 1.17 \times 10^{-5}$. Since $\tau$ is inversely
proportional to $m_N^5$, larger the mass, smaller will be the lower bound on the mixing angle.

\begin{figure}[htb!]
$$
 \includegraphics[height=5cm]{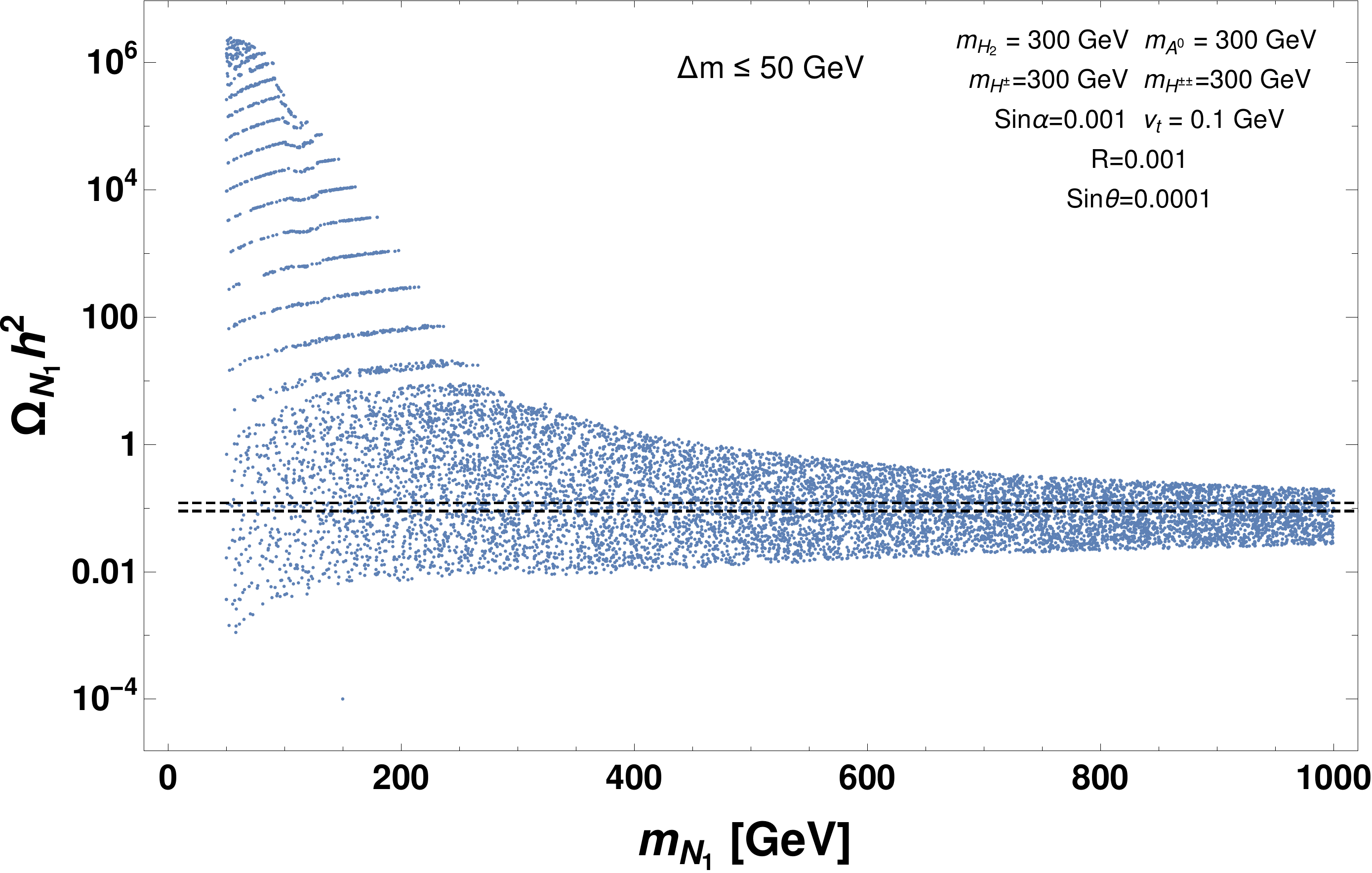}
 \includegraphics[height=5cm]{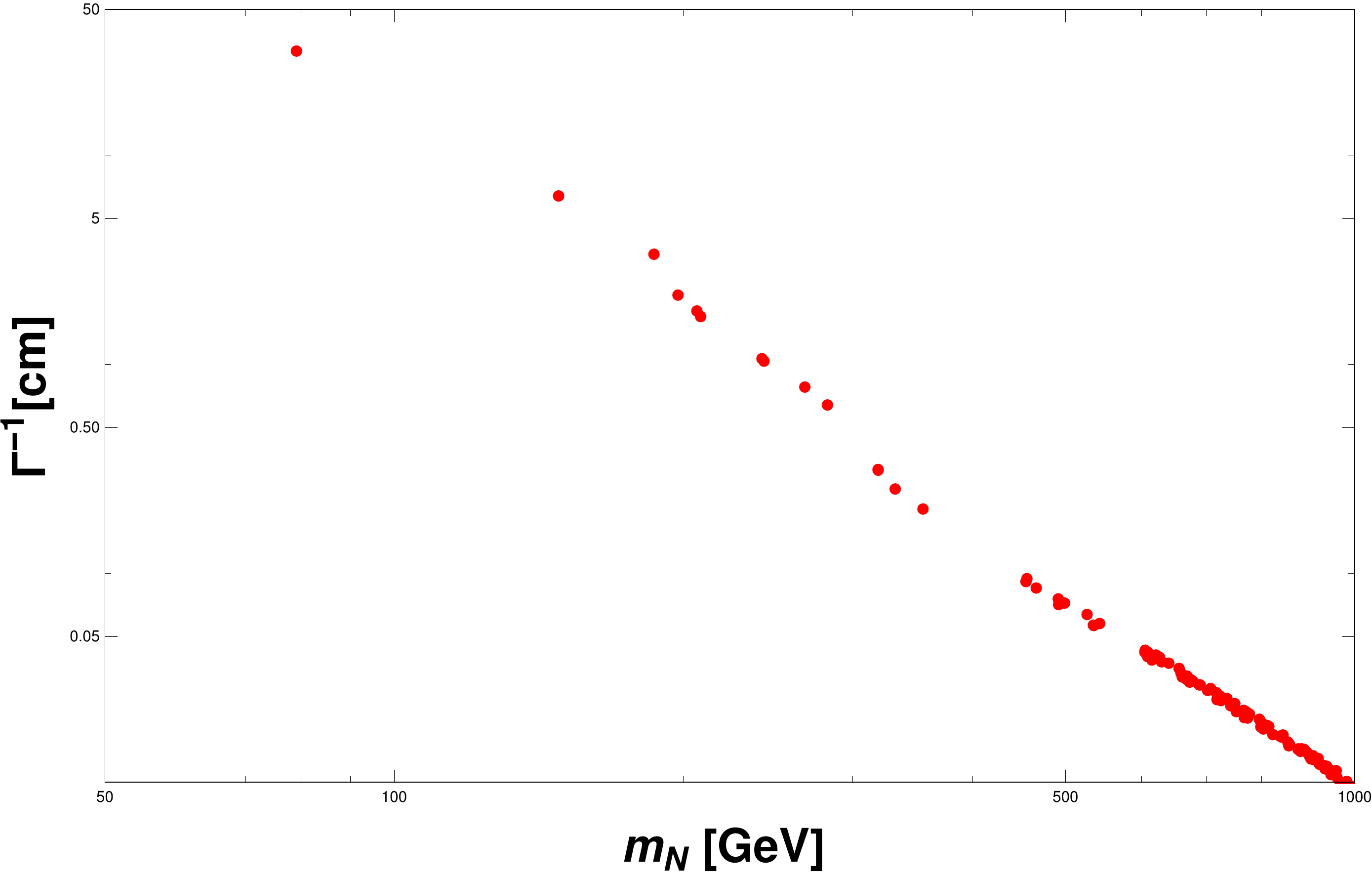}
$$
 \caption{Variation of relic density with DM mass $m_{N_1}$ keeping fixed $\Delta m \leq 50 $ GeV (left panel). Black dashed lines correspond to measured value of relic density by PLANCK.  Displaced vertex ($\Gamma ^{-1}$) is plotted as a function of $m_N$ (right panel).  For the displaced vertex we choose the set of parameters satisfying relic density from the left panel figure. }
 \label{fig:dv_rel}
\end{figure}

  To explore more whether we can get the relic abundance and displaced vertex simultaneously, we have shown in Fig.~\ref{fig:dv_rel} relic abundance as a function of DM mass keeping the mass splitting $\Delta M \leq 50$ GeV and $\sin \theta =10^{-4}$.  In this small mixing angle limit there are coannihilation channels (see Fig.~\ref{co-ann-3}) which are independent of $\sin \theta$ contributes to relic density.  We choose the set of points which are giving us the correct relic density and tried to find the displaced vertex value.  We have plotted in the right panel of Fig~\ref{fig:dv_rel} displaced vertex ( $\Gamma^{-1}$ ) as a function of $m_N$.  We can see that in the large mass of $m_N$, the displaced vertex is very small as expected as  $\Gamma^{-1}$  decreases with increase in mass. For larger mixing angles displaced vertex is suppressed.  Again $\sin \theta $ can not be arbitrarily small as shown in Eq.~\ref{theta_constraint}, so $\Gamma^{-1}$ will not be very large.
  
  \subsection{Hadronically quiet dilepton signature}

\begin{figure}[htb!]
$$
 \includegraphics[height=5.0cm]{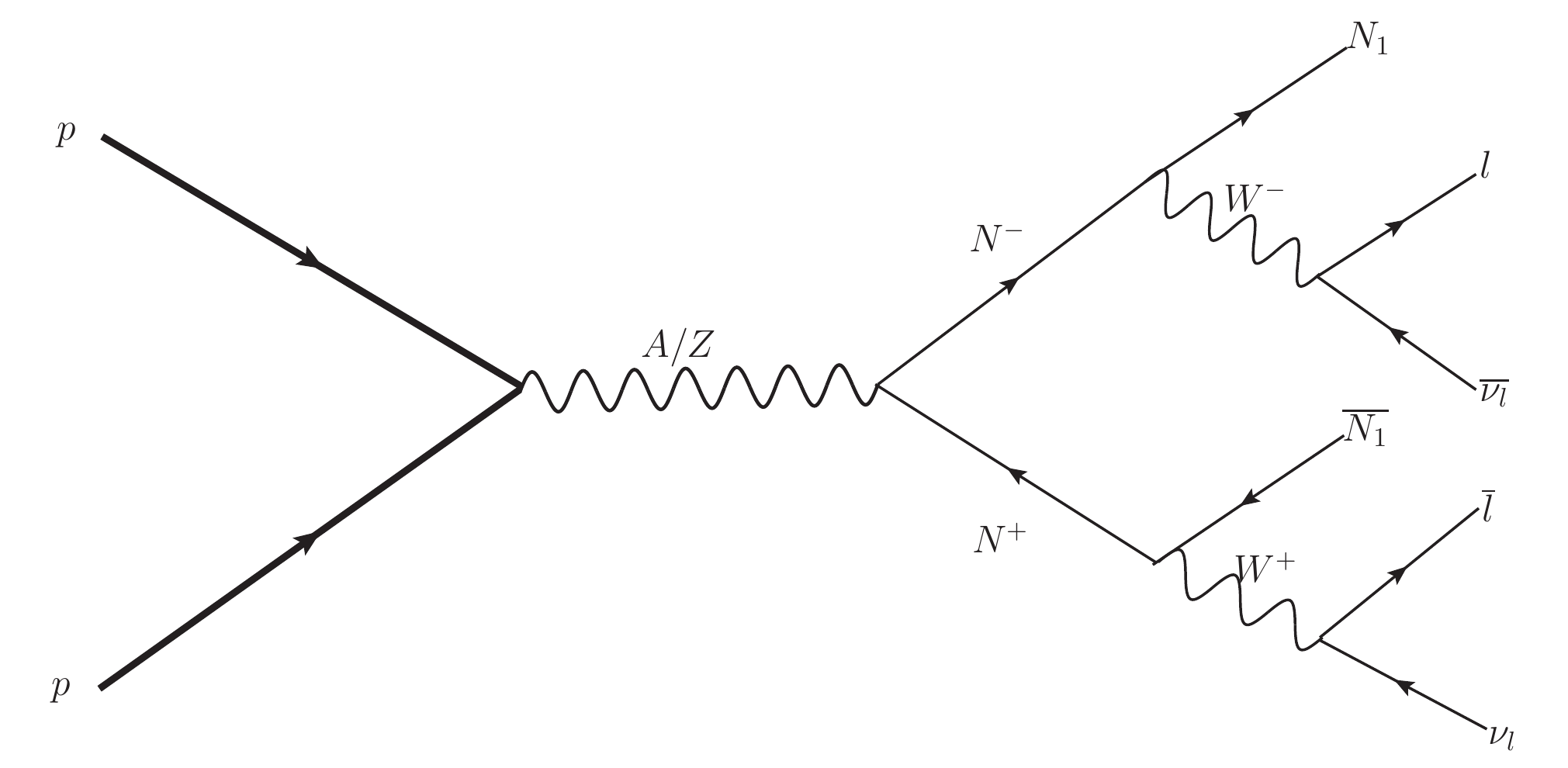}
$$
 \caption{Feynman diagram for producing hadronically quiet opposite sign dilepton plus missing energy ($\ell^+\ell^{-} +({\slashed E}_T)$) signal events at LHC. }
 \label{fig:LHC-production}
\end{figure}

Since our proposed scenario have one vector like leptonic doublet, there is a possibility of producing charge partner pair of the doublet $(N^+ ~ N^-)$ at proton proton collider (LHC). 
The decay of $N^\pm$ further produce leptonic final states through on-shell/off-shell $W^\pm$ mediator to yield opposite sign dilepton plus missing energy as is shown in 
Fig.~\ref{fig:LHC-production}. Obviously, $W$ can decay to 
jets as well, to yield single lepton plus two jets and missing energy signature or that of four jets plus missing energy signature. But, LHC being a QCD machine, 
the jet rich final states are prone to very heavy SM background and 
can not be segregated from that of the signal. We therefore refrain from calculating the other two possibilities here. A detailed analysis of collider signature of this model will be addressed in~\cite{Barman:2019tuo}.

\begin{table}[htb!]
\begin{tabular}{|p{1.0cm}|p{2.2cm}|p{0.8cm}|p{1.0cm}|p{2.0cm}|p{5.4 cm}| }
 \hline
 \hline
  BPs & {$\{~m_{N_1},\sin\theta~\}$}& $\Delta m $ &${\Omega_{N_1}} h^2  $ &$ \sigma_{N_1}^{SI}$ (in $cm^2$)  &  ~~~~~~~~~~~~~~\textrm{DM models} \\
 \hline
 \hline
 BP1&$\{~141,~0.03~\}$ & ~10~ & 0.1201 & $7.6 \times 10^{-47}$  & ~~~~~~~~~~Doublet Singlet  DM \\
 \hline
 BP2&$\{~50,~0.102~\}$ & ~147~ & 0.1165 & $1.2 \times 10^{-47}$  & Doublet Singlet DM $+$ Triplet Scalar\\
 \hline
  \hline
\end{tabular}
\caption{ DM mass, $\sin\theta$, $\Delta m = m_{N^\pm}-m_{N_1}$, relic density and SI direct search cross-sections of two benchmark points are mentioned for collider study. 
BP1 correspond to singlet doublet fermion DM scenario. BP2 depicts the case of singlet doublet DM model with an additional triplet in the picture. Note here that others parameters for 
BP2 remains same mentioned inset of the Fig.~\ref{fig:RDD_VFDM_triplet}.}
\label{tab:BPs}
\end{table}

Doublet-singlet fermion DM in absence or in presence of scalar triplet do not distinguish to yield a different final state from that of $\ell^+\ell^{-} +({\slashed E}_T)$ 
shown in Fig.~\ref{fig:LHC-production}. However, there is an important distinction 
that we discuss briefly here. $N^+ ~~ N^-$ production cross-section depends on the charge lepton masses and nothing else, 
leptonic decay branching fraction is also fixed. However, the splitting between DM ($N^0$) and its charged partner 
$(N^\pm)$ ($\Delta m = m_{N^\pm}-m_{N^0} $) is seen in the missing energy distribution. 
The signal can only be segregated from that of the SM background when the splitting is large and it falls within the heaps of SM background when 
$\Delta m$ is small. This feature can distinguish between the two cases of singlet doublet DM in presence and in absence of scalar triplet. 
To illustrate, we choose two benchmark points from two scenarios: $i)$ doublet singlet leptonic DM (BP1) 
in absence of scalar triplet and $ii)$ doublet singlet leptonic DM in presence of triplet (BP2), shown in Table~\ref{tab:BPs}. 
For BP1, we see that $\Delta m =10$ GeV, has to be very small because relic density and direct search 
(XENON 1T 2018) put strong constraint on $\Delta m$ ($\leq 12$ GeV). 
On the other hand, presence of triplet in this scenario can relax the situation to some extent, and one may choose large $\Delta m \sim 150$ GeV 
for low DM mass ($\sim 50$ GeV) and obey both relic density and direct search constraint, as indicated in BP2. Again, we note here, that such a low DM 
mass is still allowed by the Invisible Higgs data due to small $\sin\theta$ that we have taken here.

To study the collider signature of the model, we first implemented the model in {\bf FeynRule}~\cite{Alloul:2013bka}. 
To generate events files, we used {\bf Madgraph}~\cite{Alwall:2011uj} and further passed to {\bf Pythia} ~\cite{Sjostrand:2006za} 
for analysis. We have imposed further selection cuts on leptons ($\ell=e,\mu$) and jets as follows to mimic the actual collider environment: 

\begin{itemize}
 \item Lepton isolation: Leptons are the main constituent of the signal. We impose transverse momentum cut of 
 $p_T > 20$ GeV, pseudorapidity of $|\eta| < 2.5$ and separation cut $\Delta R \geq 0.2$ for separating from other leptons. 
 Additionally, $\Delta R \geq 0.4$ is required to separate the leptons from jets. The definition of separation in 
 azimuthal-pseudorapidity plane is $\Delta R = \sqrt{(\Delta \eta)^2+(\Delta \phi)^2}$.
 \item Jet formation and identification is performed in Pythia. We use cone-algorithm and impose that the jet initiator parton must have 
 $p_T \geq 20$ GeV and forms a jet within a cone of $\Delta R \leq 0.4$. Jets are required to be defined for our events as to have zero jets. 
 \end{itemize}

Using above basic cuts, we have studied hadronically quite opposite signed dilepton final states :  

\bea\label{collproc}
\nonumber
\rm{Signal ::}~~ {\ell^+\ell^{-} +({\slashed E}_T)} : ~~~~ p~p \rightarrow N^+ ~~ N^- , (N^- \rightarrow ~\ell^- ~\overline{\nu_\ell} ~ N_1), (N^+ \rightarrow \ell^+ ~{\nu_\ell} ~\overline{N_1}),~ ~~ \textrm where ~~~\ell = e ,\mu ~~ .
\eea
    
\begin{figure}[htb!]
$$
 \includegraphics[height=5.0cm]{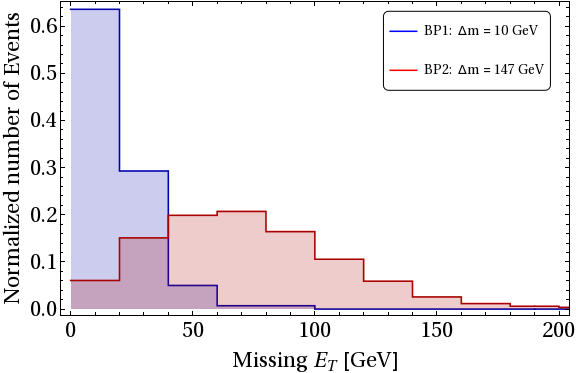}
 \includegraphics[height=5.0cm]{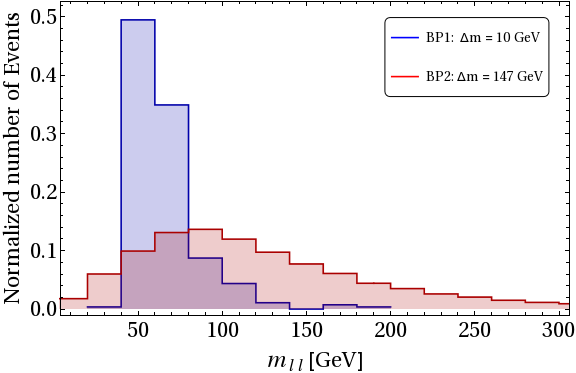}                             
$$
$$
\includegraphics[height=6.0cm]{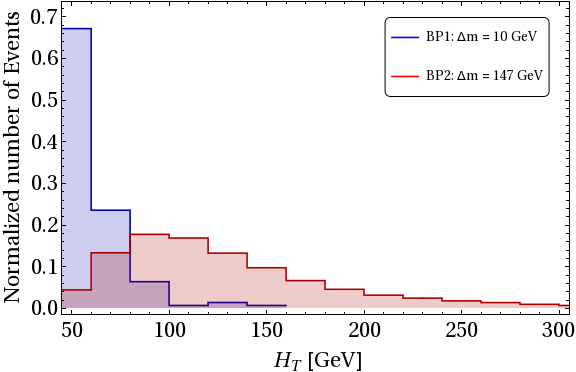} 
$$
 \caption{Missing energy ($\slashed{E}_T$), invariant mass of dilepton ($m_{ \ell \ell}$) and transverse mass ($H_T$) distributions of the hadronically quite dilepton signal events ($\ell^+\ell^{-} +{\slashed E}_T$) for C.O.M. energy, $\sqrt{s} = 14 \rm ~TeV$ at LHC . }
 \label{fig:signal-dist}
\end{figure}

The distribution of signal events with respect to missing Energy ($\slashed{E}_T$), invariant mass of OSD ($m_{\ell \ell}$) and effective momentum ($H_T$) is shown in 
Fig.~\ref{fig:signal-dist} respectively top left, top right and  bottom panel. 
We see that each of the distribution becomes flatter and the peak is shifted to higher energy value with larger $\Delta m$. 
As we have already mentioned that SM background yields a very similar distribution to that of BP1 and therefore can not be 
segregated from small $\Delta m$ cases. For details of background estimate and distribution, see for example, ~\cite{Bhattacharya:2018cgx}. 
Without further selection cuts, the signals constitute a very tiny fraction of hadronically quiet dilepton 
channel at LHC. To reduce SM background, further selection cuts must be employed: 

\begin{itemize}
 \item $  m_{\ell \ell} < |m_z - 15|$ and  $m_{\ell \ell} > |m_z + 15|$,
 \item $H_T > 100,~200$ GeV,
 \item $\slashed{E}_T > 100, ~200$ GeV.
\end{itemize}

We see that the signal events for BP1, after such a cut is reduced significantly, while for BP2, we are still left with moderately large number of events as shown in Table \ref{tab:Sig}.  

\begin{table}[htb!]
\begin{center}
\begin{tabular}{|c|c|c|c|c|c|c|c|c|}
\hline
BPs & $\Delta m$ (GeV) & $\sigma_{pp \to N^+N^-}$ (fb) & $\slashed{E}_T$ (GeV)  & $H_T$ (GeV) & $\sigma^{\text{OSD}}$(fb) & $N^{\text{OSD}}_{\text{eff}}=\sigma^{\text{OSD}}\times \mathcal{L}$ 
\\
\hline\hline 
 
 BP1 & 10 & 12.01 & $>$100  & $>$100 & $<$~ 0.0003 &$<~1$\\
 
 \cline{4-7}
 \hline 

& & & $>$100 & $>$100  & 0.711 &  71 \\

&&&& $>$200 & 0.250 & 25\\

\cline{4-7}

BP2 & 147 & 33.11 & $>$200  & $>$ 100    & 0.040  & 4\\

&&&& $>$200 & 0.039 & $4$ \\

\cline{4-7}
\hline  
\hline
\end{tabular}
\end{center}
\caption {Signal events for above mentioned  benchmark points  with $\sqrt{s}$ = 14 TeV at the LHC for the luminosity $\mathcal{L} = 100~fb^{-1}$ after $\slashed{E}_T$, $H_T$ and $m_{\ell\ell}$ cuts.} 
\label{tab:Sig}
\end{table}  

To summarize, we point out that singlet-doublet fermion DM can possibly yield a displaced vertex signature out of the charged fermion decay, thanks to small mass splitting and small 
$\sin\theta$, while due to the same reason, 
seeing an excess in leptonic final state will be difficult. On the other hand, the model where singlet-doublet fermion DM is extended with a scalar triplet satisfy relic density and 
direct search with a larger mass splitting between 
the DM and charged companion which allows such a case to yield a lepton excess to be probed at LHC, 
but the displaced vertex signature may get subdued due to this. The complementarity of the two cases will be elaborated in~\cite{Barman:2019tuo}. 
We also note that scalar triplet extension do not allow the fermion DM to have any mass to also accommodate large $\Delta m$. 
This is only possible in the vicinity of Higgs resonance. We can however, earn a freedom on 
choosing a large $\Delta m$ at any fermion mass value in the presence of a second DM component and see a lepton signal excess as has been pointed out in~\cite{Barman:2019tuo}.

\section{Conclusions and future directions}
Vector like leptons stabilised by a symmetry, provide a simple solution to DM problem of the universe. The relic density allowed parameter space provide a wide class of phenomenological 
implications to be explored in DM direct search experiments and in collider searches through signal excess or displaced vertex. In this article, 
we have provided a thorough analysis of different possible models in such a category. The results have been illustrated with parameter space scans, 
taking into account the constraints coming from non-observation of DM in present direct search data, constraints from electroweak precision tests, vacuum stability, invisible decay widths 
of Higgs and $Z$ etc. to see the allowed region where the model(s) can be probed in upcoming experiments.

We first reviewed the possibility of vector-like leptonic singlet $\chi$, doublet $N$ and their combination $\chi-N$ as viable 
candidates of DM. First we discussed about a vector-like singlet leptonic DM $\chi$. In this case, the DM can only couple to visible sector 
through non-renormalisable dimension-5 operator $\overline{\chi}\chi H^\dagger H/\Lambda$, where $\Lambda$ denotes a new physics (NP) scale. 
We find relic density allowed parameter space of the model requires $\Lambda$ to be 500 GeV or less for DM mass ranging between 100 GeV to 
500 GeV. However, the direct search cross-section for such $\Lambda$ is much larger than the constraints obtained from XENON1T data. Therefore, a singlet lepton is almost ruled out 
being a viable candidate of DM.    

We then discussed the possibility of neutral component of a vector-like inert lepton doublet (ILD) $N^0$ to be a viable DM acndidate. Since the doublet has only 
gauge interaction, the correct relic abundance can be obtained only at heavy DM mass around $\sim$ TeV. Again, the doublet DM suffers a stringent 
constraint from $Z$-mediated elastic scattering at direct search experiments. The relic density allowed parameter space therefore lies 
way above than the XENON1T bound of not observing a DM in direct search experiment. Therefore, an ILD DM alone is 
already ruled out. However, we showed that in presence of a scalar triplet, an ILD DM can be reinstated by forbidding the Z-mediated 
elastic scattering with the nucleons thanks to pseudo Dirac splitting. Due to additional interaction of ILD in presence of a scalar triplet, the mass of ILD DM is pushed 
to a higher side to achieve correct relic density. Moreover, the scalar triplet mixes with the SM doublet Higgs and paves a path for 
detecting the ILD DM at terrestrial laboratories. The presence of scalar triplet also yield a non-zero neutrino mass to three flavors of 
active neutrinos which are required by oscillation data. However, we noticed that the parameter space of an ILD DM is very limited to a 
very high mass due to its gauge coupling. 

We then searched for a combination of singlet $\chi$ and neutral component of doublet $N=(N^-, N^0)$ being a viable candidate of DM. 
This is possible if both of the fermion fields possess same $Z_2$ symmetry. They mix after electroweak symmetry breaking. In fact, we found 
that the appropriate combination of a singlet-doublet can be a viable DM candidate in a large parameter space spanning DM mass between Z resonance to 
$\sim$ 700 GeV. The singlet-doublet mixing plays a key role in deciding the relic abundance of DM as well as detecting it in terrestrial laboratories. 
In fact, we found that a large singlet component admixture with a small doublet component is an appropriate combination to be a viable candidate for DM,
particularly to meet direct search bounds ($\sin\theta \le 0.05$). However, it is difficult for a DM with large singlet component to yield correct relic density
due to small annihilation cross-section. Therefore, it has to depend heavily on co-annihilation to make up the small annihilation cross section, 
which in turn requires a small mass difference between the DM $N_1$ and its partners $N^\pm, N_2$. 
In particular, if the mixing angle is very small (around $\sin\theta \sim 10^{-4}$), the decay of NLSP ($N^- \to N_1 +l^-+\bar{\nu}_l$) gives a measurable displaced 
vertex signature at LHC, aided by a small mass difference of $N^-$ with the 
DM ($N_1$). However, this typical feature makes it difficult to identify 
any signal excess from production of the NLSP at LHC. 

The situation becomes more interesting in presence of a scalar triplet. The latter, not 
only enhances the allowed parameter space of singlet-doublet mixed DM (by allowing a larger mixing $\sin\theta \lsim 0.2$ and also a larger mass splitting between NLSP and DM), 
but also generates masses to three flavors active neutrinos via type-II seesaw. 
Presence of scalar triplet may also pave the path to discover this model through hadronically quiet dilepton channel at LHC. We would also like to add here that 
if the DM (singlet-doublet admixture) including the SM particles are charged under an additional flavour symmetry, 
say $A_4$, then non-zero value of $\theta_{13}$ can be obtained from the flavour charge of DM, which has been elaborated in \cite{Bhattacharya:2016rqj,Bhattacharya:2016lts}.  

\section*{Acknowledgement}
We thank Basabendu Barman for helpful discussions. SB would like to acknowledge the support from DST-INSPIRE faculty grant IFA-13-PH-57 at IIT Guwahati. PG also like to thank MHRD,Government of India for research fellowship.


\appendix
\section*{\large Appendix-A: Couplings of ILD dark matter with scalar triplets and SM particles}
\label{appn:int}
Trilinear vertices involving ILD and triplet Scalar:
\begin{align}
 \overline{(N^-)^c} N^- H^{++} ~~~~&:&~~~~~\sqrt{2} f_N \nonumber \\
 \overline{(N^-)^c} N^0 H^{+} ~~~~&:&~~~~f_N \nonumber \\
 \overline{(N^0)^c} N^0 H_1 ~~~~&:&~ - \sin\alpha ~f_N \nonumber \\
 \overline{(N^0)^c} N^0 H_2 ~~~~&:&~ - \cos\alpha ~f_N \nonumber \\
  \overline{(N^0)^c} N^0 A^0 ~~~~&:&~ - i ~f_N \nonumber \\
\end{align}

Trilinear vertices involving triplet scalars:
 \begin{align}
  H^{++}~H^{-}~H^{-} &:& \sqrt{2} v_t \lambda_3  ~~~~\propto ~~~~ 1/v_t    \nonumber \\
   H^{++}~H^{--}~H_2 &:& -\Big( 2 \cos\alpha~ v_t \lambda_2 - \sin\alpha~ v~\lambda_1 \Big) \nonumber \\  
                     &\xrightarrow{\sin\alpha \to 0}& -2 v_t \lambda_2 ~~~~\propto ~~~~ 1/v_t    \nonumber \\
  H^{++}~H^{--}~H_1 &:& -\Big(\cos\alpha~ v  \lambda_1 + 2 \sin\alpha~ v_t~\lambda_2 \Big) \nonumber \\                     
 \end{align}

Trilinear vertices involving ILD and SM particles:
\begin{align}
  \overline{N^0} N^- W^+ ~~~~&:&~~~  \frac{e_0}{\sqrt{2}\sin\theta_W} \gamma^\mu \nonumber \\
  N^- N^+  Z ~~~~&:&~~~ - \frac{e_0}{\sin2\theta_W}\cos2\theta_W \gamma^\mu \nonumber \\
  N^- N^+  A ~~~~&:&~~~ - e_0 \gamma^\mu \nonumber \\
  \overline{N^0} N^0 Z ~~~~&:&~~~  \frac{e_0}{\sin2\theta_W} \gamma^\mu 
\end{align}


\end{document}